\documentclass[a4paper,12pt]{article}

\usepackage{amsmath,amssymb}
\usepackage[latin1]{inputenc}
\usepackage{theorem}
\usepackage{latexsym}
\usepackage{amsfonts}
\usepackage{eufrak}
\usepackage{syntonly}
\usepackage[mathcal]{euscript}
\usepackage{pb-diagram}

\numberwithin{equation}{section}

    \newcommand{\no}{\nonumber}

    \def\be{\begin{equation}}
    \def\ee{\end{equation}}
    \def\pb{\bar\partial}

    \def\({\left(}
    \def\){\right)}
    \def\<{\left\langle\,}
    \def\>{\, \right\rangle}
    \def\[{\left[}
    \def\]{\right]}
\def\d{\text{d}}

\def\a{\alpha}
\def\b{\beta}
\def\g{\gamma}
\def\l{\lambda}
\def\o{\omega}

\def\lb{\bar\lambda}
\def\de{\delta}
\def\r{\rho}
\def\e{\epsilon}
\def\t{\theta}

    \def\p{\partial}

   \def\-{^}

\def\g{\gamma}

\def\r{\rho}

\def\l{\lambda}

\def\o{\omega}

\def\t{\theta}

\def\L{\Lambda}

\def\p{\partial}
\def\pb{\overline\partial}

\begin{document}

    \vskip1.1truein
    \centerline{ }
    \vskip 2.2cm
    
    \centerline{\Large{\bf A New Proposal for the Picture Changing
    Operators}} 
    \centerline{\Large{\bf in the Minimal Pure Spinor Formalism}}
    \vskip 1.3cm
     \centerline{Oscar A.
     Bedoya$^{*,}$\footnote{abedoya@fma.if.usp.br} and Humberto
Gomez$^{\dagger,}$\footnote{humgomzu@ift.unesp.br}}
\vskip 25pt
    \centerline{\sl 
$^{*}$Instituto de F\'isica-Universidade de S\~ao Paulo}
\centerline{\sl Caixa Postal 66.318 CEP
05314-970 S\~ao Paulo, SP, Brazil}
   \vskip 25pt
    \centerline{\sl 
$^{\dagger}$Instituto de F\'isica Te\'orica
UNESP - Universidade Estadual Paulista}
\centerline{\sl Caixa Postal 70532-2
01156-970 S\~ao Paulo, SP, Brazil}

     \vskip 1.3cm
    
\begin{abstract}
Using a new proposal for the ``picture lowering'' operators, we compute
the tree level scattering amplitude in the minimal pure spinor
formalism by performing the integration over the pure spinor space
as a multidimensional Cauchy-type integral. The amplitude will be
written in terms of the projective pure spinor variables, which turns
out to be useful to relate rigorously the minimal and non-minimal versions of 
the pure spinor formalism. The natural language for relating these
formalisms is the \v{C}ech-Dolbeault isomorphism. Moreover, the Dolbeault
cocycle corresponding to the tree-level scattering amplitude must be
evaluated in $SO(10)/SU(5)$ instead of the whole pure spinor space,
which means that the origin is removed from this space. 
Also, the \v{C}ech-Dolbeault language plays a key role for proving the 
invariance of the scattering amplitude under BRST,
Lorentz and supersymmetry transformations, as well as the decoupling
of unphysical states. We also relate the Green's function for the
massless scalar field in ten dimensions to the
tree-level scattering amplitude and comment about the scattering amplitude at higher
orders. In contrast with the traditional picture lowering operators, with our 
new proposal the tree level scattering amplitude is independent of 
the constant spinors introduced to define them and 
the BRST exact terms decouple without 
integrating over these constant spinors.

\end{abstract}

\newpage

\tableofcontents

\section{Introduction}

For more than a decade a manifestly super-Poincar\'e covariant formulation
for the superstring, known as the pure spinor formalism
\cite{BerkovitsCQS},
has shown to be a powerful framework in two branches. The first one is 
the computation of scattering amplitudes and the second one is the
quantization of the superstring in curved backgrounds
which can include Ramond-Ramond flux. The strength of the pure spinor
formalism resides precisely in the 
fact that it can be quantized in a manifestly super-Poincar\'e manner,
so this covariance is not lost neither in the scattering amplitudes
computation nor in the quantization of the superstring in curved
backgrounds.

Since the present paper is about the first branch, we will give a
brief description of what has been done in scattering amplitudes, not attempting to give a complete 
list of references.\\
One key ingredient in this formalism is a bosonic ghost $\lambda^\alpha$,
constrained to satisfy Cartan's pure spinor condition in 10 space-time 
dimensions \cite{Cartan}
\footnote{Even before  pure spinor were incorporated in the description for the
superstring, Howe showed that integrability along
pure spinor lines allowed to find the super Yang-Mills and
supergravity equations of motion in ten dimensions \cite{Howe}.}. 
The prescription for computing multiloop amplitudes was given in 
\cite{nathan minimal pure spinor}, where as in the RNS
formalism, it was necessary to introduce picture changing operators
(PCO's) in order 
to absorb the zero-modes of the pure spinor variables. Up to two-loops, various
amplitudes were computed in \cite{nathan 2 loop},
\cite{Carlos and Nathan equivalence} and \cite{Carlos four pt one loop}.
Later on, by introducing a set of non-minimal variables $\bar \l_\a$
and $r_\a$, an equivalent prescription
for computing scattering amplitudes was formulated in \cite{nathan
topological} and \cite{nathan nikita multiloops}. This last superstring description is known as the ``non-minimal'' pure
spinor formalism, in order to distinguish it from the former
``minimal'' pure
spinor formalism. With the non-minimal formalism, also were computed
scattering amplitudes up to two-loops \cite{Berkovits Mafra Anomaly},
\cite{mafra stahn}. Because of its topological nature, in the non-minimal 
version it is not necessary to introduce PCO's. Nevertheless, it is necessary 
to use a regulator.
The drawback of having to introduce this
regulator appears beyond two-loops, since it gets more complicated
due to the divergences coming from the poles contribution of the $b$
ghost \cite{grassi&vanhove} \cite{nathan&yuri}.

In this paper we will make a new proposal for the 
lowering picture changing operators, so in the
following, we will discuss some facts which led us to them. First of all,
the pure spinor condition defines a space, also called the pure spinor
cone. In the geometric treatment by 
Nekrasov \cite{beta gamma system} it was found that the pure spinor
space has non-vanishing first Pontryagin class, as well as
non-vanishing first Chern class; leading to anomalies in the pure spinor 
space diffeomorphism and worldsheet conformal symmetry respectively.
Nevertheless, the careful analysis in \cite{beta gamma system}
shows that these anomalies are cancelled by removing
the tip of the cone i.e the point $\lambda^\alpha =0$. Therefore, in order to have a
well defined theory, one should remove this point from the pure spinor
space. Secondly, according to Berkovits' prescription for computing
scattering amplitudes \cite{nathan minimal pure spinor}, in order to match the $11$ pure spinor
zero-modes in the minimal formalism, one should introduce $11$ lowering PCO's 
defined by $Y^I_{Old} =
C_\alpha ^I \theta^\alpha \delta(C_\alpha ^I \lambda^\alpha)$, for
$I=1{\ldots} 11$, where
$C^I _\alpha$ are constant spinors. In this definition, $\theta^\alpha$ are the fermionic
superspace coordinates. With the measure element also given in
\cite{nathan minimal pure spinor}, the integration over the pure
spinor zero-modes is performed without removing 
the point $\l^\a=0$\footnote{Here is worth to mention that the
geometric treatment of \cite{beta gamma system} was posterior to the multiloop
scattering amplitude prescription of \cite{nathan minimal pure spinor}.}. 
A third consideration that suggests for another treatment for the PCO's
comes from the higher
dimensional twistor transform using pure spinor; which allowed to
obtain higher-dimensional scalar Green's functions \cite{Hughston},
\cite{pure spinor like twistors}. As shown in
\cite{pure spinor like twistors}, in order to
integrate over the projective pure spinor space when $d>6$, it
was necessary to develop integration techniques because of the
non-linearity of the pure spinor conditions. Those integrations are always
integrations over cycles. These three considerations lead us to define a
new lowering $PCO$, given by $Y^I _{New}=$ ${C_\alpha ^I \theta^\alpha }\over{C
_\alpha ^I \lambda^\alpha}$. In this way the integration over the
pure spinor zero-modes is performed as a multidimensional Cauchy
integral, where the integration contours go around the anomalous point
$\l^\a =0$. As we will discuss in this paper, the new
PCO fulfill our requirement and as a bonus, allows to establish elegant 
relationships between the minimal and non-minal formalisms, as well as 
between the minimal formalism and the twistor space. Furthermore,
as was shown explicitly by tree and one-loop computations, given the
distributional definition of the PCO's $Y_{Old}^I$, the scattering amplitudes
depends on the constant spinors $C^I$; so for some choices of these
$C^I$'s, the theory is non-Lorentz invariant and the unphysical
states do not decouple \cite{skenderis}. 
These issues were solved by 
integrating over the $C^I$'s \cite{skenderis}, \cite{decouplingII}.
In contrast, with our PCO's proposal there is no need to integrate over
them. We will also formally prove that at tree level the unphysical
states decouple and that the scattering amplitude does not depend
on the constant spinors $C^I$'s.\\  Although we only consider tree-level scattering amplitudes in this
paper, we hope to make some progress at the loop level in the future,
by also redefining the raising PCO's.

The organization of this paper is as follows. In section $2$ we
briefly review the minimal pure spinor formalism, where we focus in
introducing the basic notation in order to write down the tree-level
scattering amplitude prescription of \cite{nathan minimal pure
spinor}.\\
In section $3$ we make our proposal
for the new set of PCO's and discuss the restriction that must be
imposed in order to have a well defined multidimensional Cauchy-type integral,
which will result in the condition that the integration cycles go
around the anomalous point of the theory $\l^\alpha =0$. It happens
that this condition is related to the specific choice of the constant
spinors $C^I$'s; so we will give two examples, one where the $C^I$'s
choice does not allows to define contours around the origin and
another one which does. It turns out that the first choice is the same
made in \cite{skenderis}, which will allow to make some comparisons.\\
In section $4$ we will compute the tree-level scattering amplitude. We
start by formally defining the integration contours. Then, we proceed
to write the amplitude using the projective pure spinor
coordinates. Using these coordinates we analyze the poles structure and
express the result of the scattering amplitude in terms of the {\it
degree} of the projective pure spinor space, which is useful 
to relate the minimal and non-minimal formalism. Although in 
\cite{nathan topological} was argued that taking the large scale limit
for a  regulator of the
non-minimal pure spinor formalism, the scattering amplitude behaves like the scattering
amplitude in the minimal formalism using the old PCO's and in
\cite{hoogeveen&skenderis} was shown that fixing the gauge of a
topological theory of gravity coupled to the worldsheet, the old PCO's
are equivalent to a particular regulator in the non-minimal side, we present
here a rigorous equivalence at tree level, in which the PCO's do not
correspond to any particular regulator in the non-minimal side.  Computations
of the kinematical factors in one and two-loops \cite{Berkovits Mafra Anomaly}
give evidence of the equivalence of the scattering amplitudes
prescription for the two formulations, as well as the equivalence obtained
in \cite{hoogeveen&skenderis}, so it
will be interesting to generalize the arguments presented in this
paper at the
loop level. The relationship between the minimal and non-minimal
formalisms will be established using the \v{C}ech-Dolbeault language; for
that reason we include a subsection about this subject.  \\
In section $5$ we show that the scattering amplitude is
invariant under BRST, Lorentz and supersymmetry transformations. Also
we show the decoupling of unphysical states. The Dolbeault formulation 
will be extremely useful, both for proving the invariances as well as the decoupling of
unphysical states. \\
In section $6$ we prove that the scattering amplitude is independent
of the constant spinors $C^I$'s. First we consider the simplest
non-trivial case, i.e pure spinor in four dimensions. Then, we
proceed to consider the ten dimensional case. The two cases are studied
differently; in four dimensions is straightforward and it teaches us what
should be done. Extending the four dimensional proof to ten dimensions
would be difficult, so we present a more elegant demonstration using the
\v{C}ech-Dolbeault language.\\
In section $7$ we will establish a direct relation between
pure spinor scattering amplitudes and Green's functions for massless 
scalar fields in ten dimensions.\\
In section $8$ we will comment about what should be done in order to
have a genus $g$ formulation for the scattering amplitude. In particular, 
we define a product for \v{C}ech cochains which would allow to get
a well defined scattering amplitude from the \v{C}ech point of view.
\\
Finally, we present some conclusions. The appendix contains several
simple examples cited through the paper, as well as some demonstration
of statements.

\section{Review of Minimal Pure Spinor Formalism}

In this section we will review the tree-level N-point amplitude
prescription given in \cite{nathan minimal pure spinor}. As noted in
\cite{skenderis}, the picture changing operators are not BRST closed
inside the correlators, leading to a more careful treatment for
decoupling the unphysical states.

In the pure spinor formalism, the type IIB superstring action is given by
\begin{equation}\label{action}
S=\frac{1}{2\pi}\int_{\Sigma_{g}}\d^{2}z\,\,\( {1\over2}\p
x^{m}\bar\p
x_{m}+p_{\a}\bar\p\theta^{\a} + \hat
p_{\a}\p\hat \t^{\a}-\omega_{\a}\bar\p\lambda^{\a} -
\hat \omega_{\a}\p\hat\l^{\a}
\),                                                            
\end{equation}
where $(x^m , \t^\a , \hat\t^\a)$ are coordinates for the type IIB ten-dimensional 
superspace. So, the indices run as follows: $m =
0{\ldots} 9,$ and $\a = 1,{\ldots} 16$. $(p_\a, \hat p_\a , \o_\a,
\hat\o_\a)$ are the
conjugate momenta to $(\t^\a, \hat \t^\a, \l^\a, \hat\l^\a)$, while $\l^{\a}$ and $\hat \l^\a$ 
satisfy the pure spinor condition in $d=10$
\begin{equation}
\l^\a (\g^{m})_{\a\b}\l^\b=0\, ,\qquad \hat\l^\a (\g^m)_{\a\b}\hat\l^\b = 0,
\end{equation}
where the matrices $\g^{m}$ are generators of the Clifford algebra in 
$\mathbb{R}^{10}$. From now on, we will focus on the left moving
variables in (\ref{action}), keeping in mind that all the subsequent
treatment is analogous for the right moving variables. 
From (\ref{action}) we can find easily the OPE's
\begin{equation} \label{mopes}
x^m (y) x^n (z) \to - \eta^{mn}\ln|y-z|^2 , \qquad p_\a (y) \t^\b (z) 
\to  {\de_\a ^\b \over (y-z)}.
\end{equation}
Nevertheless, the pure spinor condition does not allow a direct
computation of the OPE among $\l^\a$ and $\o_\a$. As discussed in
\cite{BerkovitsCQS}, the pure spinor spinor constraint must be solved, 
expressing $\l^\a$ in terms of $11$ unconstrained $U(5)$ variables $(\l^+ , \l_{ab} , \l^a)$,
where $a = 1,{\ldots} 5$ and $\l_{ab} = - \l_{ba}$. Although those
$U(5)$ fields are not manifestly Lorentz invariant, their OPE's are
equivalent to Lorentz invariant OPE's involving $\l^\a$, the pure
spinor Lorentz current $N_{mn} = {1\over 2}\o_\a (\g_{mn})^\a {}_\b\l^\b$ and the pure
spinor ghost number current $J = \o_\a \l^\a$. Furthermore, note that
because of the pure spinor condition, there is a gauge invariance
$\de \o_\a = \L_m (\l^m \l)_\a$, so $(\l^\a , \o_\a)$ must appear
precisely in the gauge invariant combinations $N_{mn}$ and $J$.The OPE's involving the pure spinor are 

\be \label{gopes} N_{mn}(y) \l^\a (z) \to {1\over 2}{(\g_{mn}\l)^\a (z)\over {y-z}},
\qquad J(y) \l^\a (z) \to {\l^\a (z) \over y-z},
\ee
\be\nonumber N^{mn} (y) N^{pq}(z) \to -3 {\eta^{q[m}\eta^{n]p} \over
(y-z)^2} + {\eta^{p[n}N^{m]q}(z) - \eta^{q[n}N^{m]p}(z)\over y-z},
\ee
\be\nonumber
J(y) J(z) \to {-4 \over (y-z)^2} , \qquad J(y) N^{mn}(z) \to regular,
\ee
\be\nonumber
N_{mn}(y) T(z) \to {N_{mn}(z) \over (y-z)^2}, \qquad J(y) T(z) \to {-8
\over(y-z)^3} + {J(z)\over (y-z)^2},
\ee
where $T$ is the energy momentum tensor 
\be T =  {1\over 2} \p x^m \p x_m +  p_\a \p
\t^\a -  \o_\a \p \l^\a.
\ee
Note that the ghost number current and pure spinor Lorentz current
have levels $-4$ and $-3$ respectively. Furthermore, the ghost number
current has anomaly $-8$, which should be kept in mind for defining
scattering amplitudes.

Besides the pure spinor $\l^\a$, another key ingredient in this
formalism is the BRST charge $Q = \oint dz \l^\a d_\a$, where 
\be\label{dconstraint}
d_\a = p_\a - {1\over 2} (\g^m \t)_\a \p x_m - {1\over 8} (\g^m
\t)_\a (\t \g_m \p \t)
\ee
is the supersymmetric Green-Schwarz constraint. Given the
supersymmetric combination $\Pi^m = \p x^m + {1\over 2} (\t \g^m \p
\t)$, $d_\a$ has the following OPEs
\be \label{dopes}
d_\a (y) d_\b(z) \to  {\g^m _{\a\b}\Pi_m \over y-z},
\qquad d_\a (y) \Pi^m (z) \to {\g^m _{\a\b} \p \t^\b \over (y-z)},
\ee
\be
d_{\a} (y) f(x(z),\t(z)) \to {D_\a f(x(z),\t(z)) \over y-z},
\ee
where $D_\a = {\p \over \p \t^\a} + {1\over 2} (\t \g^m)_\a {\p \over
\p x^m}$ is the supersymmetric derivative. From the first OPE in
(\ref{dopes}) it can easily be checked that $Q^2 =0$ because of the
pure spinor condition. 

Vertex operators in the pure spinor formalism for the massless states are given
by ghost number one and conformal weight zero objects. $V = \l^\a
A_\a (x,\t)$ is the most general object satisfying both conditions.
Since $V$ must be in the cohomology of $Q$, then  

\be D_{(\a } A_{\b)}  = \g^m _{\a\b} A_m , \label{symeom}\ee
where the indices on the left hand side are symmetrized and $A_m
(x,\t)$ is some superfield. The gauge invariance $\de V = Q \L$
implies that $\de A_\a = D_\a \L$ and $\de A_m = \p_m \L$, which are the
gauge invariance for super-Yang-Mills, while the equation
(\ref{symeom}) is the super-Yang-Mills equation of motion. In order to
define scattering amplitudes, the integrated version of the vertex
operators $U$ is also needed. In the case of the massless vertex operator,
through $\p V = Q U$ the explicit form of $U$ is found
\be
U = \p \t^\a A_\a (x,\t) + \Pi^m A_m (x,\t) + d_\a W_\a (x,\t) +
{1\over 2}N^{mn} F_{mn}(x,\t),
\ee
where $W^\a$ and $F_{mn}$ are the spinor and vector super-Yang-Mills
superfield strengths respectively.

Since the pure spinor $\l^{\a}$ have 11 zero
modes\footnote{the spinors $\o_{\a}$, $p_{\a}$ and
$\theta^{\a}$ have $11g$, $16g$ and $16$ zero modes respectively for the 
Riemann  surface of genus $g$} in any Riemann surface $\Sigma_g$, it is necessary
to absorb them when computing scattering amplitudes. The manner that they are
absorbed is by introducing 11 PCO's
\begin{equation}\label{PCO}
Y_{C}^I= C^I_{\a}\t^{\a} \delta(C^I_{\a}\l^{\a}), \,\,\, I =
1,{\ldots} 11, 
\end{equation}
inside the scattering amplitude \cite{nathan minimal pure spinor},
which for N-points at tree level is 
\begin{equation}\label{scattering prescription}
\mathcal{A} = \langle V_1 (z_1) V_2 (z_2) V_3 (z_3) \int dz_4 U_4
(z_4) {\ldots}  \int d z_N U_N (z_N) Y_C ^1 (y_1) {\ldots}  Y_C ^{11}
(y_{11})\rangle.
\end{equation}
So, to perform this computation the OPE's (\ref{mopes}), (\ref{gopes})
and (\ref{dopes}) are used to integrate over the non-zero modes,
remaining an integral over the zero modes of the pure spinor and
$\t^\a$\footnote{The integration over the fields $x$ is treated in
detail in D'Hoker and Phong \cite{dhokerperturbation}, we will not focus in those integrals.}.
Note that $\delta(C_{\a}\l^{\a})$ is a Dirac's delta function and
$C^I_{\a}$ is a constant projective spinor, which can be thought as a
point in the $\mathbb{C}P^{15}$ space. Although in 
\cite{nathan minimal pure spinor} it was argued that the scattering
amplitude was independent of the constant spinors $C_\a ^I$, it was
later found in \cite{skenderis} that indeed the amplitude depends on
the choice of $C_\a ^I$ and also that $Q$ exact states do not decouple.
In the next section, we propose a new picture operators, which does not 
have that disadvantage.

\section{The New Picture Changing Operators}

In this section we introduce the new lowering picture changing operators.
In particular, we will discuss why with this new
proposal for the picture changing operators, the origin must be removed from 
the pure spinor space. This will allow to write the tree-level scattering 
amplitude in terms of the projective pure spinor variables in the
following section, and also, 
to find a relationship with the twistor space in section \ref{relation
with twistor}.
In the end of the present section we give examples of choices for the
constant spinors $C^I$'s and discuss their implications.


\subsection{The New Proposal for the PCO's}

In this subsection we discuss some motivations which led us to define
new lowering PCO's.

The bosonic spinor $\lambda^\alpha$, constrained to satisfy the pure
spinor condition $\lambda \gamma^m \lambda =0$, constitutes an
interesting and non-trivial complex space, which will be denoted through this paper
as the pure spinor space $PS$ or the pure spinor cone. Since the coordinates 
$\l^{\a}$ of such space are holomorphic, the integral 
\begin{equation}
\int [\d\l] \delta(C\l) f(\l) 
\end{equation}
is only well defined if the domain of integration, i.e the cycles
around which we integrate are known. 
Moreover, as shown by Nekrasov \cite{beta gamma system}, the tip of
the pure spinor cone $\l^{\a}=0$ introduces anomalies. Then, by
removing this point of the pure spinor space, the theory is anomaly
free \footnote{This is the unique singular point of the pure spinor space because it is a complex cone over the smooth manifold $SO(10)/U(5) \subset \mathbb{C}P^{15}$.}. 
This is simple to see if one computes the de-Rham cohomology of the
pure spinor minus the origin space 
\begin{equation}
H^{i}\(PS\smallsetminus\{0\}\)=\mathbb{R},\quad \text{for }i=0,6,15,21,
\end{equation}
so the first Chern class and second Chern character both vanish,
$c_{1}(PS\smallsetminus\{0\})=ch_2(PS\smallsetminus\{0\})=0$ and therefore the
theory is anomaly free. This motivates us to make a new proposal for
the PCO's, in such a way that the tip of the cone is naturally
excluded. Furthermore, Skenderis and Hoogeveen \cite{skenderis}
showed that the scattering amplitude, as formulated in
\cite{nathan minimal pure spinor}, depends on the choice of the constant
spinors $C^{I}_\a$, having to integrate over them in order to obtain a
manifestly Lorentz invariant prescription. Nevertheless, as
we will show in section \ref{independence of Cs}, the scattering amplitude 
will not depend on the constant spinors using the new PCO's.

Our proposal, which seems to be the most natural, is to define the
PCO's as
\begin{equation}\label{nPCO}
Y_C^I=\frac{C^I_\a\t^\a}{C_\a^I\l^\a}, \qquad I = 1,{\ldots} 11,
\end{equation}
where $C^{I}_\a$ are again constant spinors. Just like the standard
PCO's (\ref{PCO}), this new PCO's are not
manifestly Lorentz invariant. Also, since $Q Y^I _C =1$, they are not
BRST closed. Using these PCO's it will be necessary to modify the usual 
BRST charge of the minimal formalism in order to have a global
description, as will be done in section \ref{symmetries of the scattering amplitude}. Then, we will be able to
show in that section that the scattering amplitude is BRST, Lorentz
and supersymmetric invariant.

Since we want to integrate over the pure spinor zero modes, basically as a
multi-dimensional Cauchy's integral, we will start by
considering the analogous of the poles. This role will be played by
the denominators of the PCO's, so we start by defining the functions
\begin{equation}
f^{I}(\l)\equiv C^{I}_\a\l^{\a},
\end{equation}
which map the pure spinor space to the complex numbers for each value
of $I=1,...,11$, i.e  $f^{I}:PS\,\,\rightarrow\,\,\mathbb{C}$. 
Given these functions, secondly we define the hypersurface
``$D_I$'' as the subspace $f^I=0$
\begin{equation}
D_I=\{\l^\a\in PS: \,C^I_\a\l^\a=0\}.
\end{equation}
In order to have
a well defined integration over the pure spinor space inside the 
scattering amplitude, it is necessary to impose the condition  that the
intersection between the $D_I$'s satisfies $D_1\cap D_2...\cap
D_{11}=\{$finite number of points$\}$ in order to have a Cauchy like
integral over $PS$. Just to be more explicit,
using the $U(5)$ decomposition \cite{BerkovitsCQS} for writing the pure 
spinor constraint, we
require that the 16 equations
\begin{equation}\label{5constrains1}
f^I=0, \quad \text{and}\quad
\chi^a =\l^{+}\l^{a}-\frac{1}{8}\e^{abcde}\l_{bc}\l_{de}=0,\quad \text{with}\quad
I=1,2,...,11;\quad a,b,c,d,e=1,2,...,5,
\end{equation}
intersect in a finite number of points. However, the five equations
$\chi^a =0$ must be taken carefully
because with only this condition, there are more singular points besides 
$\l^{\a}=0$. Therefore, a second set of equations $\zeta_a = \l^b
\l_{ba}=0$,
must be taken into account. Both set of conditions 
$\chi^a =0$ and
$\zeta_a = 0$ come from the $U(5)$ decomposition of the pure spinor
condition \cite{BerkovitsCQS}. Although the first one implies in the
second one when $\l^+ \neq 0$, as will be explained with one example
in appendix \ref{singularities appendix}, disregarding the
second one could lead to a not well defined tangent space at every
point of $PS$. Therefore, both conditions will be considered when we
construct an example for the $C^I$'s in subsection \ref{examplesCs}.

To demand that the constant          
spinors $C^I$'s are linearly independent in $\mathbb{C}^{16}$ is not enough 
to obtain an intersection in a finite number of points. However, clearly the 
origin $\{0\}$ is a common point in the intersection of all the
hypersurfaces $D_I$. We claim that the only common point
between the hypersurfaces $D_{I}$'s is the origin because precisely, it is 
the unique anomalous or singular point of the theory. Therefore, the 
integration contours are those that go around the origin of the pure
spinor space. 

In the following subsection we give an example for
the constant spinors $C^I _\a$ which allow for such a type of
intersection.

\subsection{Some Examples for the Constant Spinors $C^I
_\alpha$}\label{examplesCs}
In this subsection we will consider two examples. One where the
$C^I$'s are linearly independent, although do not allow for an
intersection of the hypersurfaces $D_I$ in a finite
number of points. In the second example, we construct a set of $C^I$'s
which intersect just in the origin.

\paragraph{First Example} We will make the same choice for the $C^I$'s
as in \cite{skenderis}, so we consider this
example basically to establish a comparison with this reference. Let the $C^I_\a$'s be
in the $U(5)$ representation:
$$
C^I_\a =  (C^{I}_+,C^{I,ab},C^I_a) \qquad a,b=1,...,5,
$$
where $ C^{I,ab}=-C^{I,ba}$. Making the choice of \cite{skenderis}
\begin{equation}\label{choice C}
C^1_\a = \de^+_\a,\quad C^{2,ab}=\delta^{[a}_{1}\de^{b]}_{2}\,, \cdot\cdot\cdot\,,
C^{11,ab}=\delta^{[a}_{4}\de^{b]}_{5}, \quad\text{all other}\quad C^{I}_\a=0,
\end{equation}
the functions $f^I$'s are
\begin{equation}
f^1=\l^+ \,,\quad f^2=\l_{12}\,, \quad f^3=\l_{13}\,,\cdot\cdot\cdot\,,\,f^{11}=\l_{45}.
\end{equation}
With the conditions $f^I=0$ the pure spinor constraints are
satisfied identically, but the parameters 
$\l^a$'s are free, therefore the intersection is the space $\mathbb{C}^5$, in
contrast with our requirement of intersecting just in the origin.
With this choice we can ``naively'' compute the three point tree
level amplitude only locally ($\l^+\neq 0$), obtaining the same result as in \cite{skenderis}
as we will review below. The answer will not be Lorentz invariant. For
$3$-points the computation is as follows:

\begin{eqnarray} \nonumber
\mathcal{A} &=&  \langle \l^\a A_{1\a} (z_1) \l^\b A_{2\b} (z_2) \l^\g
A_{3\g} (z_3) Y^1 (z) {\ldots} Y^{11} (z)\rangle \\
&=& \int_\Gamma [\d\l] \int \d^{16} \theta \l^\a \l^\b \l^\g
f_{\a\b\g}(\t) \frac{C^1 \t}{C^1 \l }  {\ldots}\frac{C^{11} \t}{C^{11}
\l }\nonumber \\
&=& \int_\Gamma [\d\l] \int \d^{16} \theta \l^\a \l^\b \l^\g
f_{\a\b\g}(\t) \frac{\t^{+}}{\l^{+}}\frac{\t_{12}}{\l_{12}}
{\ldots}\frac{\t_{45}}{\l_{45}}\nonumber \\
&=& \int_\Gamma \frac{\d\l^{+}\wedge \d\l_{12} \wedge {\ldots} 
\wedge\d\l_{45}}{(\l^{+})^3} \int \d^{16} \theta \l^\a \l^\b \l^\g
f_{\a\b\g}(\t) \frac{\t^{+}}{\l^{+}}\frac{\t_{12}}{\l_{12}}
{\ldots}\frac{\t_{45}}{\l_{45}}.\nonumber \\
\end{eqnarray}
where $\Gamma$ is defined as $\Gamma=\{\l\in PS: |f^I|=\e^I,\, I=1,...,11
,\,\e^I\in\mathbb{R}^+\}$ and $[\d\l]=\d\l^{+}\wedge \d\l_{12} \wedge 
{\ldots} \wedge\d\l_{45}/(\l^{+})^3$ \cite{nathan minimal pure spinor}.
Note that naively $\l^+ =0$ is a singularity, but we do not have
access to it since we are on the patch $\l^+ \neq 0$. So, for this coordinate is
possible that the cycle of integration is not well defined. Formally,
we should choose a patch which allows to access the singularity. In
this particular example, the singularity is $\mathbb{C}^5$, which is a
non-compact and infinite space, that can not be contourned with a
compact space defined by some cycle $\Gamma$. Therefore, in the
Cauchy's sense this is a not well defined integral. That is what we
meant with naively computing the integral. \\
The only contribution to the integral above will come from $\a = \b = \g = +$. In our case,
in contrast with \cite{skenderis}, there are no subtleties with the 
integrals coming from the other choices, which are of the form $\int_\Gamma
\d\l_{ab}\frac{\l_{ab}}{\l_{ab}}$. For example, 
\begin{equation}
\int_\Gamma [\d\l] (\l^{+})^2
\l_{cd}\frac{1}{\l^+}\frac{1}{\l_{12}}{\ldots} \frac{1}{\l_{45}} =
\int_\Gamma \d\l^+ \d^{10}\l_{ab} \frac{\l_{cd}}{\l^+} \frac{1}{\l^+}
\frac{1}{\l_{12}}{\ldots} \frac{1}{\l_{45}}
\end{equation}
will give zero because there is a double pole in $\l^+$ and any choice
of $\l_{cd}$ will kill one of the poles $\l_{ab}$. 
Choosing $\a = \b = \g = +$ we obtain
\begin{equation}
\mathcal{A} = \int \d^{16} \t f_{+++}(\t)\t^+ \t_{12} {\ldots} \t_{45},
\end{equation}
which is exactly the same answer found by Skenderis and Hoogeveen in
\cite{skenderis}, as in their case, it is not Lorentz
invariant. Now we give a geometrical explanation of why it is not
Lorentz invariant. Remember that the intersection between the
hypersurfaces is $
\mathbb{C}^5$, $D_1\cap...\cap D_{11}=\mathbb{C}^5$, so the scattering 
amplitude is defined on the space
\begin{equation}
PS\smallsetminus \mathbb{C}^5. 
\end{equation}
Since the $SO(10)$ group acts transitively up to scalings on the the pure 
spinor space $PS$, then it is always possible to have an element 
$g\in SO(10)$ such that if $\l\in (PS\smallsetminus \mathbb{C}^5)$,
then $(g\l)\notin (PS\smallsetminus \mathbb{C}^5)$, i.e $(g\l)\in
\mathbb{C}^5$. This argument implies that the scattering amplitude is
not Lorentz invariant, since it is not invariant under $SO(10)$, and it
is not globally defined on $PS$, because we can make a transformation
from  $(PS\smallsetminus \mathbb{C}^5)$ to $PS$ where the scattering
is not defined. In the appendix \ref{integrals appendix} we give further simple examples.
\\
Note that the origin is the only fixed point under $SO(10)$
transformations acting on the pure 
spinor space\footnote{This is because the origin is the unique singular point 
in $PS$.}, this means that the condition for the intersection of the
$D_I$'s in the origin, $D_1\cap...\cap D_{11}=\{0\}$, it is not just a 
sufficient condition, but actually it is necessary condition in order
to get a well defined scattering amplitude, i.e that the scattering
amplitude is invariant under the BRST, supersymmetry and Lorentz
transformations, see section \ref{symmetries of the scattering amplitude}.
Summarizing, we showed that this
specific choice for the $C^I$'s is not allowed, since it does not obey our
requirement of the hypersurfaces intersecting at the origin.

\paragraph{Second Example} Now, we show how to construct a set of
$C^I$'s which allow to satisfy $D_1 \cap{\ldots} \cap
D_{11} = \{0\}$. This geometrical construction is as follows: Take eleven points
satisfying the conditions $\chi^a =0$ and $\zeta_a =0$. Then, evaluate
each one of the $10$ gradient vectors $V^a$ and $A_a$, corresponding to
$\chi^a$ and $\zeta_a$ respectively, at each one of those eleven
points (see the appendix \ref{singularities appendix} for more details).
With this vectors, we construct $11$ planes through the origin, such
that at each one of the $11$ points in $PS$, the $10$ gradients belong to the planes.
We present the answer as an  $11\times 16$ matrix

\begin{equation}\label{Cmatrix}
C = 
\left( 
{\begin{array}{cccccccccccccccc}
1 & 2 & 1 & -1 & 1 & 0 & 0 & 0 & 0 & 0 & 1 & 0 & 0 & 0 & 0 & 1 \\
0 & -4 & -1 & 2 & 1 & 1 & 0 & 0 & 0 & 0 & 0 &- 1 & 0 & 0 & 0 & -1 \\
0 & 0 & 3 & -2 & 4 & 1 & -2 & 0 & 0 & 0 & 0 & 0 & 0 & 4 & 4 & 0 \\
0 & 0 & 0 & -1 & -1 & 3 & 1 & 1 & 0 & 0 & 0 & 0 & 0 & 0 & 2 & 1 \\
0 & 0 & 0 & 0 & -2 & 3 & 4 & 3 & 2 & 0 & 0 & 0 & 0 & 1 & 0 & 1 \\
1 & 0 & 0 & 0 & 0 & 1 & 2 & 3 & 2 & 0 & 0 & 0 & -1 & 0 & 1 & 0 \\
0 & 1 & 0 & 0 & 0 & 0 & -1 & -3 & 0 & -3 & 0 & 0 & -3 & 0 & 0 & 2 \\
0 & 0 & 1 & 0 & 0 & 0 & 0 & -1 & 1 & 2 & 0 & 0 & 1 & 1 & 0 & 0 \\
0 & 0 & 0 & 1 & 0 & 0 & 0 & 0 & -1 & -1 & 2 & 1 & -1 & 0 & 1 & 0 \\
0 & 0 & 0 & 0 & 1 & 0 & 0 & 0 & 0 & -2 & -2 & 2 & -1 & 0 & 1 & 0 \\
0 & 0 & 0 & 0 & 0 & 1 & 0 & 0 & 0 & -1 & 1 & 2 & -1 & 1 & 0 & 1 \\
\end{array} }
\right).
\end{equation}
We computed $C^I _\alpha \lambda^\alpha$ and using Mathematica,
we found the intersections of the $11$ planes with the pure spinor
condition 
\begin{equation}\nonumber
\chi^a =\l^{+}\l^{a}-\frac{1}{8}\e^{abcde}\l_{bc}\l_{de}=0.
\end{equation}
The answer is $12$ times the tip of the cone: $\lambda^\alpha =0$.
This number $12$ is the {\it multiplicity} or number of times the
hypersurfaces intersect. This will be further discussed in the next
section. Nevertheless, there are $5$ 
additional non-zero 
solutions\footnote{For these non-zero solution $\l^+=0$. Those are
precisely the points for which the constrains $\chi_a=0$ are not enough to 
describe the 
pure spinor space.}. This is not an issue, since this non-zero 
solutions are not in the remaining pure spinor equations
$\zeta_a=\l^{b}\lambda_{ba}=0$, therefore, we can discard them safely.
Note that the 11 
$C^I$'s form a $\mathbb{C}^5$ space in $\mathbb{C}^{16}$, 
which is invariant by $U(5)$ group, so applying elements of $U(5)$ to the 
matrix $C^I_\a$ (\ref{Cmatrix}) we get an infinite numbers of $C^{\prime I}$'s,
for which the intersection with $PS$ is the origin.\\
Instead of computing the scattering amplitude in this second example as we
did in the first one, we will show in the next section how to find the
answer without an explicit form for the $C^I$'s. In conclusion, what
we wanted to show with this example is that we can indeed find a set of
constant spinors fulfilling our requirement of intersection of the
planes and $PS$ only at the origin.


\section{The Tree Level Scattering Amplitude and \v{C}ech-Dolbeault Equivalence in the Pure Spinor Formalism}
\label{tl amplitude}

In the present section we will compute the scattering amplitude in a
covariant way. We start by defining the scattering amplitude and the
integration contours. Then, we proceed to perform the scattering
amplitude computation in the projective pure spinor space coordinates,
 where the singular point is explicitly removed. This scattering
 amplitude computation will become important in the rest of the paper.
 For instance, this computation will introduce the notion of 
 degree of the projective pure spinor space, which will be
 useful to relate in a simple way the minimal and non-minimal formalisms.
 Actually, the framework in which we relate both formalisms is given
 by the \v{C}ech-Dolbeault language. This is not surprising because the PCO's are defined
locally, so, the \v{C}ech language is a natural formalism to describe the
scattering amplitude because it is a description in terms of
patches. That is the reason why we include a subsection for reviewing
the \v{C}ech-Dolbeault language. \\
At the end of this section we will argue that our picture
changing operators are not related to any particular regulator.

\subsection{Integration Contours} \label{integration contours}
Before attempting to compute the tree level scattering amplitude, we must
discuss which are the integration contours. This will allow to have 
a well defined amplitude.
\\
The contours will be given by the homology cycles. In our case, they are
naturally defined as 
\begin{equation}\label{cycle}
\Gamma=\{\l^\a\,\in\, PS:\,|f^{I}(\l)|= |C^I \lambda| =\varepsilon^I\}.
\end{equation}
Clearly, $\Gamma$ is an $11$-cycle, i.e it has real dimension $11$. Except 
for the integration contour, the tree level scattering amplitude
corresponding to the zero modes 
has the same form as in the first example in the sub-section \ref{examplesCs} 
\begin{equation}\label{integral for W}
\mathcal{A}=\int\d^{16}\t\int_{\Gamma} W,
\end{equation}
where $W$ is given by
\begin{equation}\label{11forma}
W =[\d\l]Y^1_C...Y^{11}_C \l^\a\l^\b\l^\g\,\,f_{\a\b\g}(\t) 
\end{equation}
and $Y^I_C$'s are the new PCO's (\ref{nPCO}).
Since  the integrand $W$ satisfies $\d(W)=(\p+\bar\p)(W)=0$ then it
belongs to the de-Rham cohomology group $H^{11}_{DR}(PS\smallsetminus D)$, 
where $PS\smallsetminus D$ is 
the space in which the $W$-form is defined, i.e $D$ is the hypersurface on $PS$ given by $D = D_1\cup {\ldots} \cup D_{11}$. 
Then, the cycle $\Gamma$ belongs to the homology group
$H_{11}(PS\smallsetminus D,\mathbb{Z})$. We will illustrate this with the following example.
Consider for instance the integral $\int_{\g}\d z / z$, where $\g$ is the circle 
$\g=\{z\in\mathbb{C}:|z|=\e\}$. So any circle $C$ around the origin is 
related to $\gamma$ since $\gamma-C$ is the boundary of some 
annulus $U$, i.e $\p(U)=\gamma-C$, therefore $\g$ is an element of the homology 
group $H_1(\mathbb{C}\smallsetminus\{0\},\mathbb{Z})$ and by the Stokes 
theorem $\int_{\g}\d z/z=\int_{C}\d z/z$. In $\mathbb{C}^2$ we have an
analogous situation, for example consider the integral
$\int_{\varphi} \d z_1\d 
z_2/( z_1 z_2)$. Here the torus $\varphi$, defined by $\varphi
=\{(z_1,z_2)\in\mathbb{C}^2:|z_1|=\e_1,|z_2|=\e_2\}$ is an element of the 
homology group $H_2(\mathbb{C}^2\smallsetminus\{(0,z_2)\,\text{and}\,(z_1,0): 
z_1,z_2\in\mathbb{C}\},\mathbb{Z})= H_2((\mathbb{C}\smallsetminus\{0\})
\times(\mathbb{C}\smallsetminus\{0\}),\mathbb{Z})$ and the integral 
depends only of the class of the torus $\varphi$. The same is true for
the integral (\ref{integral for W}).
Therefore, the integral (\ref{integral for W}) will depend only on the homology class cycle and the
cohomology class cocycle. This is the principle that will allow us to 
show that the scattering amplitude is independent of the $C^I$'s, which 
will be discussed in the section \ref{independence of Cs}.

\subsection{The Scattering Amplitude as an Integral over the Projective Pure 
Spinor Space}
\label{sa in pps}

In the last subsection we have defined the integration contours in the
pure spinor space in order to have a well defined tree level scattering 
amplitude. Now, in this subsection we will proceed to write the coordinates 
for the pure spinor space 
in terms of the projective pure spinor coordinates. Then, we will compute
the tree-level scattering amplitude in this new
coordinates. 

As we will show in the next sub-subsection, in the
projective coordinates we can make a simple analysis of the poles in
the scattering amplitude integral. The cycle $\Gamma$ previously defined
will be used to obtain the integration contours in the projective  
pure spinor space.

\subsubsection{Contours for the Amplitude in the Projective Pure
Spinor Coordinates}
\label{contours in pps}

We can write the pure spinor coordinates as
$\l\-{\a}=\g\,\tilde{\l}\-{\a}$, where $\g\in \mathbb{C}$ and
$\tilde{\l}\-{\a}$ are global coordinates for the $SO(10)/U(5)$ 
space\footnote{Actually $\g$ is the fiber 
of the $\mathcal{O}(-1)$ line bundle over $SO(10)/U(5)$ \cite{beta gamma 
system}.}.
That is, $\tilde{\l}^\a$ satisfies the constraints
$\tilde{\l}\g\-{m}\tilde{\l}=0$ and has the equivalence relation
$\tilde{\l}\-{\a}\,\sim\,c\tilde{\l}\-{\a}$, where $c\in \mathbb{C}\-{*}$.
When $\g=0$ then $\l^{\a}=0$,  but $\tilde\l^\a$ can take any value in the
projective pure spinor space, i.e $SO(10)/U(5)$, also known as the twistor space 
\cite{pure spinor like twistors}. 
In these coordinates the poles take the form
\begin{eqnarray}
&&\g\tilde f^1\equiv\g\,C^{1}_{\a}\tilde\l\-{\a}=0,\\
&&\g\tilde f^2\equiv\g\,C^{2}_{\a}\tilde\l\-{\a}=0,\nonumber\\ 
&&\qquad\qquad\qquad\cdot\nonumber\\
&&\qquad\qquad\qquad\cdot\nonumber\\
&&\qquad\qquad\qquad\cdot\nonumber\\
&&\g\tilde f^{11}\equiv\g\,C^{11}_{\a}\tilde\l\-{\a}=0\nonumber.
\end{eqnarray}
When $\g\neq 0$, we have 11 constraints and 10
degrees of freedom for the projective pure spinor space,
 so, it is not possible to find a
solution for the 11 constraints. On the other hand,
when $\g=0$, naively all the constraints behave as being
zero. Nevertheless, we must consider this case inside the scattering
amplitude. In the numerator of $W$ there are $7$ $\gamma$'s coming 
from the
integration measure plus $3$ coming from the vertex operators,
contributing in total $\g^{10}$ in the numerator. Therefore, only
one of the $11$ $\g$'s will remain in the denominator of $W$.
This remaining $\g$ kills one of the $11$ functions $\tilde f^I$.
Therefore, now the cycle $\Gamma$ is given by 
\begin{equation}
\Gamma=C\times\tilde{\Gamma},
\end{equation}
where $C$ is the cycle $C=\{\g\in\mathbb{C}: |\g|=\e\}$ and $\tilde
\Gamma$ is a $10$-cycle which we define in the following.
After integrating around the contour
$|\g|=\varepsilon$, which belongs to $\Gamma$ and excludes the origin of the 
space, the denominator of $W$
will have 11 $\tilde f^{I}$'s. However, remember that one of the
$\tilde f^I$'s was killed by $\gamma$. Therefore, the cycle
$\tilde\Gamma$ must be given by 
\begin{equation}\label{10cycle}
\tilde\Gamma=\{|\tilde f^i|=\e_i, \text{where the $i$'s are ten numbers 
between 1 to 11}\}.
\end{equation}
After this simple analysis, now we proceed to compute the scattering amplitude.

\subsubsection{The Tree Level Scattering Amplitude in the Projective Pure spinor Space}

The tree-level scattering amplitude has the form 
\begin{eqnarray}\label{general amplitude}
\mathcal{A}=\int_{\Gamma}[\d\l]\int\d^{16}\theta Y_{C}^{1}...Y_{C}^{11}\l^{\a}\l^{\b}\l^{\g}f_{\a\b\g}(\t).
\end{eqnarray}
As discussed in \cite{nathan minimal pure spinor}, the term 
$\l^{\a}\l^{\b}\l^{\g}f_{\a\b\g}(\t)$ can always be written
in the following form 
\begin{equation}
\l^{\a}\l^{\b}\l^{\g}f_{\a\b\g}(\t)\propto(\l\g^{m}\t)(\l\g^{n}\t)(\l\g^{p}\t)(\t\g_{mnp}\t)K,
\end{equation}
up to BRST exact and global terms, which  are decoupled as we will
show later in section \ref{symmetries of the scattering amplitude}. $K$ is 
the kinematic factor, which is a function of the 
polarizations and momenta\footnote{In general, when there are more
than $3$ vertex operators in the scattering amplitude, 
it must include integrals of the worldsheet coordinates $(z,\,\bar z)$.
However, we are not taking care of those terms.}.  Then the amplitude
takes the form 
\begin{eqnarray}\label{amplitude}
\mathcal{A}&=&\int_{\Gamma}[\d\l]\int\d^{16}\theta\frac{C^1\t}{C^1\l}...\frac{C^{11}\t}{C^{11}\l}
(\l\g^{m})_{\a_1}(\l\g^{n})_{\a_2}(\l\g^{p})_{\a_3}(\g_{mnp})_{\a_4\a_5}\t^{\a_1}\t^{\a_2}\t^{\a_3}
\t^{\a_4}\t^{\a_5}K\no\\
&=&\int_{\Gamma}[\d\l]\int\d^{16}\theta\frac{\e^{\a_{1}...\a_{5}\b_1...\b_{11}}C^1_{\b_1}...C^{11}_{\b_{11}}}{C^1\l..
.C^{11}\l}
(\l\g^{m})_{\a_1}(\l\g^{n})_{\a_2}(\l\g^{p})_{\a_3}(\g_{mnp})_{\a_4\a_5}\t^{1}...\t^{16}K. 
\end{eqnarray}
In the coordinates $\l^\a=\g\tilde{\l}^{\a}$, 
we can choose the following parametrization for the projective pure
spinor in the patch $\tilde{\l}^{+}\neq 0$
\begin{equation}\label{parametrization pps}
\tilde{\l}^{\a}=(1,u_{ab},\frac{1}{8}\e^{abcde}u_{bc}u_{de}).
\end{equation}
So, as shown in \cite{beta gamma system}, the integration measure becomes $[\d\l]=\g^{7}\d\g\wedge\d
u_{12}\wedge...\wedge\d u_{45}$ and the amplitude locally can be
written as 
\begin{eqnarray}
\!\!\!\!\!\!\!\!  \mathcal{A}&=&
\int_{\Gamma}\frac{\d\g}{\g}\wedge\frac{\d u_{12}\wedge...\wedge\d
u_{45}\e^{\a_{1}...\a_{5}\b_1...\b_{11}}C^1_{\b_1}...C^{11}_{\b_{11}}}{ C^1\tilde{\l}..
.C^{11}\tilde{\l}}
(\tilde{\l}\g^{m})_{\a_1}(\tilde{\l}\g^{n})_{\a_2}(\tilde{\l}\g^{p})_{\a_3}(\g_{mnp})_{\a_4\a_5}K,
\,\,\,
\end{eqnarray}
where the $\t^\a$ variables have been integrated. The integral around
the contour $|\g|=\varepsilon$ is trivial, then 
\begin{eqnarray}\label{amplitude1}
\!\!\!\!\!\!\! \mathcal{A}&=&(2\pi i)
\int_{\tilde{\Gamma}}\frac{\d u_{12}\wedge...\wedge\d
u_{45}\e^{\a_{1}...\a_{5}\b_1...\b_{11}}C^1_{\b_1}...C^{11}_{\b_{11}}}{ C^1\tilde{\l}..
.C^{11}\tilde{\l}}
(\tilde{\l}\g^{m})_{\a_1}(\tilde{\l}\g^{n})_{\a_2}(\tilde{\l}\g^{p})_{\a_3}(\g_{mnp})_{\a_4\a_5}K,
\end{eqnarray} 
where the contour $\tilde{\Gamma}$ was defined in (\ref{10cycle}). Note
that up to a sign, the scattering amplitude is independent of the choice of 
the $10$-cycle out of the $11$ possibilities. To illustrate
that we can consider the simplest and non trivial case of the
projective pure spinor space, i.e the projective pure spinor space in
$d=4$, in this case the integral is (see also the appendix \ref{CD appendix})
\begin{equation}\label{example cp1}
\int_{\tilde\g}\frac{\e_{ab}\tilde\l^a\d\tilde\l^b
\e^{cd}C^1_cC^2_d}{(C^1\tilde\l)(C^2\tilde\l)},
\end{equation}
where $\tilde\l^a=(\tilde\l^1,\tilde\l^2)$ are the homogeneous coordinates 
of $\mathbb{C}P^1$. In this case we have two choices. First we can take 
$\tilde\g=\{\tilde\l\in \mathbb{C}P^1
: |C^1\tilde\l|=\e\}$ and for simplicity we set $C^1=(1,0)$ and 
$C^2=(0,1)$. In the patch $\tilde\l^2\neq0$ we have the parametrization 
$\tilde\l^a=(u,1)$, therefore, the contour $\tilde\gamma$ is well
defined and the integral (\ref{example cp1}) is     
\begin{equation}
\int_{|u|=\e}\frac{\d u}{u}.
\end{equation}
Note that in the patch $\tilde\lambda^1 \neq 0$ the cycle $\gamma$ is
not well defined. The second choice is $\tilde\g=\{\tilde\l\in
\mathbb{C}P^1: |C^2\tilde\l|=\e\}$. Here we must take the patch
$\tilde\l^1\neq 0$, where the parametrization is given by 
$\tilde\l^a=(1,v)$. Then, the integral (\ref{example cp1}) becomes
\begin{equation}
-\int_{|v|=\e}\frac{\d v}{v}. 
\end{equation}
So, we have shown for the $d=4$ projective pure spinor space, that
different choices of the cycle $\tilde\g$ results just in changing 
the sign of (\ref{example cp1}). The same argument holds for the ten
dimensional projective pure spinor space.

The integration measure in (\ref{amplitude1}), $\d u_{12}\wedge...\wedge\d u_{45}$, is the same found in a covariant manner by Berkovits and Cherkis in
\cite{pure spinor like twistors}.
Therefore, we have the following identity.


\paragraph{ Identity}

If $\tilde \l^{\a}$ is an element of the projective pure spinor space in 10
dimensions, i.e. if $\tilde\l^\a \in SO(10)/U(5)$, then the integration
measure $[\d\tilde\l]$ defined by \cite{pure spinor like twistors}
\begin{equation}\label{holomorphic measure}
[\d\tilde\l](\tilde\l\g^m)_{\a_1}(\tilde\l\g^n)_{\a_2}(\tilde\l\g^p)_{\a_3}(\g_{mnp})_{\a_4\a_5}=\frac{2^3}{10!}\e_{
\a_1...\a_5\b_1...\b_{11}} \d\tilde\l^{\b_1}\wedge...\wedge\d\tilde\l^{\b_{10}}
\tilde\l^{\b_{11}},
\end{equation}
written in the parametrization $\tilde\l^{\a}=(\tilde\l^{+}, \tilde\l_{ab}, \tilde\l^{a})=(1, u_{ab}, 
\frac{1}{8}\e^{abcde}u_{bc}u_{de})$ is
\begin{equation}\label{iden}
[\d\tilde\l]=\d u_{12}\wedge...\wedge\d u_{45}.
\end{equation}
This identity is proved in the appendix \ref{proof of identity}

With this identity in mind, the amplitude (\ref{amplitude1}) can be written in
a covariant manner with respect to $SO(10)/U(5)$ 
\begin{eqnarray}\label{amplitude2}
\mathcal{A}&=&(2\pi i)
\int_{\tilde{\Gamma}}\frac{[\d\tilde{\l}]\e^{\a_{1}...\a_{5}\b_1...\b_{11}}C^1_{\b_1}...C^{11}_{\b_{
11 } } } { C^1\tilde{\l}..
.C^{11}\tilde{\l}}
(\tilde{\l}\g^{m})_{\a_1}(\tilde{\l}\g^{n})_{\a_2}(\tilde{\l}\g^{p})_{\a_3}(\g_{mnp})_{\a_4\a_5}K.
\end{eqnarray} 
This integral is the same as the 10 dimensional integral found in 
\cite{pure spinor like twistors}, so it is possible to have a twistor
type version for the scattering amplitude at tree level. In \cite{pure
spinor like twistors} the integral is solved up to a proportionality
factor. However, we will find a rigorous solution.  Using (\ref{iden}) 
in (\ref{amplitude2}) we get
\begin{eqnarray}\label{amplitude3}
\mathcal{A}&=&(2\pi i)2^3
\int_{\tilde{\Gamma}}\frac{1}{10!} \d\tilde\l^{\b_1}\wedge...\wedge\d\tilde\l^{\b_{10}}
\tilde\l^{\b_{11}}\e_{
\a_1...\a_5\b_1...\b_{11}}\frac{\e^{\a_{1}...\a_{5}\g_1...\g_{11}}C^1_{\g_1}...C^{11}_{\g_{
11 } } } { C^1\tilde{\l}..
.C^{11}\tilde{\l}}
K\no\\
&=&(2\pi i)2^3
\int_{\tilde{\Gamma}}\frac{5!}{10!} \d\tilde\l^{\b_1}\wedge...\wedge\d\tilde\l^{\b_{10}}
\tilde\l^{\b_{11}}\de_{[\b_1}^{\g_1}\de_{\b_2}^{\g_2}...\de_{\b_{11}]}^{\g_{11}}\frac{C^1_{\g_1}...C^{11}_{\g_{
11 } } } { C^1\tilde{\l}..
.C^{11}\tilde{\l}}
K.
\end{eqnarray}
Without loss of generality,
we take $C^1\tilde{\l},...,C^{10}\tilde{\l}$ to define
$\tilde{\Gamma}$, then (\ref{amplitude3}) becomes
\begin{eqnarray}\label{amplitude4}
\mathcal{A}&=&(2\pi i)2^3
\int_{\tilde{\Gamma}} 5! \frac{(\d C^1\tilde\l)\wedge...\wedge(\d C^{10}\tilde\l)
(C^{11}\tilde\l) } { C^1\tilde{\l}..
.C^{11}\tilde{\l}}
K\\
&=&
(2\pi i)2^3 5!
\int_{\tilde{\Gamma}} \frac{(\d\tilde  f^1)\wedge...\wedge(\d\tilde  f^{10})} {\tilde  f^1..
.\tilde  f^{10}}
K\no
\end{eqnarray} 
where $\tilde f^I=C^I\tilde\l$. The others terms, like
$$\int_{\tilde{\Gamma}} \frac{( C^1\tilde\l)(\d C^2\tilde\l)\wedge...
\wedge(\d C^{10}\tilde\l)\wedge
(\d C^{11}\tilde\l) } { C^1\tilde{\l}..
.C^{11}\tilde{\l}}
 $$
do not contribute since one of the poles 
($C^1\tilde{\l},...,C^{10}\tilde{\l}$) is canceled, in this case, $(C^1\tilde{\l})$.
Another choice of the $C^I$'s just change the sign of
(\ref{amplitude4}). \\
Naively, it can be thought that the integral in (\ref{amplitude4})
gives $(2\pi i)^{10}$. However, remember that $\tilde f^I$ are functions
over the projective pure spinor space and $\tilde\Gamma$ is a
$10$-cycle in the projective pure spinor space. Therefore, this
integral is non-trivial as in the flat space. Despite that the answer
will just differ from this trivial case by a number, to know the
formal answer will be extremely useful for relating the minimal and
non-minimal pure spinor formalisms.

Before establishing the equivalence between the minimal and non-minimal pure 
spinor formalism for the tree level scattering amplitude, it is needed to 
give a short introduction to the \v{C}ech 
and Dolbeault language, which will be very useful to understand that 
correspondence.


\subsection{\v{C}ech and Dolbeault Language}\label{CDLanguage}

Due to the behavior ($1/\l$) in the new lowering picture changing
operators, they are defined locally in the pure spinor space. However,
it will be interesting to have a global description, i.e patch
independent, which can be achieved by introducing the \v{C}ech language.
In this section  we give a simple introduction to the \v{C}ech formalism and 
the \v{C}ech-Dolbeault isomorphism, which turns out to be useful for
relating the minimal and non-minimal pure spinor formalism from the tree 
level scattering amplitude as we will show in subsection 
\ref{equivalence mps and nmps}, and to 
check the BRST, Lorentz and SUSY symmetries in the section \ref{symmetries
of the scattering amplitude}.

Given the new formulation for the PCO's
\begin{equation}
Y^I_C=\frac{C^I\t}{C^I\l}, \qquad I=1,...,11,
\end{equation}
it is clear that $Y^I_C$ is just defined in the patch $
PS\smallsetminus D_I$ where $D_I$ is the hypersurface given by 
$f^I=C^I_\a\l^\a=0$. Because 11 PCO's are needed in order to compute the 
tree level scattering amplitude, it is sufficient to have 11 patches to 
cover the pure spinor space at this order. Each patch is defined by the 
denominator of the picture operator, i.e we define the patch $U_I$ as 
\begin{equation}\label{patch U_I}
U_I=PS\smallsetminus D_I,\qquad  D_I=\{\l\in PS: f^I\equiv C^I_\a\l^\a=0\}. 
\end{equation}
The set $\underline{U}=\{U_I\}$ is a cover of the pure spinor space without 
the origin since we claimed that $D_1\cap...\cap D_{11}=\{0\}$. This means
\begin{equation}
PS\smallsetminus\{0\} = U_1\cup...\cup U_{11}= \bigcup_{I=1}^{11}U_I. 
\end{equation}
This is as desired because the singular point is removed from the theory. 
Note that in the papers \cite{nathan nikita multiloops}\cite{yuri aldo nathan 
nikita} the authors take the patches $\mathcal{U}_\a=PS\smallsetminus 
\mathcal{D}_\a$ where $\mathcal{D}_\a=\{\l\in PS: \l^\a= 0\}$, $\a=1,...,16$. 
Clearly these $\mathcal{D}_\a$'s satisfy $\mathcal{D}_1\cap...\cap \mathcal{D}
_{16}=\{0\}$, therefore $PS\smallsetminus\{0\}=\mathcal{U}_1\cup...\cup
\mathcal{U}_{16}$ and we can define the PCO's as
\begin{equation}
Y_\a=\frac{\t^\a}{\l^\a}, \qquad \a=1,...,16. 
\end{equation}
Actually, for tree level scattering amplitudes this notation is not very 
convenient, as is explained in the appendix \ref{another cover for the pure spinor space}.

Now, we introduce the \v{C}ech cochains because this terminology will be
extensively used in this paper. 
The \v{C}ech $k$-cochain, denoted by $\psi_{I_1...I_{k+1}}$, is an 
holomorphic $p$-form in the intersection 
$U_{I_1...I_{k+1}}=U_{I_1}\cap U_{I_2}\cap...\cap U_{I_{k+1}}$. 
I.e $\psi_{I_1...I_{k+1}}\in\Omega^p(U_{I_1...I_{k+1}})$ where $\Omega^p(U)$ 
is the abelian group of the holomorphic $p$-forms over $U$.  We choose the 
abelian group of $p$-forms
because it will be the group used in this paper. 
The \v{C}ech cochains must be antisymmetric in the \v{C}ech labels, 
for instance,
$\psi_{I_1..I_i...I_j...I_{k+1}}=-\psi_{I_1..I_j...I_i...I_{k+1}}$.
This is related to the orientation of the manifold, which in our case
is $PS\smallsetminus \{0\}$. \\ We define the set of the $0$-cochains on $PS\smallsetminus\{0\}$ with values in the holomorphic $p$-forms as 
\begin{equation}
C^{0}(\underline{U},\Omega^p)=\bigoplus_{I=1}^{11}\Omega^p(U_I).
\end{equation}
Similarly the 1-cochains are elements of the set
\begin{equation}
C^{1}(\underline{U},\Omega^p)=\bigoplus_{I<J}\Omega^p(U_{IJ})
\end{equation}
and so on. 
We define the \v{C}ech operator as the map $\de:C^{k}(\underline{U},\Omega^p)\rightarrow C^{k+1}(\underline{U},\Omega^p)$ given by 
\begin{equation}\label{cech operator}
(\de\psi)_{I_1....I_{k+2}}\equiv \psi_{I_2I_3....I_{k+2}}-\psi_{I_1I_3....I_{k+2}}+...+(-1)^{k+1}\psi_{I_1I_2....I_{k+1}}. 
\end{equation}
It is easy to show that $\de$ is a nilpotent operator, $\de^2=0$. 
If $(\de\psi)_{I_1....I_{k+2}}=0$ then $\psi_{I_1....I_{k+1}}$ is called a 
cocycle and the set of all cocycles in  $C^{k}(\underline{U},\Omega^p)$ is an 
abelian subgroup denoted by $Z^{k}(\underline{U},\Omega^p)$. 
If $\psi_{I_1....I_{k+1}}=(\de\rho)_{I_1....I_{k+1}}$ then 
$\psi_{I_1....I_{k+1}}$ is called a coboundary and the set of all coboundary 
in  $C^{k}(\underline{U},\Omega^p)$ is denoted by 
$B^{k}(\underline{U},\Omega^p)$. Clearly every coboundary is a cocycle 
since $\de^2=0$, then we can define the coset 
\begin{equation}
H^{k}(PS\smallsetminus \{0\},\Omega^{p})=\frac{Z^{k}(\underline{U},\Omega^p)}{B^{k}(\underline{U},\Omega^p)} 
\end{equation}
known as the $k$-\v{C}ech cohomology group with values in the Abelian group of holomorphic $p$-forms $\Omega^p$ on $PS\smallsetminus\{0\}$. We
refer the reader to  \cite{harris}\cite{bott} for more details about this topic.

Note that the PCO's are elements of $C^{0}(\underline{U},\mathcal{O})$, where
$\Omega^0\equiv\mathcal{O}$ is a group of holomorphic functions, for instance 
\begin{equation}
Y^I_C=\frac{C^I\t}{C^I\l}\in \mathcal{O}(U_I) 
\end{equation}
is an holomorphic function on the patch $U_I$. It is easy to see that $Y^I_C$ is not a cocycle 
\begin{equation}
(\de Y_C)^{IJ}=\(\frac{C^J\t}{C^J\l}-\frac{C^I\t}{C^I\l}\)\Bigg|_{U_{IJ}}=-\frac{C^{[I}\t C^{J]}\l}{(C^I\l) (C^J\l)}\Bigg|_{U_{IJ}}\neq 0 
\end{equation}
and therefore $Y^I_C$ is not in the \v{C}ech cohomology. The PCO's have the 
particular property that the product of different PCO's is a \v{C}ech cochain, for example
\begin{equation}
\zeta^{I_1...I_k}\equiv Y^{I_1}_C...Y^{I_k}_C=\frac{C^{I_1}\t...C^{I_k}\t}{(C^{I_1}\l)...(C^{I_k}\l)}\in \mathcal{O}(U_{I_1...I_k}), \qquad k\leq11 
\end{equation}
is an element of $C^{k-1}(\underline{U},\mathcal{O})$ because $\zeta^{I_1...I_k}$ is antisymmetry in its \v{C}ech labels. This happens because the variables $\t^\a$ are grassmann numbers, $\t^\a\t^\b=-\t^\b\t^\a$. When $k=11$ we have
\begin{equation}
\zeta^{I_1...I_{11}}=\e^{I_1...I_{11}}\frac{C^{1}\t...C^{11}\t}{(C^{1}\l)...(C^{11}\l)}\in \mathcal{O}(U_{1}\cap...\cap U_{11}). 
\end{equation}
This element is important because it is inside to the scattering amplitude. 
Since the cover $\underline{U}$ just has 11 patches and $(\delta\zeta)^{I_1...I_{11}I_{12}}$ is antisymmetric in all its \v{C}ech labels then 
\begin{equation}
(\delta\zeta)^{I_1...I_{11}I_{12}}=0 
\end{equation}
so $\zeta^{I_1...I_{11}}$ belongs to \v{C}ech cohomology.

\subsubsection{\v{C}ech-Dolbeault Isomorphism}\label{CD isomorphism}
Now we give a simple explanation about the \v{C}ech-Dolbeault isomorphism, 
which as we will show in section \ref{equivalence mps and nmps}, is the base to obtain the relationship 
between the minimal and the non-minimal pure spinor formalisms. There is a simple 
way to relate the \v{C}ech and Dolbeault cocycles using the so called the 
partition of unity \cite{harris}\cite{bott}\cite{nathan nikita multiloops}. We can take the partition of unity as
\begin{equation}\label{partition of unity1}
\rho_I = \frac{f^I\bar f_I}{(|f^1|^2+....+|f^{11}|^2)},\qquad I=1,...,11 
\end{equation}
where $f^I=C^I\l$, $\bar f_I$ is its complex conjugate: $\bar
f_I=\bar C_I\lb$, and $\bar \l_\a = (\l^\a)^*$. It is clear that this 
partition of unity is subordinated to the cover $\underline{U}$, i.e, 
$\rho_I\neq 0$ only when $\l^\a\in U_I$, outside of the patch $U_I$ the 
partition of unity is identically zero. Obviously this partition of unity 
satisfies the condition 
\begin{equation}
\sum_{I=1}^{11}\rho_I = 1. 
\end{equation}
 
Let $\psi_{I_1...I_{k+1}}$ be a $k$-\v{C}ech cocycle 
($\psi_{I_1...I_{k+1}}\in Z^{k}(\underline{U},\Omega^p)$),  then we define 
the corresponding $\eta_{\psi}$ Dolbeault cocycle of type $(p,k)$ as 
\begin{equation}\label{cech dolbeault map}
\eta_\psi = \frac{1}{k!}\sum_{I_1...I_{k+1}=1}^{11}\psi_{I_1...I_{k+1}}\rho_{I_1}\wedge\bar\p\rho_{I_2}\wedge...\wedge\bar\p\rho_{I_{k+1}}. 
\end{equation}
Note that $\eta_{\psi}$ is a $(p,k)$ form, which is $p$ holomorphic
and $k$ antiholmorphic. As expected, $\bar\p\eta_\psi
=\d\lb_\a\wedge \frac{\p}{\p\lb_\a}\eta_\psi=0$ because
$\psi_{I_1{\ldots} I_{k+1}}$ is a cocycle. Also $\psi_{I_1...I_{k+1}}$ is a coboundary, $\psi_{I_1...I_{k+1}}=(\de\tau)_{I_1...I_{k+1}}$, then $\eta_\psi$ is $\bar\p$-exact, i.e $\eta_\psi=\bar\p\eta_\tau$, where $\eta_\tau$ is the corresponding Dolbeault cochain to $\tau_{I_1...I_{k}}$
\begin{equation}
\eta_\tau= \frac{1}{(k-1)!}\sum_{I_1...I_{k}=1}^{11}\tau_{I_1...I_{k}}\rho_{I_1}\wedge\bar\p\rho_{I_2}\wedge...\wedge\bar\p\rho_{I_{k}},
\end{equation}
i.e $\eta_\psi=\eta_{(\de\tau)}=\bar\p\eta_\tau$. Therefore we have a
map between the \v{C}ech and Dolbeault cohomology groups $H^{k}(PS\smallsetminus
\{0\},\Omega^{p})$ and $H^{(p,k)}_{\bar\p}(PS\smallsetminus \{0\})$.
Actually this map is an isomorphism but we do not show that statement here \cite{harris}\cite{bott}. In particular we can consider the \v{C}ech cocycle
\begin{equation}\label{integral final}
\beta_{I_1...I_{11}}=\e_{I_1...I_{11}}\frac{\d(C^1\l)\wedge...\wedge\d(C^{11}\l)}{(C^1\l)...(C^{11}\l)}\in \Omega^{11}(U_1\cap...\cap U_{11}) 
\end{equation}
which will appear in the section \ref{equivalence mps and nmps}. Clearly $\beta_{I_1...I_{11}}$ is an element of $H^{10}(PS\smallsetminus \{0\},\Omega^{11})$ so we can find its corresponding $\eta_\b\in H^{(11,10)}_{\bar\p}(PS\smallsetminus \{0\})$. Applying the map (\ref{cech dolbeault map}) to $\beta_{I_1...I_{11}}$ we get      
\begin{eqnarray}\label{corresponding dolbeault form}
\eta_\b &=& \frac{1}{10!}\sum_{I_1...I_{11}=1}^{11}\b_{I_1...I_{11}}\rho_{I_1}\wedge\bar\p\rho_{I_2}\wedge...\wedge\bar\p\rho_{I_{11}}\no\\
&=&(-1)^{i-1} \frac{\d(C^1\l)\wedge...\wedge\d(C^{11}\l)\wedge \bar\p\rho_{1}\wedge...\wedge\widehat{\bar\p\rho_{i}}\wedge...\wedge\bar\p\rho_{{11}}}{(C^1\l)...(C^{11}\l)}
\end{eqnarray}
where $\widehat{\bar\p\rho_{i}}$ means that it must be removed from
(\ref{corresponding dolbeault form}). The $C^I$ dependence is eliminated by a global
transformation from the projective pure spinor space to itself, as
will be done in the section \ref{independence of Cs}.
\footnote{We recommend to see the example in the appendix \ref{CD appendix}
 to get more information about this computation.}. 

Since the pure spinor space without the origin ($PS\smallsetminus\{0\}$) is 
contractible to $SO(10)/SU(5)$, i.e $PS\smallsetminus\{0\}$ is deformed 
to $SO(10)/SU(5)$ \footnote{For example the space
$\mathbb{C}\smallsetminus \{0\}$ can be deformed to $S^1$.}, where one can 
think of $SO(10)/SU(5)$ as the boundary of the $PS\smallsetminus\{0\}$ space, 
then the topological invariants of these two spaces are the same 
\cite{massey}, in particular the following two groups are isomorphic
\begin{equation}\label{isomorphism of cohomology}
H^{(11,10)}_{\bar\p}(PS\smallsetminus 
\{0\})\approx H^{21}_{DR} (SO(10)/SU(5)) 
\end{equation}
where $DR$ means the de-Rham 
cohomology \cite{bott}. For the purposes of this paper it is enough to show that the map 
\begin{equation}\label{map between the cohomology groups}
i^{*}\,:H^{(11,10)}_{\bar\p}(PS\smallsetminus 
\{0\})\longrightarrow H^{21}_{DR} (SO(10)/SU(5))  
\end{equation}
is an injective homomorphism, i.e for any element $\eta\in H^{(11,10)}_{\bar\p}(PS\smallsetminus 
\{0\})$ there is just one element $i^*(\eta)\in H^{21}_{DR} (SO(10)/SU(5))$, where ``$i$'' is the map which embeds the 
$SO(10)/SU(5)$ space in the $PS\smallsetminus\{0\}$ space and  ``$i^{*}$'' is the pull back of the differential forms. 

\textit{Proof}

Let $\l=(\l^1,...,\l^{16})=(\l^\a)\in\mathbb{C}^{16}$ be a point of the pure 
spinor space, $PS\smallsetminus\{0\}$, i.e $\l\g^m\l=0$ and $\l\neq0$, then 
the $SO(10)/SU(5)$ space is embedded in $PS\smallsetminus\{0\}$ by 
\begin{equation}\label{so(10)/su(5) space}
 SO(10)/SU(5)=\{(\l^{\a})\in PS\smallsetminus\{0\}: \l^\a\lb_\a=r^2\}
 ,\qquad\text{ $r$ is a positive constant, } r\in\mathbb{R}^+,
\end{equation}
where $\lb_\a$ is the conjugate complex of $\l^\a$ \footnote{Note that when 
$r\rightarrow\infty$ we can think the $SO(10)/SU(5)$ space like the boundary 
of the $PS\smallsetminus\{0\}$.}. Therefore (\ref{so(10)/su(5) space})
defines the injective map
\begin{equation}
i:\, SO(10)/SU(5)\,\,\longrightarrow\,\, PS\smallsetminus\{0\}. 
\end{equation}
Now we must prove two statements in order to show that (\ref{map between 
the cohomology groups}) is an injective homomorphism:
\begin{enumerate}
\item First, we need to verify that the map (\ref{map between the cohomology groups}) is well defined, in others words,
 if $\eta$ is an (11,10)-form on $PS\smallsetminus\{0\}$ which is $\bar\p$ 
 closed, i.e $\bar\p\eta=0$, then the 21-form on $SO(10)/SU(5)$ given by ``$i^{*}\eta$'' is ``$\d$'' closed, i.e $\d (i^{*}\eta)=0$.

Since the exterior derivate operator $\d$ commutes with pull back, then we have
\begin{equation}
\d (i^{*}\eta)=i^{*}(\d\eta)=i^{*}[(\p+\bar\p)\eta]. 
\end{equation}
Remember that $\eta$ is a (11,10)-form, this means $\p\eta=0$ because the
complex dimension of the pure spinor space is 11, dim$_\mathbb{C}(PS\smallsetminus\{0\})=11$, so we have $\d (i^{*}\eta)=i^{*}(\bar\p\eta)$. As $\bar\p\eta=0$ then we have shown $\d (i^{*}\eta)=0$.

\item Finally, we must show that the homomorphism $i^*$ is injective.
To show this, it is sufficient to prove that $i^*$ maps the zero to the zero. 
In others words, if $\eta$ is a (11,10)-form on $PS\smallsetminus\{0\}$ 
which is $\bar\p$ exact, i.e $\eta=\bar\p\tau$, where $\tau$ is a (11,9)-form on $PS\smallsetminus\{0\}$, then the 21-form on $SO(10)/SU(5)$ given by ``$i^{*}\eta$'' is ``$\d$'' exact, i.e $ (i^{*}\eta)=\d (i^{*}\tau)$.

Since $\tau$ is a (11,9)-form then $\eta=\bar\p\tau=(\p+\bar\p)\tau=\d\tau$, because $\p\tau=0$. So we have
\begin{equation}
i^*\eta=i^*(\d\tau)=\d(i^*\tau). 
\end{equation}
Therefore we showed that the map (\ref{map between the cohomology groups}) is an injective homomorphism.
\end{enumerate}
To see more information about this topic we refer to \cite{harris}\cite{massey}.

This isomorphism will be very 
useful to obtain the equivalence between the minimal and non-minimal pure spinor 
formalism and to show that the scattering amplitude is invariant under
BRST, Lorentz and supersymmetry transformations.

\subsection{Equivalence Between the Minimal and Non-Minimal Formalism}
\label{equivalence mps and nmps}

In the previous subsection we gave the basic tools for writing the Dolbeault cocycle corresponding 
to the scattering amplitude. Using that, we will relate the minimal and non-minimal 
pure spinor formalisms. Specifically, we must find the Dolbeault cocycle 
corresponding to the scattering amplitude (\ref{amplitude}) using the 
isomorphism $H^{10}(PS\smallsetminus\{0\},\Omega^{11})\approx H^{(11,10)}
_{\bar\p}(PS\smallsetminus\{0\})$, which was explained in the
previous subsection.\\
Since the elements of the group $H^{(11,10)}_{\bar\p}(PS\smallsetminus\{0\})$ 
are (11,10)-forms, they can not be evaluated in the whole space of the pure
spinor minus the origin. However, $PS\smallsetminus\{0\}$ can be contracted
to the space $SO(10)/SU(5)$, which can be thought as the boundary 
in the infinite of the $PS\smallsetminus\{0\}$ space. Then, by the
isomorphism (\ref{map between the cohomology groups}), the
elements of $H_{\bar\partial}^{(11,10)}(PS\smallsetminus\{0\})$ can be 
evaluated in the $SO(10)/SU(5)$ space. As will be explained in this section, this fact means that the picture lowering 
operators are not related to any particular regulator. 

Now we show how to get the Dolbeault cocycle corresponding to (\ref{amplitude}).         
The scattering amplitude (\ref{amplitude}) can be written as
\begin{eqnarray}\label{beta cech cochain}
\mathcal{A}
&=&\int_{\Gamma}[\d\l]\frac{\e^{\a_{1}...\a_{5}\b_1...\b_{11}}C^1_{\b_1}...C^{11}_{\b_{11}}}{C^1\l..
.C^{11}\l}
(\l\g^{m})_{\a_1}(\l\g^{n})_{\a_2}(\l\g^{p})_{\a_3}(\g_{mnp})_{\a_4\a_5}K\\
&=&\frac{1}{11!}\sum_{I_1...I_{11}}\e_{I_1...I_{11}}\int_{\Gamma}[\d\l]\frac{\e^{\a_{1}...\a_{5}\b_1...\b_{11}}C^{I_1}_{\b_1}...C^{I_{11}}_{\b_{11}}}{C^{I_1}\l..
.C^{I_{11}}\l}
(\l\g^{m})_{\a_1}(\l\g^{n})_{\a_2}(\l\g^{p})_{\a_3}(\g_{mnp})_{\a_4\a_5}K\no\\
&\equiv&\frac{1}{11!}\sum_{I_1...I_{11}}\e_{I_1...I_{11}}\int_{\Gamma}
\b^{I_1...I_{11}}=
\int_{\Gamma}\b^{1,...,{11}}
\no 
\end{eqnarray}
where the $\t^\a$'s have been integrated. Clearly $\b^{I_1...I_{11}}$ 
is a \v{C}ech cochain\footnote{In \cite{beta gamma system} was shown that 
the measure $[\d\l]$ is defined globally on $PS\smallsetminus\{0\}$,
so, the \v{C}ech indices come only from the PCO's. }
\begin{equation}
\b^{I_1...I_{11}}\in C^{10}(\underline{U},\Omega^{11}) 
\end{equation}
where $\underline{U}$ is the cover of the $PS\smallsetminus\{0\}$ space, 
which was defined in the subsection \ref{CDLanguage}, i.e  $\underline{U}=\{U_I\},\,I=1,...,11,$ 
and the patches $U_I$'s are given by $U_I=PS\smallsetminus D_I$, where 
$D_I$ is the hypersurface $D_I=\{\l^\a\in PS: C^I_\a\l^\a=0\}$. Remember 
that $PS\smallsetminus\{0\}=U_1\cup...\cup U_{11}$. Since there are $11$ 
patches to cover $PS\smallsetminus\{0\}$ then $\b^{I_1...I_{11}}$ is in 
the \v{C}ech cohomology because $C^{11}(\underline{U},\Omega^{11})=\{0\}$ 
and $(\de \b)^{I_1...I_{12}}\in C^{11}(\underline{U},\Omega^{11})$, so 
$(\de \b)^{I_1...I_{12}}=0$, so we can write
\begin{equation}
\b^{I_1...I_{11}}\in H^{10}(PS\smallsetminus\{0\},\Omega^{11}). 
\end{equation}
Now, using the partition of unity (\ref{partition of unity1}) we can
find the Dolbeault cocycle, $\eta_\b$, given by (\ref{cech dolbeault map}) and (\ref{corresponding dolbeault form})  
\begin{eqnarray}\label{corresponding dolbeault form2}
\eta_\b=\frac{1}{10!}\sum_{I_1...I_{11}=1}^{11}\b^{I_1...I_{11}}\rho_{I_1}\wedge\bar\p\rho_{I_2}\wedge...\wedge\bar\p\rho_{I_{11}}.
\end{eqnarray}
Note that, since $\b^{I_1...I_{11}}$ is an element 
of $H^{10}(PS\smallsetminus\{0\},\Omega^{11})$, 
then $\eta_\b\in H^{(11,10)}_{\bar\p}(PS\smallsetminus\{0\})$, as was 
explained in the sub-subsection \ref{CD isomorphism}. The Dolbeault cohomology group 
$H^{(11,10)}_{\bar\p}(PS\smallsetminus\{0\})$ was computed in 
\cite{beta gamma system}, $H^{(11,10)}_{\bar\p}(PS\smallsetminus\{0\})
=\mathbb{C}$, so it only has one generator which is $\eta_\b$. 
The computation 
(\ref{corresponding dolbeault form2}) is not straightforward because it is 
needed to make a non-trivial global transformation from $SO(10)/U(5)$ to itself
in order to find an expression for the Dolbeault cocycle $\eta_\beta$ 
independent of the constants $C^I$'s. See the simple example given in the 
appendix \ref{CD appendix}. To avoid this difficulty, we will introduce the concept
of the degree of the projective pure spinor space, in order to obtain
$\eta_\beta$ in a simpler way.

\subsubsection{The Projective Pure Spinor Degree}
\label{The Projective Pure Spinor Degree}
The last step (\ref{amplitude4}) in the computation of the scattering
amplitude with the projective pure spinor space variables was
\begin{eqnarray}\label{amplitude5}
\mathcal{A}&=&
(2\pi i)2^3
\int_{\tilde{\Gamma}} 5! \frac{(\d\tilde  f^1)\wedge...\wedge(\d\tilde  f^{10})} {\tilde  f^1..
.\tilde  f^{10}}
K,
\end{eqnarray} 
This integral is known \cite{harris}\cite{intersection} and its result is given by the intersection theory 
\begin{eqnarray}\label{integral}
\int_{\tilde{\Gamma}} \frac{(\d\tilde  f^1)\wedge...\wedge(\d\tilde  f^{10})} {\tilde  f^1..
.\tilde  f^{10}} = (2\pi i)^{10} \sum_{\nu}(\tilde D_1,...,\tilde D_{10})_{p_{\nu}}
\end{eqnarray} 
where $\nu$ is the number of points $p_\nu$ where the hypersurfaces
$\tilde D_I$ were defined by $\tilde D_I=\{\tilde\l^\a\in SO(10)/U(5):]
C^I \tilde\l=0,
\,I=1,{\ldots} 10 \}$  and 
$(\tilde D_1,...,\tilde D_{10})_{p_{\nu}}\equiv m_\nu$ is the
multiplicity\footnote{The multiplicity can be understood in the same
way as in the solutions of a system of algebraic equations.} in $p_\nu$.
Remember that the coordinates $\tilde\l^\a,\,\a=1,...,16$ can be
thought as coordinates of $\mathbb{C}^{16}\smallsetminus\{0\}$ with
the equivalence relation $\tilde\l^\a\sim c\tilde\l^\a,\,c\neq 0 
\in\mathbb{C}$, satisfying the constraints $\tilde\l\g^m\tilde\l=0$, so the 
projective pure spinor space $SO(10)/U(5)$ is embedded in 
$\mathbb{C}P^{15}=\mathbb{C}^{16}\smallsetminus\{0\}/(\tilde\l^\a\sim 
c\tilde\l^\a),\,c\in\mathbb{C^*}$. Therefore the hypersurface 
$\tilde D_I\subset SO(10)/U(5)$ is the intersection between the linear 
subspace $C^I_\a\tilde\l^\a=0$ and $SO(10)/U(5)$, where now 
$\tilde\l^\alpha \in \mathbb{C}P^{15}$, 
i.e $\tilde D_I=\{\{C^I_\a\tilde\l^\a=0\}\cap SO(10)/U(5), \text{ where } 
\tilde\l^\a\in \mathbb{C}P^{15}\}$. Note that the intersection of the 10 
linear subspaces $C^I\tilde\l=0$, $I=1,...,10$ in $\mathbb{C}P^{15}$  is the
linear subspace $\mathbb{C}P^5$ embedded in $\mathbb{C}P^{15}$, therefore 
the intersection of the hypersurfaces $\tilde D_I$'s is 
just the intersection between $\mathbb{C}P^{5}$ and  $SO(10)/U(5)$, this 
means 
\begin{equation}\label{intersection CP5 and PPS}
\tilde D_1\cap...\cap \tilde D_{10}= \mathbb{C}P^{5}\cap SO(10)/U(5)\Big|_{\mathbb{C}P^{15}}.  
\end{equation}
Since $SO(10)/U(5)$ is a smooth manifold on $\mathbb{C}P^{15}$, the
multiplicity in each intersection point of (\ref{intersection CP5 and
PPS}) is one \cite{harris}. So, the sum of the multiplicity at each
intersection point $p_\nu$ is the number of intersection points among
$\mathbb{C}P^5$ and $SO(10)/U(5)$, denoted by 
$^{\#}(SO(10)/U(5)\cdot\mathbb{C}P^{5})$
\begin{equation}\label{degree1}
\sum_{\nu}(\tilde D_1,...,\tilde
D_{10})_{p_{\nu}}=\,\,^{\#}(SO(10)/U(5)\cdot\mathbb{C}P^{5}),
\end{equation}
This number is called the degree of the projective pure spinor space\\
$\text{deg}(SO(10)/U(5)) \equiv
\,^{\#}(SO(10)/U(5)\cdot\mathbb{C}P^{5})$. In \cite{humberto one loop} it was 
shown that the degree of this space is given by
\begin{equation}\label{degree}
\text{deg}(SO(10)/U(5))=\int_{SO(10)/U(5)}\frac{\o^{10}}{(2\pi i)^{10}},
\end{equation}
where $\o$ is 
\begin{equation}\label{generator of cohomology}
\o=-\p\pb\ln(\tilde\l\tilde\lb),
\end{equation}
and $\tilde\l^\alpha$ is an holomorphic coordinate for $SO(10)/U(5)$. Therefore we have that 
\begin{eqnarray}\label{amplitude and omega}
\mathcal{A}&=&
\int_{\Gamma}[\d\l]\frac{\e^{\a_{1}...\a_{5}\b_1...\b_{11}}C^1_{\b_1}...C^{11}_{\b_{11}}}{C^1\l..
.C^{11}\l}
(\l\g^{m})_{\a_1}(\l\g^{n})_{\a_2}(\l\g^{p})_{\a_3}(\g_{mnp})_{\a_4\a_5}K\no\\
&=&\int_{SO(10)/U(5)}(2\pi i)\,2^3\,\,5!\,\,\o^{10}\, K
\end{eqnarray}

Notice that using the pure spinor measure \cite{humberto one loop}
\begin{equation}\label{holomorphic measure2}
[\d\l](\l\g^m)_{\a_1}(\l\g^n)_{\a_2}(\l\g^p)_{\a_3}(\g_{mnp})_{\a_4\a_5}
=\frac{2^3}{11!}\e_{
\a_1...\a_5\b_1...\b_{11}} \d\l^{\b_1}\wedge...\wedge\d\l^{\b_{11}},
\end{equation}
and replacing this measure in the amplitude (\ref{beta cech cochain}) we 
obtain \footnote{The normalization factor $2^3$ in the measure comes from the 
fact that\\
$(\l\g^m)_{\a_1}(\l\g^n)_{\a_2}(\l\g^p)_{\a_3}(\g_{mnp})_{\a_4\a_5}(\g^q\lb)^
{\a_1}(\g^r\lb)^{\a_2}(\g^s\lb)^{\a_3}(\g_{qrs})^{\a_4\a_5}=2^6
5!(\l\lb)^3$.\label{normalization}}
\begin{eqnarray}\label{amplitude6}
\mathcal{A}&=&
2^3
\int_{{\Gamma}} 5! \frac{(\d  f^1)\wedge...\wedge(\d  f^{11})} {  f^1..
.f^{11}}
K
\end{eqnarray} 
where $f^I=C^I\l$, $I=1,..,11$. In the same way as (\ref{amplitude5}) this 
integral is given by the intersection theory \cite{harris}\cite{intersection} 
\begin{eqnarray}\label{integral2}
\int_{{\Gamma}} \frac{(\d  f^1)\wedge...\wedge(\d  f^{11})} {  f^1..
.f^{11}}
=(2\pi i)^{11}(D_1,...,D_{11})_{\{0\}},
\end{eqnarray} 
where the origin is the unique point of intersection between the hypersurfaces $D_I$'s given by (\ref{patch U_I}), $D_I=\{\l^\a\in PS: f^I=0\}, I=1,...,11$, i.e  
$D_1\cap...\cap D_{11}=\{0\}$ as we claimed, and $(D_1,...,D_{11})_{\{0\}}$ 
means the multiplicity of this intersection. So using
(\ref{amplitude5}), (\ref{integral}), (\ref{degree1}) and
(\ref{integral2}) we can conclude
\begin{equation}
(D_1,...,D_{11})_{\{0\}}=\text{deg}(SO(10)/U(5)).
\end{equation}
As computed in \cite{humberto one loop}, the degree of the projective pure 
spinor space is 12. That explains why the multiplicity in the 
intersection point, i.e the origin,  between the matrix (\ref{Cmatrix}) and $PS$ is 12.

\subsubsection{The Dolbeault Cocycle}

Now, using the degree of the projective pure spinor space we
can compute easily the Dolbeault cocycle corresponding to the
form $\b^{I_1...I_{11}}$. From (\ref{amplitude and omega}) 
we have that the scattering amplitude is
\begin{eqnarray}\label{integral3}
\mathcal{A}=\int_{\Gamma}\b^{1,...,11}
=(2\pi i)\,\,2^3\,\,5! \int_{SO(10)/U(5)}\o^{10}\,\,K.
\end{eqnarray}
Writing $\omega$ in coordinates as in the appendix \ref{proof of identity}, we have
\begin{eqnarray}\label{integral4}
\int_{\Gamma}\b^{1,...,11}
&=&(2\pi i)\,2^3\,5! \int_{SO(10)/U(5)}\o^{10}\,K\no\\
&=&(2\pi i)\,\,2^3\,\,5!\int_{\mathbb{C}^{20}}\frac{(10!)\,\,\,\bigwedge_{a<b,\,c<d}\d u_{ab}\d\bar
u^{cd}}{(1+\frac{1}{2}u_{ab}\bar
u^{ab}+\frac{1}{8^{2}}\epsilon^{abcde}\epsilon_{afghi}u_{bc}u_{de}\bar
u^{fg}\bar u^{hi})^{8}}\,K\nonumber\\
&=&\,\,2^3\,\,5!\int_{\mathbb{C}^{20}}\int_{0}^{2\pi}\frac{i(10!)\,\,\,\d\phi\bigwedge_{a<b,\,c<d}\d
u_{ab}\d\bar
u^{cd}}{(1+\frac{1}{2}u_{ab}\bar
u^{ab}+\frac{1}{8^{2}}\epsilon^{abcde}\epsilon_{afghi}u_{bc}u_{de}\bar
u^{fg}\bar u^{hi})^{8}}\,\,K.
\end{eqnarray}
So (\ref{integral4}) is a 21-form evaluated locally on the $SO(10)/SU(5)$ 
space given by (\ref{so(10)/su(5) space}). This can be seen in the following 
simple way: the variables $u_{ab}$ parametrize the projective pure spinor 
space in the patch $\l^+\neq 0$, i.e $\tilde{\l}^{\a}=(1,u_{ab},\frac{1}{8}
\e^{abcde}u_{bc}u_{de})$, and $\phi$ parametrizes the circle
$\g=e^{i\phi}$. So we have locally the space
$SO(10)/U(5)\Big|_{\l^+\neq 0}\,\times\, U(1)$. Since $U(5)=U(1)\times
SU(5)$ then we get the space $SO(10)/U(5)\Big|_{\l^+\neq 0}\,\times\,
U(1)=SO(10)/SU(5)\Big|_{\l^+\neq 0}$.  Note that we have done just a
local analysis. Actually, it is impossible to write globally the space 
$SO(10)/SU(5)$ as the product between the projective pure spinor space and 
the circle, $SO(10)/SU(5)\neq SO(10)/U(5)\,\times\, U(1)$.     

The expression (\ref{integral4}) means that we found the Dolbeault 
cocycle $\eta_\b$ evaluated in the space $SO(10)/SU(5)$ locally, i.e we
got $(i^*\eta_\b)\Big|_{\l^+\neq 0}$, where $i$ is the embedding
$i:SO(10)/SU(5)\rightarrow PS\smallsetminus\{0\}$, explained in the
sub-subsection \ref{CD isomorphism}. In the following, we are going to obtain
$\eta_\b$ in a covariant way in the $PS\smallsetminus\{0\}$ space.

Remember that the holomorphic pure spinor measure $[\d\l]$ was given in
(\ref{holomorphic measure2}). We define a new antiholomorphic 10-form  in the $PS\smallsetminus\{0\}$ space as
\begin{eqnarray}\label{antiholomorphic 10-form}
\[\d\lb\]^{\prime}(\lb\g^m)^{\a_1}(\lb\g^n)^{\a_2}(\lb\g^p)^{\a_3}(\g_{mnp})^{\a_4\a_5}&=&\frac{2^3}{10!}\e^{
\a_1...\a_5\b_1...\b_{11}}
\d\lb_{\b_1}\wedge...\wedge\d\lb_{\b_{10}}\lb_{\b_{11}},
\end{eqnarray}
where $\lb_{\a}$ is a pure spinor, $\lb_\a(\g^m)^{\a\b}\lb_\b=0$. Note that (\ref{antiholomorphic 10-form}) has the same algebraic 
expression as in (\ref{antiholomorphic measure1}), with the difference that 
in this case $\lb_\a$ belongs to the $PS\smallsetminus\{0\}$ space
while the one in (\ref{antiholomorphic measure1}) it is a projective pure spinor.  It is easy to see that 
in the parametrization on the patch $\l^{+}\neq 0$
\begin{eqnarray}\label{parametrization1}
\l^{\a}=\g(1,u_{ab},\e^{abcde}u_{bc}u_{de}/8),\qquad \bar\l_{\a}=\bar\g(1,\bar u^{ab},\e_{abcde}\bar u^{bc}\bar u^{de}/8),
\end{eqnarray}
the (11,10)-form $[\d\l]\wedge[\d\lb]^{\prime}$ becomes
\begin{equation}
[\d\l]\wedge[\d\lb]^{\prime}=\g^7\bar\g^8\d\g\wedge\d u_{12}\wedge...\wedge\d u_{45}\wedge\d\bar u^{12}\wedge...\wedge\d\bar u^{45}. 
\end{equation}
The $SO(10)/SU(5)$ space given in (\ref{so(10)/su(5) space}) is
parametrized on the patch $\l^+\neq 0$ in the following way
\begin{eqnarray}\label{parametrization1}
\l^{\a}=r\,e^{i\phi}(1,u_{ab},\e^{abcde}u_{bc}u_{de}/8), 
\qquad\text{ where $r$ is positive constant}.
\end{eqnarray}
So, we can write the 21-form of (\ref{integral4}) as
\begin{eqnarray}
\frac{[\d\l]\wedge[\d\lb]^{\prime}}{(\l\lb)^{8}}\Bigg|_{SO(10)/SU(5)\Big|_{\l^+\neq0}}=
\frac{i\,\,\,\d\phi\bigwedge_{a<b,\,c<d}\d u_{ab}\d\bar
u^{cd}}{(1+\frac{1}{2}u_{ab}\bar
u^{ab}+\frac{1}{8^{2}}\epsilon^{abcde}\epsilon_{afghi}u_{bc}u_{de}\bar
u^{fg}\bar u^{hi})^{8}}.
\end{eqnarray}

Using the pure spinor constraint it is not hard to verify that the (11,10)-
form $[\d\l]\wedge[\d\lb]^{\prime}/(\l\lb)^{8}$ is $\bar\p$ closed on 
$PS\smallsetminus\{0\}$:
\begin{equation}
\bar\p\(\frac{[\d\l]\wedge[\d\lb]^{\prime}}{(\l\lb)^{8}}\)=0.
\end{equation}
Therefore, the (11,10)-form $[\d\l]\wedge[\d\lb]^{\prime}/(\l\lb)^{8}$ belongs
to cohomology group $H^{(11,10)}_{\bar\p}(PS\smallsetminus\{0\})$ and the 
pull back $i^{*}$ is just the restriction
\begin{equation}\label{generator so(10)/su(5)}
i^*\(\frac{[\d\l]\wedge[\d\lb]^{\prime}}{(\l\lb)^{8}}\)=\frac{[\d\l]\wedge[\d\lb]^{\prime}}{(\l\lb)^{8}}\Bigg|_{SO(10)/SU(5)},
\end{equation}
which is an element of the de-Rham cohomology group
$H^{21}_{DR}(SO(10)/SU(5))$. Finally, we found the Dolbeault cocycle $\eta_\b$
corresponding to $\b^{1,...,11}$ 
\begin{equation}
\b^{1,...,11}\,\,^{\underrightarrow{\quad\text{\v{C}ech-Dol}\quad}}\,\,\eta_\b\equiv 2^3\,5!\,(10!)\frac{[\d\l]\wedge[\d\lb]^{\prime}}{(\l\lb)^{8}}\,K, 
\end{equation}
and (\ref{integral4}) in a covariant way is given by
\begin{equation}\label{integral5}
\int_{\Gamma}\b^{1,...,11}=\int_{SO(10)/SU(5)}\eta_\b\Big|_{SO(10)/SU(5)}\equiv 2^3\,5!\,\int_{SO(10)/SU(5)}(10!)\frac{[\d\l]\wedge[\d\lb]^{\prime}}{(\l\lb)^{8}}\Bigg|_{SO(10)/SU(5)}\, K. 
\end{equation}

Using the \v{C}ech-Dolbeault isomorphism we have gone from a theory in
an 11-cycle $\Gamma$ to a theory in the whole $SO(10)/SU(5)$ space.
Furthermore, notice that  since the non-minimal pure spinor formalism is
defined in the whole pure spinor space $PS\smallsetminus\{0\}$, which is
a non-compact space, then there are an infinite number of global functions 
on it such that the amplitude does not change. These functions are called 
 regulators. This is in contrast with the $SO(10)/SU(5)$ space, which is a 
 compact manifold whose unique generator is given by (\ref{generator so(10)/su(5)}).

Note that integrating the non compact direction of the $PS\smallsetminus\{0\}$ space we get the space
$SO(10)/SU(5)$. This means that for any regulator in the non-minimal
formalism after integrating the non compact direction of the 
$PS\smallsetminus\{0\}$ space, one must get the expression
(\ref{integral5}). We will be more explicit by using coordinates in
the following. If $\l^\a$ is a pure spinor, then it can be written as 
$\l^\a=\g\tilde\l^\a$, where $\g\in\mathbb{C}^*= U(1)\times \mathbb{R}^+$ 
and $\tilde\l^\a$ is a projective pure spinor. So, setting $\g=\r\, e^{i\phi}$, where $e^{i\phi}\in U(1)$
and $\r\in\mathbb{R}^+$ and integrating by $\r$ in the non-minimal formalism
we must get (\ref{integral5}) for any regulator. This implies that our picture 
changing operators does not correspond to any particular regulator and 
therefore we believe that the scattering amplitude prescription with the 
new picture operators is perhaps more fundamental than the prescription with 
regulators.

\subsubsection{A Particular Regulator}

In this sub-subsection we would like to illustrate what we said in the last paragraph
with a particular regulator.

The most useful regulator in the non-minimal pure spinor formalism for 
computing the tree level scattering amplitude is 
\begin{equation}
\mathcal{N}=\exp(-\lb_\a\l^\a-r_\a\t^\a),
\end{equation}
as given in \cite{nathan topological},
where $r_\a$ is a spinor such that $r_\a(\g^m)^{\a\b}\lb_\b=0$. After
integrating the variables $r_\a$ and $\t^\a$ we get \cite{humberto and carlos}
\begin{equation}
\mathcal{A}=\,2^3\,5!\,\int_{PS}[\d\l]\wedge[\d\lb]\,e^{-(\l\lb)}\,(\l\lb)^{3}\,K 
\end{equation}
where the measure $[\d r]$ was given in \cite{humberto one loop}\cite{nathan topological}
\begin{equation}
[\d  r]=\frac{1}{2^3\, 5! \,11!}(\lb\g^m)^{\a_1}(\lb\g^n)^{\a_2}(\lb\g^p)^{\a_3}(\g_{mnp})^{\a_4\a_5}\e_{\a_1,...,\a_5\b_1...\b_{11}}\p_r^{\b_1}....\p_r^{\b_{11}}, 
\end{equation}
the factor $2^3$ comes from a normalization explained in the footnote
\ref{normalization}. We replaced the vertex
operators in the amplitude by $(\l\g^m)_{\a_1}(\l\g^n)_{\a_2}(\l\g^p)_{\a_3}(\g_{mnp})_
{\a_4\a_5}\,K$, where $K$ is the kinematic factor. Using the coordinates
$\l^\a=\g\tilde\l^\a=\r\,e^{i\phi}\tilde\l^\a$, which were explained 
previously, the integration measure is \cite{beta gamma system}
\begin{eqnarray}\nonumber
[\d\l]\wedge[\d\lb]=
(\g\bar\g)^7\,\d\g\wedge\d\bar\g\wedge[\d\tilde\l]\wedge[\d\tilde\lb]&=&
-2\,i\,(\r^2)^7\,\r\,\d\r\wedge\d\phi\wedge[\d\tilde\l]\wedge[\d\tilde\lb]\\
&=&
-2\,i\,(\r^2)^7\,\r\,\d\r\wedge[\d\l]\wedge[\d\lb]'\Big|_{SO(10)/SU(5)\Big|_{r=1}},
\end{eqnarray}
where $SO(10)/SU(5)|_{r=1}$ means that the space $SO(10)/SU(5)$ has size
$r=1$ (see (\ref{so(10)/su(5) space})).
So, integrating the non-compact variable $\r$ from $r_0$ to $r$ we get  
\begin{eqnarray}
&& 2^3\,5!\,\int_{PS}[\d\l]\wedge[\d\lb]\,e^{-(\l\lb)}\,(\l\lb)^{3}\,K\\
&=& 2^3\,5!\int_{SO(10)/SU(5)}(10!)[\d\l]\wedge[\d\lb]^{\prime}
\left(
\frac{ e^{-\r^2 (\tilde\l\tilde\lb) }}{(\tilde\l\tilde\lb)^8 } + \frac{\r^2 e^{-\r^2 (\tilde\l\tilde\lb) }}{
 (\tilde\l\tilde\lb)^7 } + \frac{\r^4 e^{-\r^2 (\tilde\l\tilde\lb) }}{2 (\tilde\l\tilde\lb)^6} + \right.  \no\\
&&\quad+ 
\frac{
 \r^6 e^{-\r^2 (\tilde\l\tilde\lb) }}{3!(\tilde\l\tilde\lb)^5 } +
\frac{ \r^8 e^{-\r^2 (\tilde\l\tilde\lb)}}{4!
 (\tilde\l\tilde\lb)^4 }+\frac{ \r^{10} e^{-\r^2 (\tilde\l\tilde\lb) }}{5! (\tilde\l\tilde\lb)^3 }
\frac{
  \r^{12} e^{-\r^2 (\tilde\l\tilde\lb)}}{6! (\tilde\l\tilde\lb)^2 } +\frac{ \r^{14} e^{-\r^2 (\tilde\l\tilde\lb)}}{7!
 (\tilde\l\tilde\lb) } + \no\\
&&\quad+\left.
\frac{ \r^{16} e^{-\r^2 (\tilde\l\tilde\lb) }}{8!} +\frac{
  \r^{18} (\tilde\l\tilde\lb) e^{-\r^2 (\tilde\l\tilde\lb)}}{9!} + \frac{ \r^{20} (\tilde\l\tilde\lb)^2 e^{-\r^2 (\tilde\l\tilde\lb)}}{10!}\right)\Bigg|_{SO(10)/SU(5)|_{\r=r_0}}^{SO(10)/SU(5)|_{\r=r}}K,\no
\end{eqnarray}
Note that $SO(10)/SU(5)|_{\r=r}-SO(10)/SU(5)|_{\r=r_0}$ is the boundary of the finite pure spinor space, i.e
\begin{equation}
PS_{r_0,r}\equiv\{\l^{\a}\in\mathbb{C}^{16}: \l^\a(\g^m)_{\a\b}\l^\b=0 \text{ and } r_0^2\leq\l^\a\lb_\a\leq r^2\},  
\end{equation}
where $r_0,r$ are positive constants. In order to obtain the whole pure spinor space, $PS\smallsetminus\{0\}$, we must take the limits $r_0\rightarrow 0$ and $r\rightarrow\infty$, so we get the equivalence
\begin{equation}\label{minimo and nonminimo}
2^3\,5!\int_{PS}[\d\l]\wedge[\d\lb]\,e^{-(\l\lb)}\,(\l\lb)^{3}K=2^3\,5!\int_{SO(10)/SU(5)}(10!)\frac{[\d\l]\wedge[\d\lb]^{\prime}}{(\l\lb)^{8}}\Bigg|_{SO(10)/SU(5)}K=\int_{\Gamma}\b^{1,...,11}. 
\end{equation}
This is the reason why we say that $SO(10)/SU(5)$ is the ``boundary''
of the $PS\smallsetminus\{0\}$ space, although it is a non-compact
space. The equivalence (\ref{minimo and nonminimo}) holds 
for any regulator because the \v{C}ech-Dolbeault isomorphism relates
the minimal formalism with a formalism in $SO(10)/SU(5)$, which only has one cohomology generator given by  $([\d\l]\wedge[\d\lb]^{\prime}/(\l\lb)^{8})|_{SO(10)/SU(5)}$.

Although the equivalence between the minimal and non-minimal formalisms
is somewhat premature because in tree level we can absorb any number
in the coupling constant $e^{-2\mu}$ \cite{humberto and carlos}, the
previous result is beautiful and it will be very import for computing loop amplitudes \cite{work in progress}.


\section{Symmetries of the Scattering Amplitude}
\label{symmetries of the scattering amplitude}
In this section we analyze the symmetries of the scattering amplitude
with the new PCO's. Namely, we will show that the scattering amplitude is 
invariant under BRST, Lorentz and supersymmetry transformations. Here we 
will often use the \v{C}ech language and the \v{C}ech-Dolbeault isomorphism 
presented in the subsection \ref{CDLanguage}. 

\subsection{BRST Invariance}
\label{brst invariance}
We will show that the tree level scattering  
amplitude is BRST invariant and that the $Q$ exact states are
decoupled.\\
As we discussed in the subsection \ref{CDLanguage}, the PCO's are
defined locally because they behave like 
$1/\l$: $Y^I_C=\frac{C^I\t}{C^I\l}$ and they are well defined only in
$U_I=PS\smallsetminus D_I$. Therefore, as proposed in \cite{nathan nikita 
multiloops} one must add to the old BRST charge 
\begin{equation}
Q=\oint \d z\,\l^\a d_\a, 
\end{equation}
where $d_\a$ are the constraints (\ref{dconstraint}), the \v{C}ech operator $\delta$ 
given in (\ref{cech operator}). The $\de$ operator play an important role 
in the construction of the $b$-ghost, as we will discuss in the
section \ref{loops}. So the total BRST charge is
\begin{equation}
Q_{T}=\oint \l^{\a}d_\a + \delta \equiv Q+\delta.
\end{equation}
By definition, if the tree level scattering amplitude $\mathcal{A}$ is
physical then it must be $Q_T$ closed, i.e $Q_T\,\mathcal{A}=0$.

In the following we will show that the amplitude is $Q_T$ closed. First of 
all, remember that in the tree level scattering amplitude the vertex operators
can always be written as a global function in $PS\smallsetminus\{0\}$ given 
by $\l^\a\l^\b\l^\g\,\, f_{\a\b\g}(\t,\,k_i,\,e_i)$
\cite{nathan minimal pure spinor}, where the $k_i$'s are the momenta and the 
$e_i$'s are the polarizations of the vertex operators.
Since the tree level scattering amplitude is given by
\begin{equation}\label{amplitude general}
\mathcal{A}=\int_\Gamma[\d\l]\int\d^{16}\t \prod_{I=1}^{11}Y^I_C\,\l^\a\l^\b\l^\g\,\, f_{\a\b\g}(\t,\,k_i,\,e_i).
\end{equation}
and the measure $[\d\l]$ is globally defined on $PS\smallsetminus\{0\}$ 
\cite{beta gamma system}, then the scattering amplitude is $\de$ closed.
This was explained carefully in the subsection 
\ref{equivalence mps and nmps} (see the explanation after 
(\ref{beta cech cochain})). Now it remains to show $Q\,\mathcal{A}=0$.
Because 
\begin{equation}
Q\, Y^I_C =1,
\end{equation}
therefore we have
\begin{eqnarray}\label{q of the amplitude}
Q(\mathcal{A})&=& Q\(\int_\Gamma[\d\l]\int\d^{16}\t \prod_{I=1}^{11}Y^I_C\,\l^\a\l^\b\l^\g\,\, f_{\a\b\g}(\t,\,k_i,\,e_i)\)\\
&=&\frac{1}{11!}\sum_{I_1,...,I_{11}=1}^{11}\e_{I_1,...,I_{11}}Q\(\int_\Gamma[\d\l]\int\d^{16}\t\frac{C^{I_1}\t...C^{I_{11}}\t}{C^{I_1}\l...C^{I_{11}}\l}\,\l^\a\l^\b\l^\g\,\, f_{\a\b\g}(\t,\,k_i,\,e_i)\)\no\\
&=&\frac{1}{11!}\sum_{I_1,...,I_{11}=1}^{11}\e_{I_1,...,I_{11}}\int_\Gamma
(\de\tau)^{I_1,...,I_{11}}\no
\end{eqnarray}
where $\tau^{I_1,...,I_{10}}$ is the holomorphic 11-form 
\begin{eqnarray}
\tau^{I_1,...,I_{10}}=
[\d\l]\int\d^{16}\t\frac{C^{I_1}\t...C^{I_{10}}\t}{C^{I_1}\l...C^{I_{10}}\l}\,\l^\a\l^\b\l^\g\,\, f_{\a\b\g}(\t,\,k_i,\,e_i)\in\,C^9(\underline{U},\Omega^{11})
\end{eqnarray}
where $\underline{U}$ is the cover of the $PS\smallsetminus\{0\}$ space given 
in the subsection \ref{CDLanguage}. Clearly $(\de\tau)^{I_1,...,I_{11}}$ is a trivial element of the \v{C}ech cohomology group $H^{11}(PS\smallsetminus\{0\},\Omega^{11})$, so its corresponding Dolbeault cocycle 
\begin{equation}
\eta_{(\de\tau)}=\frac{1}{10!}\sum_{I_1,...,I_{11}=1}^{11}(\de\tau)^{I_1,...,I_{11}}\rho_{I_1}\wedge\bar\p\rho_{I_2}\wedge...\wedge\bar\p\rho_{I_{11}}, 
\end{equation}
where $\rho_I$ is partition of unity (\ref{partition of unity1}), is a trivial element of the Dolbeault cohomology group $H^{(11,10)}_{\bar\p}(PS\smallsetminus\{0\})$, i.e
\begin{equation}
\eta_{(\de\tau)}=\bar\p(\eta_\tau)\qquad \text{(11,10)-form on }PS\smallsetminus\{0\}, 
\end{equation}
where $\eta_\tau$ is the (11,9)-form given by
\begin{equation}
\eta_{\tau}=\frac{1}{9!}\sum_{I_1,...,I_{10}=1}^{11}\tau^{I_1,...,I_{10}}\rho_{I_1}\wedge\bar\p\rho_{I_2}\wedge...\wedge\bar\p\rho_{I_{10}},
\end{equation}
as explained in the sub-subsection \ref{CD isomorphism}. So we can write (\ref{q of the amplitude}) as
\begin{equation}
\int_\Gamma
(\de\tau)^{1,...,11}=\int_{SO(10)/SU(5)}i^*(\bar\p(\eta_\tau))=\int_{SO(10)/SU(5)}\d(i^*(\eta_\tau)), 
\end{equation}
where ``$i$'' is the map $i:\,SO(10)/SU(5)\rightarrow
PS\smallsetminus\{0\}$ given in the sub-subsection \ref{CD isomorphism}. Finally, applying the Stokes theorem 
\begin{equation}
\int_\Gamma
(\de\tau)^{1,...,11}=\int_{SO(10)/SU(5)}\d(i^*(\eta_\tau))=\int_{\p(SO(10)/SU(5))}i^*(\eta_\tau) 
\end{equation}
and since the $SO(10)/SU(5)$ space is a compact manifold without boundary, $\p(SO(10)/SU(5))=\varnothing$,  then we can conclude
\begin{equation}\label{q A =0}
Q(\mathcal{A})=0. 
\end{equation}
Thus, we have shown that the tree level scattering amplitude is $Q_T$ closed.

Now we will show that the global (i.e
$(\delta\Omega)^{IJ}=\Omega^J-\Omega^I=0$) and $Q$ exact (i.e
$\<Q(\Omega)\>$) functions are decoupled,
that is, they are $Q_T=Q+\de$ exact.
A $Q$ exact function inside to the scattering amplitude is given by 
\begin{equation}\label{decoupling}
\<Q(\Omega)\>=\int_{\Gamma} [\d\l] \int\d^{16}\t Y^1_C...Y^{11}_C  Q(\Omega(\l,\t,k)).
\end{equation}
Only the terms with 5 $\t$'s 
and 3 $\l$'s in $Q(\Omega)$ will contribute,
because there are $11$ $\t$'s coming from the 11 PCO's and 
the scattering amplitude must have ghost number zero. So, we focus on the global term
\begin{equation}
\Omega(\l,\t,k,e)=\l^{\a}\l^{\b}\t^{\g_1}...\t^{\g_6}f_{\a\b\g_1...\g_6}(k_i,e_i) ,
\end{equation}
where $k_i$ are the momenta and $e_i$ are the polarizations.   
We can write (\ref{decoupling}) as 
\begin{eqnarray}
\int_{\Gamma} [\d\l]\int\d^{16}\t\, Y^1_C...Y^{11}_C  Q(\Omega(\l,\t,k))&=&
-\int_{\Gamma} [\d\l]\int\d^{16}\t\, Q(Y^1_C...Y^{11}_C  \Omega(\l,\t,k))\no\\
&& +\int_{\Gamma} [\d\l]\int\d^{16}\t\, Q(Y^1_C...Y^{11}_C)  \Omega(\l,\t,k).
\end{eqnarray}
The term $Y^1_C...Y^{11}_C  \Omega(\l,\t,k)$ is identically zero because there are 17 $\t$'s. 
So
\begin{eqnarray}
\int_{\Gamma} [\d\l]\int\d^{16}\t\, Y^1_C...Y^{11}_C  Q(\Omega(\l,\t,k))&=&
\int_{\Gamma} [\d\l]\int\d^{16}\t\, Q(Y^1_C...Y^{11}_C)  \Omega(\l,\t,k)\\
&=&\frac{1}{11!}\sum_{I_1,...,I_{11}=1}^{11}\e_{I_1...I_{11}}\int_{\Gamma} [\d\l]\int\d^{16}\t\, Q(Y^{I_1}_C...Y^{I_{11}}_C)  \Omega(\l,\t,k)\no\\
&=&\frac{1}{11!}\sum_{I_1,...,I_{11}=1}^{11}\e_{I_1...I_{11}}\int_{\Gamma} (\de\kappa)^{I_1...I_{11}}\no,
\end{eqnarray}
where $\kappa^{I_1...I_{10}}$ is the holomorphic 11-form 
\begin{equation}
\kappa^{I_1...I_{10}}=[\d\l]\int\d^{16}\t\, \frac{C^{I_1}\t...C^{I_{10}}\t}{C^{I_1}\l...C^{I_{10}}\l}\,
\l^{\a}\l^{\b}\t^{\g_1}...\t^{\g_6}f_{\a\b\g_1...\g_6}(k_i,e_i)\in \, C^{9}(\underline{U},\Omega^{11}). 
\end{equation}
Note that $\delta\<Q(\Omega)\>=0$, since $(\delta(\delta\kappa))^{I_1...I_{12}}=0$. Using the same 
procedure that allowed as go from (\ref{q of the amplitude}) and to
conclude in (\ref{q A =0}) we have
\begin{equation}
\<Q(\Omega)\>=\frac{1}{11!}\sum_{I_1,...,I_{11}=1}^{11}\e_{I_1...I_{11}}\int_{\Gamma} (\de\kappa)^{I_1...I_{11}}=\int_{\p(SO(10)/SU(5))}i^*(\eta_\kappa)=0. 
\end{equation}
Therefore we have shown that every global and exact function inside to the scattering amplitude is decoupled.

For a general case we must show that the scattering amplitude decouple the 
states which are $Q_T$ exact, i.e
\begin{equation}\label{QT decouple}
\<(Q+\delta)(\Omega)\>=0 
\end{equation}
for any $\Omega$. \\
First of all, we know that the BRST operator is nilpotent  $(Q+\de)^2=0$
and we want to show that the BRST exact terms are decoupled from the
scattering amplitude. The $Q_T$ exact terms are written as 
\begin{equation}
\<(Q+\de)(\Omega)\>=\int_{\Gamma}[\d\l]\int\d^{16}\t\prod_{I=1}^{11}Y^{I}_C(Q+\de)(\Omega).
\end{equation}
However, as  the 11-form  $\,[\d\l]\prod_{I=1}^{11}Y^{I}_C$ is a 10-\v{C}ech cochain, 
$C^{10}(\underline{U},\Omega^{11})$, then it is possible that the
product of the cochains $\,[\d\l]\prod_{I=1}^{11}Y^{I}_C$ and
$\,(Q+\de)(\Omega)$ is not well defined, because the product of two
cochains is not always a cochain. For instance, let us consider the 
following 2 cochains
\begin{equation}
Y^I_C=\frac{C^I\t}{C^I\l}\in C^{0}(\underline{U},\mathcal{O}),\qquad
\Omega^J=\frac{\Lambda_{mn}(C^J\g^{mn}\t)}{(C^J\l)} \in
C^{0}(\underline{U},\mathcal{O}).
\end{equation}
Clearly
\begin{equation}\label{example of product of 2 cochain}
\Psi^{IJ}\equiv Y^I_C\,\Omega^J=\frac{(C^I\t)\,(C^J\g^{mn}\t)\Lambda_{mn}}{(C^I\l)(C^J\l)}\neq -\frac{(C^J\t)\,(C^I\g^{mn}\t)\Lambda_{mn}}{(C^I\l)(C^J\l)}\notin C^{1}(\underline{U},\mathcal{O}), 
\end{equation}
In the particular case when $\Omega$ is a global holomorphic function in $PS\smallsetminus\{0\}$ the product with any \v{C}ech cochain is well defined, 
for example the vertex operators in (\ref{amplitude general}), or as in the 
computation (\ref{decoupling}). Note also that the \v{C}ech operator
is not a derivate operator, i.e it does not satisfy the Leibniz rule.
So it is not well defined acting on the product (\ref{example of product of 2 cochain}) 
\begin{equation}\label{operador de cech leibniz}
(\de \Psi)^{IJK}\neq (\de Y)^{IJ}\Omega^K \pm Y^I (\de \Omega)^{JK}. 
\end{equation}
Therefore the expressions $(Q+\de)\<(Q+\de)(\Omega)\>$ and
$\<(Q+\de)(Q+\de)(\Omega)\>$ are not equal i.e 
$(Q+\de)\<(Q+\de)(\Omega)\>\neq\<(Q+\de)(Q+\de)(\Omega)\>$, 
and in most cases the left hand side is not defined when $\Omega$ has \v{C}ech labels. 
Therefore the expression (\ref{QT decouple}) does not make sense
unless that $\Omega$ will be a global holomorphic function, like we
assumed in (\ref{decoupling}).

From the analysis above we can conclude that for the tree level scattering 
amplitudes, the naive existence of the homotopy operator 
\cite{nathan topological}\cite{nathan nikita multiloops} given by 
\begin{equation}\label{homotopy operator}
\xi = \frac{C^I\t}{C^I\l}\Bigg|_{U_I}+\frac{C^I\t C^J\t}{C^I\l
C^J\l}\Bigg|_{U_I\cap U_J}+...+\frac{C^1\t C^2\t...C^{11}\t}{C^1\l C^2\l...C^{11}\l}\Bigg|_{U_1\cap U_2\cap...\cap U_{11}},
\end{equation}
which by definition satisfies 
\begin{equation}
(Q+\de)(\xi V_1V_2V_3 U_1...U_{N-3})=V_1V_2V_3 U_1...U_{N-3} 
\end{equation}
for $V_1V_2V_3$ unintegrated vertex operators and $U_1...U_{N-3}$ integrated vertex operators, is not allowed because
\begin{equation}
\xi V_1V_2V_3 U_1...U_{N-3}
\end{equation}
is not a global function on $PS\smallsetminus\{0\}$. Therefore at tree level 
it is sufficient to decouple the global and $Q$ exact functions, see (\ref{decoupling}).


\subsection{Lorentz and Supersymmetry Invariance}

Now we show that although the new lowering picture operators are neither
Lorentz nor supersymmetry invariant, the scattering amplitude is invariant
under both transformations.

\subsubsection{Lorentz Invariance}
It is easy to show that the action of the Lorentz generators
$M^{mn}=(1/2)\int\d z[(\o\g^{mn}\l)+(p\g^{mn}\t)]$ on the
PCO's is $Q$ exact:
\begin{equation}
M^{mn}Y_C ^I = -\frac{1}{2}Q\[\frac{(C^I\g^{mn}\t)(C^I\t)}{(C^I\l)^2}\], 
\end{equation}
then, replacing this in the scattering amplitude we get
\begin{eqnarray}\label{lorentz over amplitude}
&&M^{mn}(\mathcal{A})\\
&=&\frac{1}{2\,\,11!}\sum_{I_1,...,I_{11}=1}^{11}\e_{I_1,...,I_{11}}\int_\Gamma[\d\l]\int\d^{16}\t\sum_{i=1}^{11}(-1)^iQ\[\frac{(C^{I_i}\g^{mn}\t)(C^{I_i}\t)}{(C^{I_i}\l)^2}\]\frac{C^{I_1}\t...\widehat{C^{I_i}\t}...C^{I_{11}}\t}{C^{I_1}\l...\widehat{C^{I_i}\l}...C^{I_{11}}\l}\no\\
\,&&\,\l^\a\l^\b\l^\g\,\, f_{\a\b\g}(\t,\,k_i,\,e_i)\no\\
&=&\frac{1}{2\,\,11!}\sum_{I_1,...,I_{11}=1}^{11}\e_{I_1,...,I_{11}}\int_\Gamma[\d\l]\int\d^{16}\t\sum_{i=1}^{11}(-1)^{i-1}\frac{(C^{I_i}\g^{mn}\t)(C^{I_i}\t)}{(C^{I_i}\l)^2}Q\(\frac{C^{I_1}\t...\widehat{C^{I_i}\t}...C^{I_{11}}\t}{C^{I_1}\l...\widehat{C^{I_i}\l}...C^{I_{11}}\l}\)\no\\
\,&&\,\l^\a\l^\b\l^\g\,\, f_{\a\b\g}(\t,\,k_i,\,e_i)\no
\end{eqnarray}
where the term 
$
\frac{\widehat{C^{I_i}\t}}{\widehat{C^{I_i}\l}}
$ means it must be removed from the expression. Making an algebraic manipulation we find 
\begin{eqnarray}\label{algebraic manipulation}
\sum_{i=1}^{11}(-1)^{i-1}\frac{(C^{I_i}\g^{mn}\t)(C^{I_i}\t)}{(C^{I_i}\l)^2}Q\(\frac{C^{I_1}\t...\widehat{C^{I_i}\t}...C^{I_{11}}\t}{C^{I_1}\l...\widehat{C^{I_i}\l}...C^{I_{11}}\l}\)&\equiv& \sum_{i=1}^{11}(-1)^{i-1}\pi^{I_i}Q\(Y^{I_1}_C\,...\widehat{Y^{I_i}_C}...\,Y^{I_{11}}_C\)\no\\
& =& (\de\psi)^{I_1...I_{11}}
\end{eqnarray}
where we define
\begin{equation}
\pi^I\equiv \frac{(C^{I}\g^{mn}\t)(C^{I}\t)}{(C^{I}\l)^2}
\end{equation}
and $\psi^{I_1...I_{10}}$ is given by
\begin{eqnarray}\label{definition of psi}
\psi^{I_1...I_{10}}&=&
-\frac{1}{9!}\pi^{[I_1}\,Y^{I_2}_C\,...\,Y^{I_{10}]}_C\\
&\equiv& -\frac{1}{9!}\(\pi^{I_1}\,Y^{I_2}_C\,Y^{I_3}_C...\,Y^{I_{10}}_C-\pi^{I_2}\,Y^{I_1}_C\,Y^{I_3}_C\,...\,Y^{I_{10}}_C+\text{all permutation}\)\in C^{9}(\underline{U},\mathcal{O})\no.
\end{eqnarray}
We define the holomorphic 11-form
\begin{equation}
\Psi^{I_1...I_{10}}=[\d\l]\int\d^{16}\t\,
\psi^{I_1...I_{10}}
\,\l^\a\l^\b\l^\g\,\, f_{\a\b\g}(\t,\,k_i,\,e_i)\in C^{9}(\underline{U},\Omega^{11}).
\end{equation}
So we can write (\ref{lorentz over amplitude}) in the following way
\begin{eqnarray}
M^{mn}(\mathcal{A})
=\frac{1}{2\,\,11!}\sum_{I_1,...,I_{11}=1}^{11}\e_{I_1,...,I_{11}}\int_\Gamma(\de\Psi)^{I_1...I_{11}}.
\end{eqnarray}
With the same procedure used to show the BRST invariance of the
amplitude, i.e following the steps from (\ref{q of the amplitude}) to 
(\ref{q A =0}) we obtain
\begin{eqnarray}\label{lorentz invariant}
\sum_{I_1,...,I_{11}=1}^{11}\e_{I_1,...,I_{11}}\int_\Gamma(\de\Psi)^{I_1...I_{11}}=\int_{\p(SO(10)/SU(5))}i^*(\eta_\Psi)=0
\end{eqnarray}
since $\p(SO(10)/SU(5))=\varnothing$. Finally, we conclude the tree level scattering amplitude is Lorentz invariant
\begin{eqnarray}
M^{mn}(\mathcal{A})=0.
\end{eqnarray}

\subsubsection{Invariance under Supersymmetry}
Now we show that the tree level scattering amplitude is invariant under 
supersymmetry transformations. We call the supersymmetry generator ``$q$'', which is given by
\begin{equation}
q=\varepsilon^\a q_\a 
\end{equation}
where $\varepsilon^\a$ is a Grassmann constant spinor spinor and 
$$
q_{\a}=\int\d z\,(p_{\a}+\frac{1}{2}\g^{m}_{\a\b}\theta^{\b}\p
x_{m}+\frac{1}{24}\g^{m}_{\a\b}\g_{m\,\g\delta}\theta^{\b}\theta^{\g}\p\theta^{\delta}).
$$
It is easy to see that the action of $q$ on the PCO's is
\begin{equation}
q (Y^I_C)=\varepsilon^\a q_\a (Y^I_C) =Q\[ \frac{(\varepsilon C^I)(C^I\t)}{(C^I\l)^2}\]. 
\end{equation}
Therefore in the tree level scattering amplitude we have 
\begin{eqnarray}
&&q(\mathcal{A})\\
&=&\frac{1}{11!}\sum_{I_1,...,I_{11}=1}^{11}\e_{I_1,...,I_{11}}\int_\Gamma[\d\l]\int\d^{16}\t\sum_{i=1}^{11}(-1)^{i-1}Q\[\frac{(\varepsilon C^{I_i})(C^{I_i}\t)}{(C^{I_i}\l)^2}\]\frac{C^{I_1}\t...\widehat{C^{I_i}\t}...C^{I_{11}}\t}{C^{I_1}\l...\widehat{C^{I_i}\l}...C^{I_{11}}\l}\no\\
\,&&\,\l^\a\l^\b\l^\g\,\, f_{\a\b\g}(\t,\,k_i,\,e_i)\no\\
&=&\frac{1}{11!}\sum_{I_1,...,I_{11}=1}^{11}\e_{I_1,...,I_{11}}\int_\Gamma[\d\l]\int\d^{16}\t\sum_{i=1}^{11}(-1)^{i}\frac{(\varepsilon C^{I_i})(C^{I_i}\t)}{(C^{I_i}\l)^2}Q\(\frac{C^{I_1}\t...\widehat{C^{I_i}\t}...C^{I_{11}}\t}{C^{I_1}\l...\widehat{C^{I_i}\l}...C^{I_{11}}\l}\)\no\\
\,&&\,\l^\a\l^\b\l^\g\,\, f_{\a\b\g}(\t,\,k_i,\,e_i)\no.
\end{eqnarray}
Making a similar algebraic manipulation like in (\ref{algebraic manipulation}) we get
\begin{eqnarray}
&&[\d\l]\int\d^{16}\t\sum_{i=1}^{11}(-1)^{i}\frac{(\varepsilon C^{I_i})(C^{I_i}\t)}{(C^{I_i}\l)^2}Q\(\frac{C^{I_1}\t...\widehat{C^{I_i}\t}...C^{I_{11}}\t}{C^{I_1}\l...\widehat{C^{I_i}\l}...C^{I_{11}}\l}\)\no\\
\,&&\,\l^\a\l^\b\l^\g\,\, f_{\a\b\g}(\t,\,k_i,\,e_i)\no\\
&=&
(\de\Phi)^{I_1...I_{11}},
\end{eqnarray}
where we have the following definitions
\begin{equation}
\Phi^{I_1...I_{10}}=[\d\l]\int\d^{16}\t\,
\phi^{I_1...I_{10}}
\,\l^\a\l^\b\l^\g\,\, f_{\a\b\g}(\t,\,k_i,\,e_i)\in C^{9}(\underline{U},\Omega^{11}),
\end{equation}
\begin{eqnarray}\label{definition of phi}
\phi^{I_1...I_{10}}&=&
\frac{1}{9!}\varphi^{[I_1}\,Y^{I_2}_C\,...\,Y^{I_{10}]}_C,
\end{eqnarray}
\begin{equation}
\varphi^I=\frac{(\varepsilon C^I)(C^I\t)}{(C^I\l)^2}. 
\end{equation}
Using the same argument as in (\ref{lorentz invariant}) it is clear that
\begin{eqnarray}
q(\mathcal{A})
=\frac{1}{11!}\sum_{I_1,...,I_{11}=1}^{11}\e_{I_1,...,I_{11}}\int_\Gamma(\de\Phi)^{I_1...I_{11}}=\int_{\p(SO(10)/SU(5))}i^*(\eta_\Phi)=0.
\end{eqnarray}
So the tree level scattering amplitude is invariant under
supersymmetry transformations.

In conclusion, we succeeded in proving the invariance under the total
BRST, Lorentz and supersymmetry transformations, where the
\v{C}ech-Dolbeault language played a central role.


\section{Independence of the Constant Spinors $C^I_{\a}$'s }
\label{independence of Cs}

In this section,  our goal is to show that the scattering amplitude is 
independent of the choice of the constant spinors $C^I$'s.
This implies that they do not need to be integrated, in
contrast with the analysis presented in \cite{nathan minimal pure
spinor}\cite{skenderis}, where it did was necessary.\\
We will present an example of pure spinors in four dimensions, where
the conditions of linear independence for the $C^I$'s and the
intersection of the hyperplanes $D_I$'s in the origin are equivalent. However, in ten
dimensions it is not sufficient that the $C^I$'s are linearly
independent, so, based on the assumption that the hypersurfaces 
$D_I=\{C^I\l=0\},I=1,...,11,$ meet just in the origin, 
we will show that the scattering  amplitude is independent of the
$C^I$'s choice.

\subsection{Pure Spinors in $d=4$: A Simple Example}

Before we show the independence of the $C^I$'s for pure spinors in
ten dimensions, we give a simple example in four dimensions in order to 
understand how this can be achieved.\\ Consider the pure spinor space in 
$d=4$ dimensions, i.e the flat space $\mathbb{C}^2$. In this case the
integral corresponding to (\ref{beta cech cochain}) is given by   
\begin{equation}\label{independence of the CI for d=4}
\int_\Gamma\vartheta=\int_\Gamma
[\d\l]\frac{\e^{cd}C^1_c\,C^2_d}{(C^1\l)(C^2\l)},\qquad c,d=1,2 
\end{equation}
where $\l^a=(\l^1,\l^2)$ are the coordinates of $\mathbb{C}^2$ and the 
measure is simply $[\d\l]=2^{-1}\e_{ab}\,\d\l^a\wedge\d\l^b=\d\l^1\wedge\d\l^2$. 
Clearly, the vectors $C^j,j=1,2$ must be linearly independent in order
to obtain an integral different from zero, i.e the determinant det$(C^j_a)\neq0$. This implies that the intersection of the 
hyperplanes $C^j\l=0,\,j=1,2$ is just the origin. To compute
(\ref{independence of the CI for d=4}), firstly we consider the simple 
case when $C^j_a=\de^j_a$. Then the integral is
\begin{equation}\label{trivial integral}
\int_\Gamma\vartheta = \int_\Gamma \frac{\d\l^1\wedge\d\l^2}{\l^1 \l^2}. 
\end{equation}
Secondly we define in a natural way the 2-cycle $\Gamma$ as the torus
$\Gamma=\{\l^a\in\mathbb{C}^{2}:|\l^1|=\e^1 \text{ and }
\,|\l^2|=\e^2\}$, where $\e^1,\,\e^2$ are positive arbitrary constants. 
So (\ref{trivial integral}) is a trivial integral and its answer is 
\begin{equation}\label{answer trivial integral}
\int_\Gamma\vartheta = (2\pi i)^2.
\end{equation}
Once the answer is known, we must know what happens if we choose two 
arbitrary vectors $C^i_a$ but keep the same 2-cycle 
$\Gamma=\{\l^a\in\mathbb{C}^{2}:|\l^1|=\e^1 \text{ and }\,|\l^2|
=\e^2\}$. In other words, we want to know the answer to the question:
is (\ref{independence of the CI for d=4}) independent of the 
constants $C^j$'s?. We will show that the answer is affirmative and 
its result is the same as in (\ref{answer trivial integral}). 
From (\ref{independence of the CI for d=4}) we have
\begin{equation}\label{integral general Cs}
\int_{|\l^2|=\e^2}\int_{|\l^1|=\e^1} \frac{(a_1b_2-a_2b_1)\d\l^1\wedge\d\l^2}{(a_1\l^1+a_2\l^2)(b_1\l^1+b_2\l^2)},
\end{equation}
where $C^1=(a_1,a_2)$ and $C^2=(b_1,b_2)$. Without loss of generality, we 
can set $a_2,b_1\neq0$. To solve (\ref{integral general Cs}), first
note that since $\e^1$ is an arbitrary constant, it can be set to
a very large value such that the pole $-(b_2/b_1)\l^2$ is inside of
the cycle $|\l^1|=\e^1$, so integrating $\l^1$ we have
\begin{equation}\label{general Cs}
\int_{|\l^2|=\e^2}\int_{|\l^1|=\e^1}
\frac{(a_1b_2-a_2b_1)\d\l^1\wedge\d\l^2}{(a_1\l^1+a_2\l^2)(b_1\l^1+b_2\l^2)}=(2\pi i)\int_{|\l^2|=\e^2}\frac{(a_1b_2-a_2b_1)\d\l^2}{(a_1b_2-a_2b_1)\l^2}, 
\end{equation}
getting the same answer as in (\ref{answer trivial integral}). However, since 
the integral depends on a very large value of $\e^1$, this is
not a satisfactory way for computing, so we must look for a better analysis.\\
As det$(C^j_a)=(a_1b_2-a_2b_1)\neq0$, then we can make the following
change of variables
\begin{eqnarray}
\l^1&=& M^{-1}(b_2z^1-a_2z^2)\\
\l^2&=& M^{-1}(-b_1z^1+a_1z^2)\no 
\end{eqnarray}
where $M=(a_1b_2-a_2b_1)$. Using these new coordinates (\ref{general Cs}) becomes
\begin{equation}
\int_\Gamma \frac{\d z^1\wedge\d z^2}{z^1 z^2}. 
\end{equation}
where $\Gamma$ is the 2-cycle given by $|b_2 z^1-a_2z^2|=\e^1|M|$  and
$|a_1z^2-b_1 z^1|=\e^2|M|$. Since $\e^1$ and $\e^2$ are positive arbitrary 
constants then $|z_1|>0,\,|z_2|>0$  and applying the triangle inequality we get
\begin{eqnarray}
&&0<|z^1|\leq  (\e^1|a_1|+\e^2|a_2|)\\
&&0<|z^2|\leq  (\e^1|b_1|+\e^2|b_2|)\no 
\end{eqnarray}
where without loss of generality, we set $a_2,b_1\neq0$. Therefore the torus 
$\Gamma=\{(z^1,z^2)\in\mathbb{C}^2:|b_2 z^1-a_2z^2|=\e^1|M|,\,|a_1z^2-b_1 z^1|
=\e^2|M|\}$ can be deformed to the torus 
$\Gamma'=\{(z^1,z^2)\in\mathbb{C}^2:|z^1|=  (\e^1|a_1|+\e^2|a_2|)/2,\,|z^2|
= (\e^1|b_1|+\e^2|b_2|)/2\}$. So, we have shown that the integral (\ref{independence of the CI for d=4}) is independent of the constants $C^j$'s when we 
fix the integration cycle, because it can always be deformed to a
cycle of the type $|C^i\l|=\varepsilon^j,\,j=1,2$, for some
$\varepsilon^j,\,j=1,2$. Formally we are saying the following: remember that 
the integral (\ref{independence of the CI for d=4})  just depends of the classes of the homology cycle $[\Gamma]$ and  the cocycle of cohomology $[\vartheta]$ (see subsection 
\ref{integration contours}). So, if det$(C^j_a)\neq0$ then all the holomorphic 
2-forms $\vartheta$ are in the same cohomology class $[\vartheta]$ and
all the 2-cycle $\Gamma=\{(\l^1,\l^2)\in\mathbb{C}^2:
|C^j\l|=\e^j,\,j=1,2\}$ are in the same homology class $[\Gamma]$. Now
we must show the same for pure spinors in $d=10$.

\subsection{Pure Spinors in $d=10$}

In the previous example the conditions det$(C^j_a)\neq0$ and
$\{C^1\l=0\}\cap\{C^2\l=0\}=\{0\}$ were equivalent. However, in the
pure spinor space in $d=10$ the condition det$(C^I_\a)\neq0$ does not
make sense because $I=1,..,11$ and $\a=1,...,16$, but remember
that we have always claimed that $D_1\cap...\cap D_{11}=\{0\}$, where
$D_I=\{\l^\a\in PS: C^I_\a\l^\a=0\}$. In this case is not easy to
follow the same analysis of the previous example because $PS$
 is not a flat space. Therefore we will make use of the ideas presented 
 previously in this paper, like the \v{C}ech-Dolbeault isomorphism, to prove that
 the scattering amplitude is independent of the $C^I$'s.

From (\ref{beta cech cochain}) we have that the amplitude is given by
\begin{equation}
\mathcal{A}=\int_\Gamma \b^{1,...,11} 
\end{equation}
where the 11-cycle $\Gamma$ was defined as $\Gamma=\{\l^\a\in PS:|C^I\l|=\e^I,\,I=1,...,11\},\,\e^I\in\mathbb{R}^+$. In the sub-subsection 4.4.2 we found the Dolbeault cocycle
\begin{equation}
\eta_\b=2^3\,5!\,\frac{[\d\l]\wedge[\d\lb]'}{(\l\lb)^8}\Bigg|_{SO(10)/SU(5)} K
\end{equation}
corresponding to $\b^{1,...,11}$, thanks to the isomorphism from the group $H^{10}(PS\smallsetminus\{0\},\Omega^{11})$ to\\
$H^{21}_{DR}(SO(10)/SU(5))$ (see sub-subsection \ref{CD isomorphism}).
As the Dolbeault cocycle $\eta_\b$ is independent of the constants $C^I$'s, then,
choosing another set of constant spinors $C'^I$, $I=1,..,11$,
such that they satisfy the same condition $D'_1\cap...\cap
D'_{11}=\{0\}$, its \v{C}ech cocycle $\b'^{1,...,11}$ is in the same
cohomology class as $\b^{1,...,11}$, because  $\b'^{1,...,11}$ and $\b^{1,...,11}$ have the same corresponding Dolbeault cocycle $\eta_\b$ and the groups  $H^{10}(PS\smallsetminus\{0\},\Omega^{11})$ and $H^{21}_{DR}(SO(10)/SU(5))$ are isomorphic. It means that
\begin{equation}
\int_\Gamma \b^{1,...,11}=\int_\Gamma \b'^{1,...,11}, 
\end{equation}     
because the cohomology classes $[\b^{1,...,11}]$ and $[\b'^{1,...,11}]$ are the same.\\
So we have shown that the tree level scattering amplitude is independent of the constant spinors $C^I$'s and therefore it is not needed to integrate over them.


\section{Relation with Twistor Space}
\label{relation with twistor}

In the sub-subsection \ref{tl amplitude} we studied the tree-level scattering 
amplitude in the projective pure spinor space. In this section we 
will show that the result found there, given by  the integral
(\ref{amplitude4}), is the same found by Berkovits and Cherkis in 
\cite{pure spinor like twistors}. In that reference, the projective pure 
spinor space  
allowed to get the Green's function for a massless scalar field in
$d=10$ dimensions.

The Green's function for a massless scalar field in $d=10$ dimensions
is given by the integral 
\begin{equation}
\Phi(x)= \int_{\tilde{\Gamma}}[\d\tilde\l]\,\,
F(\tilde\l,\o)|_{\o=x\tilde\l},
\end{equation}
which is written covariantly \cite{pure spinor like twistors}. In this integral, 
$\tilde\l^\a$ is a projective pure spinor, $[\d\tilde\l]$ is the measure 
of the projective pure spinor space given by (\ref{holomorphic measure}), 
while $F(\tilde\l,\o)|_{\o=x\tilde\l}$ is given by
\begin{equation}
 F(\tilde\l,\o) =
 \frac{\e_{\a_1...\a_{11}\b_1...\b_5}A_{1}^{\a_{1}}...A_{11}^{\a_{11}}(\g^m\o)^{\b_1}(\g^n\o)^{\b_2}(\g^p\o)^{\b_3}(\g_{mnp})^{\b_4\b_5}}{\prod_{r=1}^{11}(A_{r}^{\a}\o_{\a})}, \quad \o_{\a}= (x\cdot\g\tilde\l)_\a .
\end{equation}
$A_I^\a$'s are constant spinors and the cycle $\tilde\Gamma$ is given 
by ten out of the eleven poles of $F(\tilde\l,\o)$. The measure (\ref{holomorphic 
measure}) is not suitable for obtaining the relationship between twistors and 
scattering amplitude, so we will modify it as follows. First note that  
for $|x|\neq 0$ then $(x\cdot\g\tilde\l)_{\a}$ is a pure spinor
if and only if $\tilde\l^{\a}$ is also a pure spinor. This is very
easy to show:
\begin{enumerate}
 \item In the backward direction: If $\tilde\l^\a$ is a pure spinor
 in ten dimensions, then $(x\cdot\gamma\tilde\lambda)_\alpha$ is also a pure
 spinor. So, we must prove that
 $(x\cdot\g\tilde\l)_{\a}(\g^m)^{\a\b}(x\cdot\g\tilde\l)_{\b}=0$ using
 the condition $\tilde\l\gamma\tilde\l=0$. Then, 
\begin{eqnarray}
(x\cdot\g\tilde\l)_{\a}(\g^m)^{\a\b}(x\cdot\g\tilde\l)_{\b}
&=& x^n x^p \{\tilde\l^{\de}(\g_n)_{\de\a}(\g^m)^{\a\b}(\g_p)_{\b\r}\tilde\l^\r\}\no\\
&=&
2\,x^m x^p(\tilde\l\g_p\tilde\l)- x^n x^p(\tilde\l\g^m\g_n\g_p\tilde\l)\no\\
&=&
-\frac{1}{2} x^n x^p(\tilde\l\g^m\{\g_n,\g_p\}\tilde\l)\no\\
&=&
-\frac{1}{2} x\cdot x(\tilde\l\g^m\tilde\l)=0.
\end{eqnarray}
\item Let's now make the prove in the forward direction, i.e, assuming that 
$(x\cdot\gamma\tilde\lambda)_\alpha$ is a pure spinor, then $\tilde\l^\a$ is 
also a pure spinor. We start
defining the pure spinor $\rho_\a$: $\rho_\a \equiv
(x\cdot\g\tilde\l)_{\a} $. Then, writing $\tilde\l^\a$ in terms of
$\rho_\a$ we find
\begin{equation}
 \tilde\l^{\a}=\frac{1}{x\cdot x}(\r\g\cdot x)^\a. 
\end{equation}
Since $\r_\a$ is a pure spinor, then performing a similar computation
as in the proof in the backward direction, it 
is trivial to show that $\tilde\l^{\a}$ is a pure spinor: $\tilde\l\gamma^m\tilde\l=0$.
\end{enumerate}
Using the previous property we redefine the measure $[\d\tilde\l]$ given 
in (\ref{holomorphic measure}) by
\begin{eqnarray}\label{holomorphic measure3}\nonumber
&&[\d\tilde\l]^\prime(\tilde\l\g\cdot x\g^m)^{\a_1}(\tilde\l\g\cdot
x\g^n)^{\a_2}(\tilde\l\g\cdot x\g^p)^{\a_3}(\g_{mnp})^{\a_4\a_5}=\\
&&\frac{2^3}{10!|x|^8}\e^{
\a_1...\a_5\b_1...\b_{11}} \d(\tilde\l\g\cdot x)_{\b_1}\wedge...\wedge\d(\tilde\l\g\cdot x)_{\b_{10}}
(\tilde\l\g\cdot x)_{\b_{11}}.
\end{eqnarray}
Performing a simple computation as in (\ref{demostration}) we  can
show that $[\d\tilde\l]=e^{i\phi}[\d\tilde\l]^\prime$, 
where $\phi\in\mathbb{R}$ is a constant. Then, up to a phase factor,
we write 
\begin{equation}
\Phi(x)= \int_{\tilde{\Gamma}}[\d\tilde\l]^\prime\,\, F(\tilde\l,\o)|_{\o=x\tilde\l}.
\end{equation}
Replacing $F(\tilde\l,\o)|_{\o=x\tilde\l}$ and using the measure (\ref{holomorphic measure3}) we get
\begin{equation}\label{green function}
\Phi(x)=\frac{2^3
5!}{|x|^8}\int_{\tilde\Gamma}\frac{\d(A_1x\cdot\g\tilde\l)\wedge...\wedge\d(A_{10}x\cdot\g\tilde\l)}{(A_1x\cdot\g\tilde\l)...(A_{10}x\cdot\g\tilde\l)},
\end{equation}
where without loss of generality, we chose
$\tilde\Gamma=\{\l^{\a}\in SO(10)/U(5): |(A_Ix\cdot\g\tilde\l)|
=\e_I,\,I=1,...,10\}$. From (\ref{amplitude4}) and (\ref{green function}) the 
relationship between the tree level scattering amplitude and the Green's function for the massless scalar field is clear
\begin{eqnarray}
C^I_\a\,\,&\rightarrow&\,\, (A_I\g\cdot x)_\a\no,\\
 K  \,\,&\rightarrow&\,\, \frac{1}{(2\pi i)|x|^{8}}.
\end{eqnarray}
This result was not known using the old PCO's. Although the
construction for the scattering amplitudes at the genus $g$ is in progress  we think that it is likely
to have a relationship between loops scattering amplitudes and
massless solutions for higher-spin \cite{pure spinor like twistors}.


\section{Comments About the Loop-Level}\label{loops}

Now we give a glance about the scattering amplitude at the loop-level.
The loop level in the minimal pure spinor formalism has two fundamental ingredients: the picture raising operators and the $b$-ghost.
The picture raising operators are needed to absorb the zero modes of the 
field $\o_\a$ and some of the zero modes of the field $p_\a$, given in the 
action (\ref{action}). \\
Because at the loop level the complex structure of the Riemann
surfaces have 
deformations, known as the moduli space, in order to fix these 
deformations it is necessary to introduce the $b$-ghost. In the pure spinor 
formalism the $b$-ghost is not a fundamental field 
\cite{nathan minimal pure spinor}\cite{nathan topological} and therefore its 
construction in terms of the others fields  is such that  satisfies
\begin{equation}
\{Q_T,b(z)\}=T(z),
\end{equation}
where $Q_T$ is the BRST charge and $T(z)$ is the stress-energy tensor.\\
In \cite{nathan nikita multiloops} was given the $b$-ghost for the minimal pure 
spinor formalism. Nevertheless it has not been used to compute scattering
amplitudes. One reason for not using it is the difficulty for dealing
with the \v{C}ech indices inside the scattering amplitude. In this
section we will give some directions for computing scattering
amplitudes with $b$-ghost in the minimal pure spinor formalism.  

\subsection{Product of \v{C}ech Cochains}

In this sub-subsection we define a product between the \v{C}ech
cochains such that the result will be also a \v{C}ech cochain. The aim
is to obtain a well defined scattering amplitude, i.e since the loop
level scattering amplitude includes the $b$-ghost and the lowering
and raising picture changing operators, which are mathematical objects define locally,
then it is necessary that the product  of all these objects will be a \v{C}ech
cochain, such that the BRST operator $Q_T=Q+\de$ is also well defined
allowing in this manner to establish a relationship between the minimal and non-minimal formalism.

As we have shown in the example (\ref{example of product of 2
cochain}), the product of two \v{C}ech cochains is not in general a
\v{C}ech cochain. This implies that the \v{C}ech operator $\de$ is
not defined acting on this product, because it does not satisfy the
Leibniz rule, i.e it is not a derivative operator, see the example (\ref{operador de cech leibniz}). So we are going to define a product between the \v{C}ech cochains 
and show how the \v{C}ech operator acts on them. This is a small step towards
the definition of loop-level scattering amplitudes.     

As we showed previously with the example (\ref{example of product of 2 cochain}), considering two general cochains $\psi^{I}$ and $\tau^{I}$ in the Abelian group of holomorphic function, the product in most cases is not a \v{C}ech cochain
\begin{equation}
\psi^{I}\tau^{J}\notin C^{1}(\underline{U},\mathcal{O})
\end{equation}
because $\psi^{I}\tau^{J}\neq -\psi^{J}\tau^{I}$. So, we define the following antisymmetric product
\begin{equation}\label{antisymmetry}
\psi^{I}*\tau^{J}\equiv\frac{1}{2}\(\psi^{I}\tau^{J}-\psi^{J}\tau^{I}\)\Big|_{U_I\cap
U_J}\equiv \frac{1}{2}\psi^{[I}\tau^{J]}=\pi^{IJ}\in C^{1}(\underline{U},\mathcal{O}),
\end{equation}
which looks like an exterior product. Obviously this product is 
antisymmetric in the index $I,J$, i.e
$\psi^{I}*\tau^{J}=-\psi^{J}*\tau^{I}$, however 
the exchange of the $\psi^{I}$ and $\tau^{J}$ depends if they are grassmann or bosonic variables, this means
\begin{equation}
\psi^{I}*\tau^{J}=(-)^{(deg(\psi^{I})\cdot
deg(\tau^{J}))}\tau^{J}*\psi^{I},
\end{equation}
where $deg(\psi^{I})=0$ or $1$ if $\psi^{I}$ is a bosonic or grassmann 
variable respectively. Note that if the product of the \v{C}ech cochains is 
well defined, as in the case of the product of the picture lowering operators 
$Y^I_C\,Y^J_C$, then it is in agreement with (\ref{antisymmetry}):
\begin{equation}
Y^I_C*Y^J_C = \frac{1}{2}Y^{[I}_C\,Y^{J]}_C=Y^I_CY^J_C. 
\end{equation}
Now, acting with the \v{C}ech operator on (\ref{antisymmetry}) we get
\begin{eqnarray}\label{delta operator}
(\de\pi)^{IJK}&=&[\de(\psi*\tau)]^{IJK}
=\frac{1}{2^2}(\delta\psi)^{[IJ}\tau^{K]}= -\frac{1}{2^2}\psi^{[I}(\de\tau)^{JK]} 
=\frac{3}{4}\frac{1}{3!}\((\delta\psi)^{[IJ}\tau^{K]}-\psi^{[I}(\de\tau)^{JK]}\)\no\\
&=&\frac{3}{4}\((\delta\psi)^{IJ}*\tau^{K}-\psi^{I}*(\de\tau)^{JK}\)\in C^{2}(\underline{U},\mathcal{O}).
\end{eqnarray}
Note that $\delta$ acts like the exterior derivative over elements with star product, but with a coefficient in the front. This is a beautiful property.
If we have three \v{C}ech cochains $\psi^I,\tau^J,\rho^K\in C^{0}(\underline{U},\mathcal{O})$ then we define
\begin{equation}
\psi^I*\tau^J*\rho^K= \frac{1}{3!}\psi^{[I}\tau^J\rho^{K]}=\chi^{IJK}\in C^{2}(\underline{U},\mathcal{O}).
\end{equation}
It is simple to see that this product is associative. Acting with the \v{C}ech
operator we have
\begin{eqnarray}\label{delta operator2}
(\de\chi)^{IJKL}&=&\frac{1}{3!\,2} (\de\psi)^{[IJ}\tau^K\rho^{L]}
=-\frac{1}{3!\,2} \psi^{[I}(\de\tau)^{JK}\rho^{L]}=\frac{1}{3!\,2} \psi^{[I}\tau^J(\de\rho)^{KL]}\no\\
&=&\frac{4}{3!}\frac{1}{4!}\((\de\psi)^{[IJ}\tau^K\rho^{L]}-\psi^{[I}(\de\tau)^{JK}\rho^{L]}+ \psi^{[I}\tau^J(\de\rho)^{KL]}\)\\
&=&\frac{4}{3!}((\de\psi)^{IJ}*\tau^K*\rho^{L}-\psi^{I}*(\de\tau)^{JK}*\rho^{L} +\psi^{I}*\tau^J*(\de\rho)^{KL})\in C^{3}(\underline{U},\mathcal{O})\no.
\end{eqnarray}
Again, the \v{C}ech operator acts like the exterior derivative operator over
elements with the star product, nevertheless it has a coefficient in the front. 
It is straightforward to generalize this procedure for higher cochains
with values on any Abelian group. Notice that the expressions 
(\ref{definition of psi}) and (\ref{definition of phi}) are just the $*$ product.\\
If we use this product between the homotopy operator (\ref{homotopy operator}) 
and the PCO's inside the tree level scattering 
amplitude the result vanishes because there are 11 patches to cover the pure spinor space.

\subsection{The $b$-ghost}

Unlike the tree-level scattering amplitude, in
higher orders of the genus expansion the cover
$\underline{U}=\{U_I\}$ given by the eleven patches
$U_I=PS\smallsetminus D_I,\,I=1,..,11$ (see \ref{patch U_I}) is not
enough. The explanation is simple, since the $b$-ghost is a linear
combination of  0,1,2 and 3-\v{C}ech cochains in the pure spinor space
\cite{nathan nikita multiloops}  and the product of the 11 picture
lowering operators $\prod_{I=1}^{11}Y^{I}_C\,$ is a 10-\v{C}ech
cochain then, with the antisymmetric $*$ product it is clear that the 
scattering amplitude will vanish if the number of patches is  less than $11+4(3g-3),\, g>1$, where $g$ is the genus of the Riemann surface. In the particular case when $g=1$ this number is $11+4$.
One can think 
to add more patches to the tree level scattering amplitude (see the appendix \ref{another cover for the pure spinor space}) and so to apply 
the $*$ product with the naive homotopy operator, however this product
needs to be better understood, since actually the operator $\delta$ 
is not a derivate operator strictly speaking because of those coefficients 
in the front of (\ref{delta operator}) and (\ref{delta operator2}) 
\footnote{Actually, we wish that the operator $Q_T=Q+\delta$ acts like the 
exterior derivate, however we do not succeed yet.}. Note also that the  tree level scattering amplitude must be $\de$ closed 
in contrast to the genus-$g$, as we discuss later.
  
So, in this approach the $b$-ghost is given by   
\begin{equation}\label{b ghost}
b=b_{(0)}+b_{(1)}+b_{(2)}+b_{(3)} 
\end{equation}
with
\begin{equation}
b_{(0)}^{\mu}=\frac{A^\mu_\a G^\a}{(A^\mu\l)},\quad b_{(1)}^{\mu\nu}=\frac{A^{\mu}_{\a}A^\nu_\b H^{[\a\b]}}{(A^\mu\l)(A^\nu\l)}, 
\quad b^{\mu\nu\rho}_{(2)}=\frac{A^\mu_\a A^\nu_\b A^\rho_\g K^{[\a\b\g]}}{(A^\mu\l)(A^\nu\l)(A^\rho\l)}, 
\quad b_{(3)}^{\mu\nu\rho\kappa}=\frac{A^\mu_\a A^\nu_\b A^\rho_\g A^\kappa_\de L^{[\a\b\g\de]}}{(A^\mu\l)(A^\nu\l)(A^\rho\l)(A^\kappa\l)},\no
\end{equation}
where the specific form of the numerators $(G,H,K,L)$ above can be
found in \cite{nathan topological} and the $A_\a^{\mu}$'s belong to a bigger 
set of constant spinors $A^\mu\in\{C^I,V^i\}\equiv\{C^1,...,C^{11},V^{1},...,V^{4(3g-3)}\}$, i.e $\mu\in\{I,i\}$, where the $V^{i}$'s are
 linearly independent vectors in $\mathbb{C}^{16}$ such that the hypersurfaces $P^i\equiv\{\l^\a\in PS: V^i_\a\l^\a=0\}$ satisfy  $D_1\cap..\cap D_{11}\cap P^1\cap...\cap P^{4(3g-3)}=\{0\}$. We define the
 patches $U_I=PS\smallsetminus D_I$, $U^i=PS\smallsetminus P^i$ and get the cover
 $\mathcal{U}=\{U_I,U^i\}$ where
 $PS\smallsetminus\{0\}=\bigcup_{I=1}^{11}U_I\bigcup_{i=1}^{4(3g-3)}U^{i}$.
Note that the $C^I$'s are the same as in the tree level case, so the cycle $\Gamma$ given in (\ref{cycle}) is a good definition to compute the scattering amplitude. 
Using the commutators and anticommutators given in \cite{nathan nikita multiloops}
\begin{equation}
\{Q,G^\a(z)\}=\l^\a T(z),\quad \[Q,H^{[\a\b]}\]=\l^{[\a} G^{\b]},\quad \{Q,K^{[\a\b\g]}\}= \l^{[\a}H^{\b\g]},
\end{equation}
\begin{equation}\nonumber
\[Q,L^{[\a\b\g\delta]}\]=\l^{[\a}K^{\b\g\delta]},\quad \l^{[\eta}L^{\a\b\g\delta]}=0,
\end{equation}
where $T(z)$ is the stress-energy tensor given in the section 2, it is easy to verify that the $b$-ghost (\ref{b ghost}) satisfies 
\begin{equation}\label{ Q+de b=T}
\{Q+\de,b(z)\}=T(z), 
\end{equation}
where, for instance
$$
(\de\,b_{(0)})^{\mu\nu}= \frac{A^\nu_\a G^\a}{(A^\nu\l)}-\frac{A^\mu_\a G^\a}{(A^\mu\l)}=\frac{A^\mu_\a A^\nu_\b \l^{[\a}G^{\b]}}{(A^\mu\l)(A^\nu\l)}=\frac{A^\mu_\a A^\nu_\b \[Q,H^{[\a\b]}\] }{(A^\mu\l)(A^\nu\l)}.
$$

\subsubsection{The Ghost Number Bidegree}

As in bosonic string theory, the $b$-ghost must have ghost number -1 and the BRST  charge must increase the ghost number by one unit. Note that $b_{(0)}$ has ghost number -1 but $b_{(1)},b_{(2)},b_{(3)}$ have ghost number $-2,-3,-4$ respectively (the numerators have ghost number zero, see \cite{nathan nikita multiloops}), where the ghost current is given by $J_\l=\l^\a\o_\a$. However, since the total BRST charge is $Q_T=Q+\de$ and the \v{C}ech operator increases the number of patches in one, then the number of patches is also a ghost number. So we define the total ghost number by 
$$
J_T=\int\d z J_\l+ J_\de,   
$$
$$
J_\de\equiv  (\sum_\mu \mu\p_\mu -1),
$$
where the operator $J_\de$ acts on the \v{C}ech cochains in the \v{C}ech labels, for example
$$
(\sum_\eta \eta\p_\eta -1)\,\,b_{(3)}^{\mu\nu\rho\kappa}=3\,b_{(3)}^{\mu\nu\rho\kappa}.
$$  
Therefore the $b$-ghost (\ref{b ghost}) has $J_T$ ghost number -1, as
expected. In the tree level scattering amplitude the $J_\de$ ghost number
is not relevant because this amplitude is $\de$ closed, see the
subsection \ref{brst invariance}. Nevertheless, at loop level this
ghost number become very important since the relation (\ref{ Q+de b=T}) means that the scattering amplitude is $Q_T=Q+\de$ closed up to boundary terms in the moduli space \cite{dhokerperturbation}, i.e
if we consider a loop level scattering amplitude where the $*$ product is used to insert the $b$-ghost, then we expect to get  
\begin{eqnarray}
(Q+\delta)\<....\int\d z\mu^z_{\bar z}(z)b(z)\> = \<....\int\d z\mu^z_{\bar z}(z)(Q+\de)(b(z))\>&=&\<....\int\d z\mu^z_{\bar z}(z)T(z)\>\no\\
&=&\int_\mathcal{M}\frac{\p}{\p\tau^i}\<....\>,
\end{eqnarray}
where $....$ means the global insertions, $\mu^z_{\bar z}(z)$ is the
Beltrami differential, $\tau^i$'s are the Teichm\"{u}ller parameters
and $\mathcal{M}$ is the Moduli space. Now, with the aim to see the
importance of the $J_\de$ ghost number at the loop level we can regard
the 1-loop scattering amplitude. In this amplitude we have 11 zero
modes of the pure spinor $\l^\a$ and 11 zero modes of the spinor
$\o_\a$ \cite{nathan minimal pure spinor}, so at 1 loop the zero modes
of $\l^\a$ and $\o_\a$ form the pure spinor phase space. Integrating
somehow the zero modes of the fields $\o_\a,d_\a$ and $\t^\a$ then the
scattering amplitude shall behave as
\begin{equation}\label{1 loop}
\int_\Gamma
[\d\l]\frac{A^{I_1}...A^{I_{11}}}{(A^{I_1}\l)...(A^{I_{11}}\l)}\frac{A^{I_{12}}A^{I_{13}}\l^4}{(A^{I_{12}}\l)(A^{I_{13}}\l)},
\end{equation}
where we are not being careful with the spinorial indices. Note that
this amplitude is a 12-\v{C}ech cochain and have $J_\l$ ghost number
-1. However this ghost number must always be zero, therefore it should
be compensated with one $J_\de$ ghost number\footnote{Perhaps because 
(\ref{1 loop}) is $\de$ exact.} and so we will get an 11-\v{C}ech
cochain. As the tree level scattering amplitude is a 10-\v{C}ech
cochain and it is related to the Green's function for the massless scalar
field then the 1 loop amplitude, which should be a 11-\v{C}ech
cochain, suggests that it should be related to the Green's function
for the massless higher-spin field \cite{pure spinor like twistors}.
This was only a simple and crude analysis about the $1$-loop scattering
amplitude. Actually the full analysis must be over the whole pure
spinor phase space. This is because, for instance, at two loops
 we should get a \v{C}ech cochain in the pure
spinor space bigger than 11, so its corresponding Dolbeault cochain
will be identically zero and the amplitude will vanish. Therefore it
is necessary to regard the whole pure spinor phase space, i.e the space of the $\l^\a$'s and $\o_\a$'s.         

One could think that the scattering amplitude must have
$J_T=J_\l+J_\de$ ghost number zero, but that is not true. For example,
in the tree level amplitude it is impossible to construct the picture
lowering operator such that the amplitude has $J_T$ ghost number zero
and the origin is removed from the pure spinor space. So the
conditions that the scattering amplitude has $J_\l$ ghost number zero
is necessary in order to get a physical amplitude, i.e a $(Q+\de)$ closed scattering amplitude.

This was just a glance about
the loop level scattering amplitudes, which is a work in progress \cite{work in progress}.


\section{Conclusions}

We proposed a new ``picture lowering'' operator and computed the
scattering amplitude at tree level in such a way that we eliminated
the singular point of the pure spinor space, getting in this way a
theory free of anomalies \cite{beta gamma system}. Since the new picture 
operators are defined just on each patch 
of the pure spinor space, it is necessary to introduce the \v{C}ech 
operator as part of the BRST charge in order to have a  well defined
formalism. Therefore, we have introduced the \v{C}ech 
formalism for the scattering amplitudes computation, which seems to be
the correct formulation \cite{nathan nikita multiloops}\cite{yuri aldo
nathan nikita}. Given the \v{C}ech-Dolbeault isomorphism,
we found the corresponding Dolbeault cocycle for the scattering amplitude. 
What is interesting here is that the
Dolbeault cocycle must not be evaluated in whole pure spinor space,
but in the $SO(10)/SU(5)$, which can be thought like a sphere in the 
pure spinor space. This confirms that the singular point was removed from the pure spinor
space. Moreover, since the de-Rham cohomology group of this manifold has just one generator given by
$$
\frac{[\d\l]\wedge[\d\bar\l]'}{(\l\lb)^8}\Bigg|_{SO(10)/SU(5)},
$$
i.e
$H^{21}_{DR}(SO(10)/SU(5))=\mathbb{C}$ \cite{beta gamma system},
then these picture operates do not correspond to any particular
regulator of the non-minimal formalism. This may suggest that the minimal 
pure spinor formalism is, in this sense, more fundamental than the 
non-minimal formalism since it directly involves cohomology
generators. Note that in this paper the tree level scattering
amplitude was always computed using three unintegrated vertex
operators and the rest were integrated vertex operators. In contrast
with the 
minimal formalism, in the non-minimal formulation it is possible to compute 
tree level amplitude with all the vertex operators unintegrated 
\cite{nathan topological}. The difficulty in the minimal formalism is the 
$b$-ghost. Although we gave a glance about how to treat this issue, in order 
to continue with the loop-level this subject must be further developed 
\cite{work in progress}.

Using the \v{C}ech-Dolbeault isomorphism, we also showed in an elegant
manner that the tree-level scattering amplitude is BRST, Lorentz and
supersymmetric invariant.

In contrast with the PCO's proposed in \cite{nathan minimal pure
spinor}, with the new PCO's proposed in this paper the tree level scattering 
amplitude is independent of the choice of the constants 
spinors $C^I$'s. That is because the cohomology class of the scattering 
amplitude is the same when the constants $C^I$'s satisfy the constraint 
$\{C^1\l=0\}\cap\{C^2\l=0\}\cap...\cap\{C^{11}\l=0\}=\{0\}$, for
$\l^\a$ satisfying the pure spinor condition.

Finally, we obtained a relationship between the tree-level scattering 
amplitude in the pure spinor formalism and the Green's function for the massless scalar field in the twistor formalism \cite{pure spinor like twistors}. 
We believe that perhaps there is a relationship between the loop-level 
scattering amplitudes in the pure spinor formalism and the Green's function 
for the higher-spin massless fields \cite{pure spinor like twistors}, which we
would like to explore in the future.

\section*{Acknowledgments}
We would like to thank Yuri Aisaka, Nathan Berkovits, Freddy Cachazo
and Kostas Skenderis for
valuable discussions and comments, as well as Carlos Cardona for reading the 
manuscript and useful comments. We also thank to the
school {\it Aspects of Supersymmetry} at the IAS-Princeton, where our 
collaboration was initiated. H.G is grateful to Sergey Cherkis for the 
discussions about the twistor space and to the Simons 
Center for Geometry and Physics for warm hospitality during the
initial stages of this work. H.G also thanks to the string theory 
group at the IFT-UNESP, where the 
work was presented. O.B would like to thank the Simons Workshop for
Mathematics and Physics, ICTP, University of Amsterdam, the Vrije
Universiteit Brussels and IFT-UNESP for hospitality. The work of H.G is supported by 
FAPESP Ph.D grant 07/54623-8. The work of O.B is supported by 
FAPESP grant 2009/08893-9.


\appendix

\section{Some Simple Examples}

\subsection{The Pure Spinor Condition in the $U(5)$ decomposition}
\label{singularities appendix}
We will give an example of a point $p \in PS$, for which in the $U(5)$ 
decomposition, is necessary to consider both conditions $\chi^a = 0$
and $\zeta_a = 0$ in order to have a well defined tangent space at $p$.\\

Consider for instance the 
point $p=(\l^{+}=0,\l_{ab}=0, \l^{a}=\delta^{1a})$ in the pure spinor
space. Then, the gradient vectors $V^{a}=(\l^a,-\frac{1}{4}\epsilon^{abcde}\l_{bc} , \l^{+} \delta^{ab})$, 
which generate the holomorphic tangent space to the cone given by
$$
\chi^{a}\equiv \l^+\l^a-\frac{1}{8}\epsilon^{abcde}\l_{bc}\l_{de}=0,\quad a,b,c,d,e=1,...,5,
$$ 
do not generate a tangent space of complex dimension
11 at the point $p$. This is because $V^{i}=(0,...,0)$ 
for $i=2,..,5$, i.e only $V^1$ is different from  zero at $p$, which
means that $p$ is a singular point\footnote{We say that $p$ is a singular 
point of $PS$ (or any space) if and only if it is not possible to define 
an unique tangent space in $p$ with the same dimension of $PS$.} of 
the space $\chi^a=0,\,a=1,...,5$. So $\chi^a =0$ does not describe completely
the pure spinor
space since actually $PS$ only has one singular point: $\l^{\a}=0$. 
For that reason, we must consider the rest of the pure spinor equations
\begin{equation}\label{5constrains2}
\zeta_a=\l^{b}\lambda_{ba}=0,\qquad a,b=1,...,5. 
\end{equation}
Note that $p$ is actually a point in the pure spinor space
since it satisfies both $\chi^a = 0$ and $\zeta_a = 0$. In contrast,
there exists points which do not satisfy simultaneously both set of
equations.
To the five equations $\zeta_a =0$ corresponds five gradient 
vectors $A_a=(0,\l^{b}\delta_{a}^{c} - \l^{c}\delta_{a}^{b},\l_{ba})
=(0,\l^{[b}\delta_{a}^{c]},\l_{ba})\,$. Therefore at the point $p$ we
have in addition to $V^1$, four linearly independent vectors
\begin{equation}
A_i=(0,\l^{[1} \delta_{i}^{j]} , 0), \qquad i,j =2,3,..,5, 
\end{equation}
where $[1j]$ are the components ($[12], [13],...,[15]$) of such
vectors and we have a well defined tangent space at $p$. Summarizing, with this particular example for
the point $p$, we argued the necessity of considering the conditions
$\zeta_a =0$ and now we have at every  point
of the pure spinor space, except for the origin,  a tangent space of complex dimension 11. 
In other words, the pure spinor space without 
the origin is a smooth manifold embedded in $\mathbb{C}^{16}$. Note
however that for $\l^{+}\neq 0$, the five vectors $V^a$'s are linearly 
independent and the solutions of the equations $\chi^a = 0$ satisfy trivially the equations $\zeta_a=0$. Therefore, when $\l^+\neq 0$ the five 
equations $\chi^a=0$ are enough to describe the pure spinor space.

\subsection{Another Cover For The Pure Spinor Space}
\label{another cover for the pure spinor space}

We argued that the constant spinors $C^I$'s given in (\ref{choice C})
are not a good choice because the intersection of the hypersurfaces 
$\{C^I\l=0\}, I=1,...,11$ is the non compact space $\mathbb{C}^5$. This 
means that the union of the patches $U_I=PS\smallsetminus\{C^I\l=0\}$, where 
the scattering amplitude is supported, is not the whole pure spinor space, i.e
\begin{equation}
U_1\cup...\cup U_{11}=PS\smallsetminus\mathbb{C}^5. 
\end{equation}
Then one question arises: Is it possible to complete the patches $U_I$
in such a way that they form a cover of the $PS\smallsetminus\{0\}$ space? Obviously
the answer is positive. Here we show what is the difficulty for
completing the patches for the tree level scattering amplitude. 

In \cite{nathan nikita multiloops} it was proposed the cover for the pure spinor space
$\mathcal{U}=\{U_\a\},\,\a=1,...,16$, with the patches $U_\a$'s given by
\begin{equation}
U_\a=PS\smallsetminus\mathcal{D}_\a,\qquad \mathcal{D}_\a\equiv\{\l^\a\in PS: \l^\a=0\}. 
\end{equation}
Clearly $\mathcal{U}$ is a cover of the pure spinor space without the origin
\begin{equation}
\bigcup_{\a=1}^{16}U_\a=U_1\cup...\cup U_{16}=PS\smallsetminus\{0\}. 
\end{equation}
So, we can define the following picture operators
\begin{equation}\label{PCO for a new cover}
Y^\a=\frac{\t^\a}{\l^\a}. 
\end{equation}
The first difficulty here is that there are 16 PCO's instead of 11, 
however this is not really a problem. Note that in the $U(5)$ decomposition, i.e $\l^\a=(\l^+,\l_{ab},\l^a),\,a,b=1,...,5$ and $\l_{ab}=-\l_{ba}$,  and choosing the picture operators   
\begin{equation}
Y^{+}=\frac{\t^+}{\l^+} \quad  \text{and} \quad Y_{ab}=\frac{\t_{ab}}{\l_{ab}},
\end{equation}
we fall in the first example of the
subsection \ref{examplesCs}, with the difference that now we have a cover for $PS\smallsetminus\{0\}$. So the tree level scattering amplitude is given by
\begin{equation}\label{amplitude with another cover}
\mathcal{A}=\int_\Gamma[\d\l]\int\d^{16}\t \prod_{i=1}^{11}\frac{\t^{i}}{\l^i} \l^\a\l^\b\l^\g f_{\a\b\g}(\t)
\end{equation}
where 
\begin{equation}
\frac{\t^1}{\l^1}\equiv\frac{\t^+}{\l^+},\quad\frac{\t^2}{\l^2}\equiv\frac{\t_{12}}{\l_{12}}, \,\cdot\cdot\cdot \,,\frac{\t^{11}}{\l^{11}}\equiv\frac{\t_{45}}{\l_{45}}
\end{equation}
and the cycle $\Gamma$ is given by $\Gamma=\{\l^\a\in PS: |\l^i|
=\varepsilon^i, i=1,...,11\},\,\varepsilon^i\in\mathbb{R}^+$. Now we must 
verify if (\ref{amplitude with another cover}) is a physical amplitude, i.e 
if it is $Q_T=Q+\de$ closed.\\
In the same way as in (\ref{q of the amplitude}) it is not hard to show 
that (\ref{amplitude with another cover}) is $Q$ closed. Then now we must 
show that the amplitude (\ref{amplitude with another cover}) is $\de$ closed. 
Since in this case there are 16 patches then the analysis can not be 
similar to the one presented in the subsection \ref{equivalence mps and 
nmps}. Acting with the $\de$ operator in (\ref{amplitude with another cover}) we get
\begin{equation}\label{delta of the amplitude with another cover}
(\delta\mathcal{A})^{1,...,11,j}=\int_\Gamma[\d\l]\int\d^{16}\t\( \prod_{i=2}^{11}\frac{\t^{i}}{\l^i}\frac{\t^{j}}{\l^j}-\prod_{i=3}^{11}\frac{\t^{1}}{\l^1}\frac{\t^{j}}{\l^j}\frac{\t^{i}}{\l^i}+...+\prod_{i=1}^{11}\frac{\t^{i}}{\l^i} \)\l^\a\l^\b\l^\g f_{\a\b\g}(\t),
\end{equation}
where $j$ is any number from 12 to 16. Naively  
(\ref{delta of the amplitude with another cover}) can be written as 
\begin{equation}\label{delta of the amplitude with another cover2}
(\delta\mathcal{A})^{1,...,11,j}=\int_\Gamma[\d\l]\int\d^{16}\t\, Q\( \prod_{i=1}^{11}\frac{\t^{i}}{\l^i}\frac{\t^{j}}{\l^j} \)\l^\a\l^\b\l^\g f_{\a\b\g}(\t),
\end{equation}
nevertheless that is not true. In the subsection
\ref{integration contours} we said that the scattering amplitude also
depends of the homology class of the cycle $\Gamma$. Since  the computation (\ref{delta of the amplitude with another cover2}) has 12 \v{C}ech labels and $\Gamma$ is a 11-cycle we need to be careful. From (\ref{delta of the amplitude with another cover}) we can see that just the term     
\begin{equation}
\int_\Gamma[\d\l]\int\d^{16}\t \prod_{i=1}^{11}\frac{\t^{i}}{\l^i} \l^\a\l^\b\l^\g f_{\a\b\g}(\t)
\end{equation}
contributes, since the other terms vanish because the cycle $|\l^j|=\varepsilon^j$ is not in $\Gamma$. Therefore the scattering amplitude (\ref{amplitude with another cover}) is not physical. 

Actually, we have shown that the cycle $\Gamma$ is a trivial element
of the homology group $H_{11}(PS\smallsetminus \mathcal{D})$, where
$\mathcal{D}=\mathcal{D}_1\cup...\cup \mathcal{D}_{16}$. This is
because the intersection $\mathcal{D}_1\cap...\cap \mathcal{D}_{11}$
is $\mathbb{C}^5$, so the difficulty of using the cover $\mathcal{U}$
and the PCO's (\ref{PCO for a new cover}) is to get a well defined
cycle $\Gamma$ such that we can write
$(\delta\mathcal{A})^{1,..,11,j}$ like in (\ref{delta of the amplitude
with another cover2}), i.e a non trivial element of the homology group $H_{11}(PS\smallsetminus \mathcal{D})$.  Note that if we add to the cover $\underline{U}=\{U_I\}, I=1,..,11$, where the patches $U_I$'s are given in (\ref{patch U_I}), more patches, then there is not problem. The reason is simple, since the condition $D_1\cap...\cap D_{11}=\{0\}$, for the $D_I$'s given in (\ref{patch U_I}), then the cycle $\Gamma$ (\ref{cycle}) will always be a non trivial element of the homology class, so applying the $\de$ operator to the amplitude we get something equal to (\ref{delta of the amplitude with another cover2}).       

In conclusion, for tree level scattering amplitude with 3 unintegrated
vertex operators and the remaining integrated, it is sufficient to work with 11 patches such that they cover the pure spinor space without the origin
$PS\smallsetminus\{0\}$.

\subsection{The \v{C}ech-Dolbeault Correspondence for Pure Spinor in $d=4$}
\label{CD appendix}
Our next simple example is the pure spinor space in $d=4$
dimensions, i.e $PS=\mathbb{C}^{2}$.
We choose the coordinates $\l^a=(\l^1,\l^2)$ and consider the integral
\begin{equation}\label{pure_spinor_d=4_integral}
I=\int_{\Gamma}\frac{\d(C^1\l)\wedge\d(C^2\l)}{(C^1\l)(C^2\l)}=\int_{\Gamma}\psi,
\end{equation}
where $C^i\l=C^i_a\l^a$, det$(C^i_a)\neq 0$
and $\Gamma$ is given by $\Gamma=\{\l^a\in\mathbb{C}^{2}:|C^i\l|
=\varepsilon^i\}$, $\varepsilon^i \in\mathbb{R}^+$. 
We can write (\ref{pure_spinor_d=4_integral}) as
\begin{equation}
I=\int_{\Gamma}[\d\l]\frac{\epsilon^{ab}C^1_aC^2_b}{(C^1\l)(C^2\l)},
\end{equation}
where
$[\d\l]=(1/2)\epsilon_{ab}\d\l^a\wedge\d\l^b=\d\l^1\wedge\d\l^2$.\\
Note that $\mathbb{C}^2$ can be seen as the total space of the
universal line bundle $\mathcal{O}(-1)$  over $\mathbb{C}P^1$, i.e
$\l^a=\g\tilde\l^a$ where $\g$ is the fiber and $\tilde\l^a$ are the
 coordinates of $\mathbb{C}P^1$. So, without loss of
generality we choose $\Gamma=\{\l^a\in\mathcal{O}(-1): |\g|
=\varepsilon, |C^1\tilde\l|=\varepsilon^1, \,\text{where}\,\, 
\tilde\l^a\in\mathbb{C}P^1\}\,
\varepsilon,\varepsilon^1\in\mathbb{R}^+$ (as in the sub-subsection 
\ref{contours in pps}) and the measure $[\d\l]$ is given like in
reference \cite{beta gamma system} by
$[\d\l]=\g\d\g\wedge[\d\tilde\l]$, where 
$[\d\tilde\l]=\epsilon_{ab}\d\tilde\l^a \tilde\l^b$   is the measure
for the twistor  space in $d=4$ \cite{pure spinor like twistors}. Then, integrating $\g$ we get
\begin{equation}
\int_{\Gamma}[\d\l]\frac{\epsilon^{ab}C^1_aC^2_b}{(C^1\l)(C^2\l)}= 
\int_{\Gamma}\frac{\d\g}{\g}\wedge[\d\tilde\l]\frac{\epsilon^{ab}C^1_aC^2_b}{(C^1\tilde\l)(C^2\tilde\l)} 
=(2\pi i)
\int_{|C^1\tilde\l|=\varepsilon_1}[\d\tilde\l]\frac{\epsilon^{ab}C^1_aC^2_b}{(C^1\tilde\l)(C^2\tilde\l)},
\end{equation}
where the right hand side has the same form as the Green's function for the
massless scalar field in $d=4$ using the twistor language \cite{pure
spinor like twistors}. This result in $d=4$ is analogous 
to the one obtained in $d=10$, see (\ref{amplitude4}).  

Now, using the partition of unity 
\begin{equation}
\rho_i=\frac{|C^i\l|^2}{(|C^1\l|^2+|C^2\l|^2)},\qquad i=1,2 
\end{equation}
subordinated
to the cover $\mathcal{U}=\{U_1,U_2\}$, where
$$ 
U_i=\mathbb{C}^2\smallsetminus\{C^i\l=0\},\qquad i=1,2,
$$
we find the Dolbeault cocycle corresponding to $\psi$.
Note that from the condition det$(C^i_a)\neq 0$ then  $\{C^1\l=0\}\cap \{C^2\l=0\}=\{0\}$, so we get $\mathbb{C}^2\smallsetminus\{0\}=U_1\cup U_2$.\\
Since $\psi$ is a (2,0) holomorphic form over $U_1\cap U_2$ then
$\psi$ is 1-\v{C}ech cochain, i.e $\psi\in C^{1}(\mathcal{U},\Omega^2)$, where $\Omega^2(\mathbb{C}^2\smallsetminus\{0\})$ is the abelian group of the (2,0) holomorphic forms over $\mathbb{C}^2\smallsetminus\{0\}$. So, using (\ref{cech dolbeault map}), the Dolbeault cocycle corresponding to $\psi\equiv \psi_{12}=-\psi_{21}$ is given by
\begin{equation}
\eta_\psi=\sum_{\a,\b=1}^{2}\psi_{\a\b}\r_\a\wedge\bar\p\rho_\b=\psi_{12}\rho_1\wedge\bar\p\rho_2+\psi_{21}\rho_2\wedge\bar\p\rho_1=\psi_{12}\wedge\bar\p\rho_2
\end{equation}
where $1,2$ are the \v{C}ech labels. Replacing $\psi_{12}$ and $\rho_2$ in $\eta_\psi$ we get
\begin{equation}
\eta_\psi=\frac{\d(C^1\l)\wedge\d(C^2\l)\wedge\[(\bar C^1\lb)\d(\bar C^2\lb)-(\bar C^2\lb)\d(\bar C^1\lb)\]}{(|C^1\l|^2+|C^2\l|^2)^2}.
\end{equation}
Note that this (2,1)-form is global on
$\mathbb{C}^2\smallsetminus\{0\}$.\\
Therefore from the \v{C}ech-Dolbeault correspondence we have
\begin{equation}
\int_{\Gamma}\psi_{12}=\int_{S^3}\eta_{\psi}|_{S^3}, 
\end{equation}
where $S^3$ is the sphere $|\l^1|^2+|\l^2|^2=r^2,\, r\in\mathbb{R}^+$.
Since $S^3$ is a $U(1)$-line bundle over $\mathbb{C}P^1$ space then we can write $\eta_\psi$ in the $S^3$ coordinates 
$$
\l^a=r\,e^{i\t}(1,u),
$$
where $e^{i\t}$ parametrizes the fiber $U(1)$, $u$ parametrizes the
$\mathbb{C}P^1$ space and $r$ is the size of $S^3$. So, 
\begin{equation}
\eta_\psi|_{S^3}=i\frac{|\epsilon^{ab}C^1_a C^2_b|^2}{(|C^1_1+C^1_2\,u|^2+|C^2_1+C^2_2\,u|^2)^2}\d\t\wedge\d u\wedge\d\bar u.
\end{equation}
Note that the constant $r$ does not appear and the $U(1)$ part is decoupled. Therefore we can perform a
global transformation from $\mathbb{C}P^1\,\rightarrow\,\mathbb{C}P^1$
to eliminate the $C^I$'s constants. This transformation is known as the M\"{o}bius transformation
\begin{equation}
v=\frac{C^1_1+C^1_2\,u}{C^2_1+C^2_2\,u},\,\,\text{ where }\,\,\(
\begin{matrix}
C^1_1 & C^1_2 \\
C^2_1 & C^2_2  
\end{matrix}
\)\,\in GL(2,\mathbb{C}).
\end{equation}
With this transformation we obtain
\begin{equation}\label{eta s3}
\eta_\psi|_{S^3}=i\frac{1}{(1+v\bar v)^2}\d\t\wedge\d v\wedge\d\bar v.
\end{equation}
(\ref{eta s3}) is the $d=4$ equivalent to  (\ref{integral4}) for pure 
spinors in $d=10$ and $\eta_\psi|_{S^3}$ is a generator of the de-Rham cohomology group $H^{3}_{DR}(S^3)=\mathbb{C}$ in coordinates. Integrating by $\d\t$ we have the following equality
\begin{equation}\label{degree cp1}
\int_{|C^1\tilde\l|=\varepsilon_1}[\d\tilde\l]\frac{\epsilon^{ab}C^1_aC^2_b}{(C^1\tilde\l)(C^2\tilde\l)} 
=\int_{\mathbb{C}^2}\frac{1}{(1+v\bar v)^2}\d v\wedge\d\bar v=(2\pi
i)\int_{\mathbb{C}P^1} H,
\end{equation}
where the hyperplane class $H$ is written locally as $H=(1/(2\pi i))(1+v\bar v)^{-2}\d v\wedge\d\bar v$ \cite{humberto one loop}. So (\ref{degree cp1}) is just $(2\pi i)$ times the degree of the projective complex space $\mathbb{C}P^1$, which is one.

\subsection{Global Integrals}
\label{integrals appendix}

Now we want to give a simple example with the aim to explore the global
definition of the degree of a hypersurface. 
Let us consider the following cone in $\mathbb{C}^{4}$
\begin{equation}
\chi\equiv z_1z_2-z_3z_4=0 
\end{equation}
and the integral
\begin{equation}\label{ejemplo integral}
I=\int_{\Gamma}\frac{\d f^1\wedge \d f^2 \wedge\d f^3}{ f^1 f^2 f^3}, 
\end{equation}
where $f^i=C^iZ=C^i_1 z_1+C^i_2 z_2+C^i_3 z_3+C^i_4 z_4$ and 
$\Gamma=\{Z\in\mathbb{C}^4 : \chi=0 \text{ and } ||f^i||=\varepsilon_i\},\,\varepsilon_i\in\mathbb{R}^+$. 
We choose the $C^i$'s in a similar way to (\ref{choice C}), i.e, 
$f^{1}=z_1,\,f^{2}=z_2,\,f^{3}=z_3$. 

Note that the intersection 
$$
\{\tilde f^1=0\}\cap \{\tilde f^2=0\}\cap\{\tilde\chi=0\}\Big|_{\mathbb{C}P^3}=\{[0,0,1,0],[0,0,0,1]\}
$$ 
where $\{\tilde f^i=0\}\equiv \{f^i=0\}/\sim$ and the equivalence relation is 
given by $Z\sim c Z,\, c\in\mathbb{C}^*$. The same is true for $\tilde\chi$. This means that the degree of the smooth manifold $\tilde\chi=0$ embedded in $\mathbb{C}P^3$ is deg($\tilde\chi=0$)=2. So we would expect that (\ref{ejemplo integral}) will be $(2\pi i)^3 2$ from the discussion of the sub-subsection \ref{The Projective Pure Spinor Degree}.

Now, it is important to note that the intersection 
$$
\{f^1=0\}\cap \{f^2=0\}\cap \{f^3=0\}\cap\{\chi=0\}\Big|_{\mathbb{C}^4}=\mathbb{C},
$$ 
which, as we will
explain, implies that the integral (\ref{ejemplo integral}) is not well
defined.  Replacing the $f^i$'s in (\ref{ejemplo integral}) we have an integral like in $\mathbb{C}^3$  
\begin{equation}
\int_{|z_i|=\varepsilon_i}\frac{\d z_1\wedge \d z_2 \wedge\d z_3}{ z_1
z_2 z_3}=(2\pi i)^3 ,
\end{equation}
where we have lost all the information about the cone, 
in fact we are in one chart. If we want to obtain global information,
we must write the integral in the following way
\begin{equation}\label{global integral}
I=\frac{1}{(2\pi i)}\int_{R}\frac{\d f^1\wedge \d f^2 \wedge\d f^3\wedge\d\chi}{ f^1 f^2 f^3\chi}=\frac{1}{(2\pi i)}
\int_{R}\frac{\d z_1\wedge \d z_2 \wedge\d
z_3\wedge\d(z_1z_2-z_3z_4)}{ z_1 z_2 z_3(z_1z_2-z_3z_4)},
\end{equation}
where $R$ is given by $R=\{Z\in\mathbb{C}^4 :
||f^i||=\varepsilon_i,|\chi|=\varepsilon\}$. Integrating first $z_1$
and then $z_2,z_3$ and $z_4$, we would obtain as a result $(2\pi i)^3$. Nevertheless, note that the pole $f^3$ is 
eliminated and it should be recovered from $\chi$.  This implies that the cycles $|f^3|=\varepsilon_3$ and 
$|\chi|=\varepsilon$ were mixed. To understand this better, let us
first integrate over the cycle $|f^3|=\varepsilon_3$ 
in (\ref{global integral}). We will obtain
\begin{eqnarray}
\frac{1}{(2\pi i)}
\int_{R}\frac{\d z_1\wedge \d z_2 \wedge\d z_3\wedge\d(z_1z_2-z_3z_4)}{ z_1 z_2 z_3(z_1z_2-z_3z_4)}&=&
\frac{1}{(2\pi i)}
\int_{R}\frac{\d z_3\wedge\d z_1\wedge \d z_2 \wedge(-z_3)\d z_4}{ z_1 z_2 z_3(z_1z_2-z_3z_4)}\no\\
&=&
\frac{-1}{(2\pi i)}
\int_{R}\frac{\d z_3\wedge\d z_1\wedge \d z_2 \wedge\d z_4}{ z_1 z_2 z_3(\frac{z_1z_2}{z_3}-z_4)}, 
\end{eqnarray}
so we get an infinite in the denominator and the integral is zero. Therefore we have a contraction and (\ref{ejemplo integral}) is not well defined for $f^{i}=z_i$, $i=1,2,3$. 
This contraction comes from the fact that the integral is not well
defined globally for those $f^i$'s, i.e changing the order in which
we compute the integral (\ref{global integral}) is equivalent to a change 
of chart in the cone.\\ 
In the pure spinor formalism the $f^{I}$'s, given by the constant spinors $C^I$'s (\ref{choice C}), have the same problem. Although we can not do 
the same trick as (\ref{global integral}), because  the constraints (\ref{5constrains1}) $\chi_a=0$ 
do not describe the whole pure spinor space, 
it can be useful to understand this more complicated problem. For the constrains $\chi_a=0$ we have 
\begin{eqnarray}
I&=&
\int_{{R}} \frac{(\d  f^1)\wedge...\wedge(\d  f^{11})\wedge \d(\chi_1)\wedge...\wedge \d(\chi_5)} {  f^1..
.  f^{11}(\chi_1)...(\chi_5)}
\end{eqnarray}
where $R$ goes around every pole. Integrating first by the cycle
$|\l^+|=\varepsilon$ we get an infinite in the denominator, just as in the previous example.

Now, changing $f^{3}=z_3$ by $f^3=z_3-z_4$ in the example of the cone $\chi=z_1z_2-z_3z_4=0$, we get the intersection 
$$
\{f^1=0\}\cap \{f^2=0\}\cap \{f^3=0\}\cap\{\chi=0\}\Big|_{\mathbb{C}^4}=\{0\}
$$
with multiplicity $m_{\{0\}}=2$, which comes from the equation $z_4^2=0$. So, we have the integral 
\begin{equation}\label{global integral2}
I=\frac{1}{(2\pi i)}\int_{R}\frac{\d f^1\wedge \d f^2 \wedge\d f^3\wedge\d\chi}{ f^1 f^2 f^3\chi}=\frac{1}{(2\pi i)}
\int_{R}\frac{\d z_1\wedge \d z_2 \wedge\d (z_3-z_4)\wedge\d(z_1z_2-z_3z_4)}{ z_1 z_2 (z_3-z_4)(z_1z_2-z_3z_4)}.  
\end{equation}
This integral does not have any problems and its result is the expected $(2\pi i)^{3} 2$ (which was explained in the sub-subsection \ref{The Projective Pure Spinor Degree} and matches with the Bezout theorem \cite{harris}).

\section{Proof of the Identity $[\d\tilde\l]=\d u_{12}\wedge...\wedge\d u_{45}.$}
\label{proof of identity}

Let us give again the statement that we want to proof. \\If $\tilde \l^{\a}$ is an element of the projective pure spinors space in 10
dimensions, i.e. if $\tilde\l^\a \in SO(10)/U(5)$, then the integration
measure $[\d\tilde\l]$ defined by \cite{pure spinor like twistors}
\begin{equation}\label{holomorphic measure}
[\d\tilde\l](\tilde\l\g^m)_{\a_1}(\tilde\l\g^n)_{\a_2}(\tilde\l\g^p)_{\a_3}(\g_{mnp})_{\a_4\a_5}=\frac{2^3}{10!}\e_{
\a_1...\a_5\b_1...\b_{11}} \d\tilde\l^{\b_1}\wedge...\wedge\d\tilde\l^{\b_{10}}
\tilde\l^{\b_{11}},
\end{equation}
written in the parametrization $\tilde\l^{\a}=(\tilde\l^{+}, \tilde\l_{ab}, \tilde\l^{a})=(1, u_{ab}, 
\frac{1}{8}\e^{abcde}u_{bc}u_{de})$ is
\begin{equation}
[\d\tilde\l]=\d u_{12}\wedge...\wedge\d u_{45}.
\end{equation}

{\it Proof}

Since $SO(10)/U(5)$ is a complex manifold we can write an anti-holomorphic
measure as \cite{humberto one loop}
\begin{equation}\label{antiholomorphic measure1}
[\d\tilde\lb](\tilde\lb\g^m)^{\a_1}(\tilde\lb\g^n)^{\a_2}(\tilde\lb\g^p)^{\a_3}(\g_{mnp})^{\a_4\a_5}=\frac{2^3}{10!}\e^{
\a_1...\a_5\b_1...\b_{11}} \d\tilde\lb_{\b_1}\wedge...\wedge\d\tilde\lb_{\b_{10}}
\tilde\lb_{\b_{11}},
\end{equation}
or in a more appropriate way as
\begin{equation}\label{antiholomorphic measure}
[\d\tilde\lb]=\frac{1}{2^3 5!10!}\frac{1}{(\tilde\l\tilde\lb)^{3}}(\tilde\l\g^m)_{\a_1}(\tilde\l\g^n)_{\a_2}(\tilde\l\g^p)_{\a_3}(\g_{mnp})_
{ \a_4\a_5}
\e^{\a_1...\a_5\de_1...\de_{11}}\d\tilde\lb_{\de_1}\wedge...\wedge\d\tilde\lb_{\de_{10}}\tilde\lb_{\de_{11}},
\end{equation}
where $\tilde\lb_{\a}=(\tilde\lb_{+},\tilde\lb^{ab},\tilde\lb_{a})=(1,\bar u^{ab},\frac{1}{8}\e_{abcde}\bar u^{bc}\bar
u^{de})$. From (\ref{holomorphic measure}) and (\ref{antiholomorphic measure})
it is simple to see that 
\begin{eqnarray}\label{demostration}
[\d\tilde\l]\wedge[\d\tilde\lb]&=&\frac{1}{5!(10!)^2}\frac{1}{(\tilde\l\tilde\lb)^{3}}
\e_{\a_1...\a_5\b_1...\b_{11}}\e^{\a_1...\a_5\de_1...\de_{11}}
\d\tilde\l^{\b_1}\wedge...\wedge\d\tilde\l^{\b_{10}}  \tilde\l^{\b_{11}} \wedge
\d\tilde\lb_{\de_1}\wedge...\wedge\d\tilde\lb_{
\de_{10}}\tilde\lb_{\de_{11}}\nonumber\\
&=&\frac{1}{(10!)^2}\frac{1}{(\tilde\l\tilde\lb)^{3}}
\de_{[\b_1}^{\de_1}\de_{\b_2}^{\de_2}...\de_{\b_{11}]}^{\de_{11}} \tilde\l^{\b_{11}}  \tilde\lb_{\de_{11}}
\d\tilde\l^{\b_1}\wedge...\wedge\d\tilde\l^{\b_{10}}   \wedge
\d\tilde\lb_{\de_1}\wedge...\wedge\d\tilde\lb_{
\de_{10}}\nonumber\\
&=&\frac{1}{10!}\frac{1}{(\tilde\l\tilde\lb)^{2}}
\d\tilde\l^{\b_1}\wedge...\wedge\d\tilde\l^{\b_{10}}   \wedge
\d\tilde\lb_{\b_1}\wedge...\wedge\d\tilde\lb_{
\b_{10}}\nonumber\\
&&-\frac{10}{10!}\frac{1}{(\tilde\l\tilde\lb)^{3}}
\d\tilde\l^{\b_1}\wedge...\wedge\d\tilde\l^{\b_{9}}\wedge\tilde\lb_{\a_1}\d\tilde\l^{\a_{1}}   \wedge
\d\tilde\lb_{\b_1}\wedge...\wedge\d\tilde\lb_{
\b_{9}}\wedge\tilde\l^{\a_2}\d\tilde\lb_{
\a_{2}}\nonumber\\
&=&-\frac{1}{10!}\(\frac{1}{(\tilde\l\tilde\lb)^{2}}
\p\pb(\tilde\l\tilde\lb)\wedge...\wedge \p\pb(\tilde\l\tilde\lb)\right.\nonumber\\
&&\left.-\frac{10}{(\tilde\l\tilde\lb)^{3}}
\p(\tilde\l\tilde\lb)\wedge\pb(\tilde\l\tilde\lb)\wedge\p\pb(\tilde\l\tilde\lb)\wedge...\wedge \p\pb(\tilde\l\tilde\lb)\)\nonumber\\
&=&\frac{1}{10!}\(i (\tilde\l\tilde\lb)^{8/10}\p\pb\ln (\tilde\l\tilde\lb)\)^{10}.
\end{eqnarray}
In \cite{humberto one loop} it was shown that 
\begin{equation}
\frac{\o^{10}}{10!}=\frac{\d u_{12}\wedge...\wedge\d u_{45}\wedge \d \bar
u^{12}\wedge...\wedge\d
\bar u^{45}}{(\tilde\l\tilde\lb)^{8}},
\end{equation}
where 
\begin{equation}\label{generator of cohomology}
\o=-\p\pb\ln(\tilde\l\tilde\lb) 
\end{equation}
and
$$
(\tilde\l\tilde\lb)=(1+\frac{1}{2}u_{ab}\bar
u^{ab}+\frac{1}{8^2}\e^{a_1b_1c_1d_1e_1}\e_{a_1b_2c_2d_2e_2}u_{b_1c_1}u_{d_1e_1}
\bar u^{b_2c_2}\bar u^{d_2e_2}).
$$
So, we have shown that 
\begin{equation}
[\d\tilde\l]=\exp(i\phi)\d u_{12}\wedge...\wedge\d u_{45}
\end{equation}
where $\phi\in\mathbb{R}$ is a constant. Since this phase factor does
not affect the amplitude we can set $\phi=0$ and thus the identity was proven $\blacksquare$



\begin{thebibliography}{99}

\bibitem{BerkovitsCQS}N. Berkovits, ``Super-Poincar\'e Covariant
Quantization of the Superstring,'' JHEP 04 (2000) 018, hep-th/0001035

\bibitem{Cartan} E. Cartan,
``Lecons sur la Theorie des Spineurs'', Hermann, Paris, 1937
\bibitem{Howe}
P.S. Howe, ``Pure spinor
lines in superspace and ten-dimensional supersymmetric theories,''
Phys. Lett. B258: 141 (1991).

\bibitem{nathan minimal pure spinor}
  N.~Berkovits,
  ``Multiloop amplitudes and vanishing theorems using the pure spinor
  formalism for the superstring,''
  JHEP {\bf 0409}, 047 (2004)
  [arXiv:hep-th/0406055].

\bibitem{nathan 2 loop}
  N.~Berkovits,
  ``Super-Poincare covariant two-loop superstring amplitudes,''
  JHEP {\bf 0601}, 005 (2006)
  [arXiv:hep-th/0503197].

\bibitem{Carlos and Nathan equivalence}
N. Berkovits and C.R. Mafra, ``Equivalence of two-loop superstring
amplitudes in the pure spinor and RNS formalism'', 
Phys. Rev. Lett. 96:011602 (2006), [arXiv:hep-th/0509234]

\bibitem{Carlos four pt one loop}
C.R. Mafra, ``Four-point one-loop amplitude computation in the pure
spinor formalism '', 
JHEP 0601:075 (2006), [arXiv:hep-th/0512052]


\bibitem{nathan topological}
  N.~Berkovits,
  ``Pure spinor formalism as an N = 2 topological string,''
  JHEP {\bf 0510}, 089 (2005)
  [arXiv:hep-th/0509120].

\bibitem{nathan nikita multiloops}
  N.~Berkovits and N.~Nekrasov,
  ``Multiloop superstring amplitudes from non-minimal pure spinor formalism,''
  JHEP {\bf 0612}, 029 (2006)
  [arXiv:hep-th/0609012].

\bibitem{Berkovits Mafra Anomaly}
N. Berkovits and C. Mafra, ``Some Superstring Amplitude Computations
with the Non-Minimal Pure Spinor Formalism,'' JHEP {\bf 0611} 079
(2006), [arXiv:hep-th/0607187]

\bibitem{mafra stahn}
C. Mafra and C. Stahn, ``The One-loop Open Superstring Massless
Five-point Amplitude with the Non-Minimal Pure Spinor Formalism'',
JHEP {\bf 0903} 126 (2009), [arXiv:0902.1539 [hep-th]]

\bibitem{grassi&vanhove}
P.A. Grassi and P. Vanhove, ``Higher-loop Amplitudes in the
non-minimal Pure Spinor Formalism'', JHEP {\bf 0905} 089 (2009),
[arXiv:0903.3903 [hep-th]]

\bibitem{nathan&yuri}
Y. Aisaka and N. Berkovits, ``Pure Spinor Vertex Operators in Siegel
Gauge and Loop Amplitude Regularization'', JHEP {\bf 0907} 062
(2009), [arXiv:0903.3443 [hep-th]]

\bibitem{beta gamma system}
  N.~A.~Nekrasov,
  ``Lectures on curved beta-gamma systems, pure spinors, and anomalies,''
  arXiv:hep-th/0511008.

\bibitem{Hughston}
L.P. Hughston, ``Applications of $SO(8)$ Spinors'', pp.253-287 in
Gravitation and Geometry: a volume in honor of Ivor Robinson (eds. W.
Rindler and A. Trautman), Bibliopolis, Naples (1987).

\bibitem{pure spinor like twistors}
  N.~Berkovits and S.~A.~Cherkis,
  ``Pure spinors are higher-dimensional twistors,''
  JHEP {\bf 0412}, 049 (2004)
  [arXiv:hep-th/0409243].

\bibitem{skenderis}
  J.~Hoogeveen and K.~Skenderis,
  ``Decoupling of unphysical states in the minimal pure spinor formalism I,''
  JHEP {\bf 1001}, 041 (2010)
  [arXiv:0906.3368 [hep-th]].

\bibitem{decouplingII}
N. Berkovits,  J.~Hoogeveen and K.~Skenderis,
  ``Decoupling of unphysical states in the minimal pure spinor
  formalism II,''
  JHEP {\bf 0909}, 035 (209)
  [arXiv:0906.3371 [hep-th]].

\bibitem{hoogeveen&skenderis}
J. Hoogeveen and K. Skenderis, ``BRST Quantization of the Pure Spinor
Superstring'', JHEP {\bf 0711} 081 (2007), [arXiv:0710.2598 [hep-th]].

\bibitem{dhokerperturbation}
 E.~D'Hoker and D.~H.~Phong,
 ``The Geometry of String Perturbation Theory,''
  Rev.\ Mod.\ Phys.\  {\bf 60}, 917 (1988).

\bibitem{yuri aldo nathan nikita}
  Y.~Aisaka, E.~A.~Arroyo, N.~Berkovits and N.~Nekrasov,
  ``Pure Spinor Partition Function and the Massive Superstring Spectrum,''
  JHEP {\bf 0808}, 050 (2008)
  [arXiv:0806.0584 [hep-th]].

\bibitem{harris}
Griffiths and Harris, 
``Principles of Algebraic Geometry'', [Wiley Classics Library Edition
Published 1994]

\bibitem{bott}
Raoul Bott and Loring W.Tu
``Differential Forms in Algebraic Topology'', [Springer-Verlag published 1982]

\bibitem{massey}
W.S. Massey, ``A Basic Course in Algebraic Topology'',
[Springer 1991]

\bibitem{humberto one loop}
  H.~Gomez,
  ``One-loop Superstring Amplitude From Integrals on Pure Spinors Space,''
  JHEP {\bf 0912}, 034 (2009)
  [arXiv:0910.3405 [hep-th]].

\bibitem{intersection}
William Fulton, ``Intersection theory'',
[Springer 1998]


\bibitem{humberto and carlos}
  H.~Gomez and C.~R.~Mafra,
  ``The Overall Coefficient of the Two-loop Superstring Amplitude Using Pure
  Spinors,''
  JHEP {\bf 1005}, 017 (2010)
  [arXiv:1003.0678 [hep-th]].

\bibitem{work in progress}
  Work in progress


  	


 
  
 




 
 
 





























\end{thebibliography}
\end{document}